\documentclass[aps,
prb,
twocolumn,
showpacs,
floatfix,
nobalancelastpage,
superscriptaddress,
notitlepage]{revtex4-2}
\usepackage{dcolumn}
\usepackage{bm}
\usepackage{soul}
\usepackage{mathrsfs,dsfont}
\usepackage{braket}
\usepackage{amsmath,amssymb,graphicx,float}
\usepackage{subfigure}
\usepackage{hyperref}
\hypersetup{colorlinks=true,citecolor=blue,urlcolor=blue,linkcolor=blue}
\usepackage{transparent}
\usepackage[export]{adjustbox}

\DeclareUnicodeCharacter{2009}{\,} 

\usepackage{braket}

\usepackage{tabularx}

\begin{document}

\title{Optical properties of orthorhombic germanium sulfide: Unveiling the Anisotropic Nature of Wannier Exciton}

\author{Mehdi Arfaoui}
\email{mehdi.arfaoui@fst.utm.tn}
\affiliation{Laboratoire de Physique de la Matière Condensée, Département de Physique, Faculté des Sciences de Tunis, Université Tunis El Manar, Campus Universitaire 1060 Tunis, Tunisia.}

%\alsoaffiliation[]{}
%\phone{}
%\fax{}
\author{Natalia Zawadzka}
\affiliation{Institute of Experimental Physics, Faculty of Physics, University of Warsaw, Warsaw, Poland.}
\author{Sabrine Ayari}
\affiliation{Laboratoire de Physique de l'Ecole normale sup\'erieure, ENS, Universit\'e
PSL, CNRS, Sorbonne Universit\'e, Universit\'e de Paris, 24 rue Lhomond, 75005 Paris, France.}
\author{Zhaolong Chen}
\affiliation{Institute for Functional Intelligent Material, National University of Singapore, 117575, Singapore.}
\affiliation{Department of Materials Science and Engineering, National University of Singapore, 117575, Singapore.}
\author{Kenji Watanabe}
\affiliation{Research Center for Electronic and Optical Materials, National Institute for Materials Science, 1-1 Namiki, Tsukuba 305-0044, Japan}
\author{Takashi Taniguchi}
\affiliation{Research Center for Materials Nanoarchitectonics, National Institute for Materials Science,  1-1 Namiki, Tsukuba 305-0044, Japan}
\author{Adam Babiński}
\affiliation{Institute of Experimental Physics, Faculty of Physics, University of Warsaw, Warsaw, Poland.}
\author{Maciej~Koperski}
\affiliation{Institute for Functional Intelligent Material, National University of Singapore, 117575, Singapore.}
\affiliation{Department of Materials Science and Engineering, National University of Singapore, 117575, Singapore.}
\author{Sihem Jaziri}
\affiliation{Laboratoire de Physique de la Matière Condensée, Département de Physique, Faculté des Sciences de Tunis, Université Tunis El Manar, Campus Universitaire 1060 Tunis, Tunisia.}
\author{Maciej R. Molas}
\email{maciej.molas@fuw.edu.pl}
\affiliation{Institute of Experimental Physics, Faculty of Physics, University of Warsaw, Warsaw, Poland.}

\begin{abstract}
To fully explore exciton-based applications and improve their performance, it is essential to understand the exciton behavior in anisotropic materials. Here, we investigate the optical properties of anisotropic excitons in GeS encapsulated by h-BN, using different approaches that combine polarization- and temperature-dependent photoluminescence (PL) measurements, \textit{ab initio} calculations, and effective mass approximation (EMA). Using the Bethe-Salpeter Equation (BSE) method, we found that the optical absorption spectra in GeS are significantly affected by the Coulomb interaction included in the BSE method, which shows the importance of excitonic effects besides it exhibits a significant dependence on the direction of polarization, revealing the anisotropic nature of bulk GeS. Combining \textit{ab initio} calculations and EMA methods, we investigate the quasi-hydrogenic exciton states and oscillator strength (OS) of GeS along the zigzag and armchair axes. We found that the anisotropy induces lifting of the degeneracy and mixing of the excitonic states in GeS, which results in highly nonhydrogenic features. Very good agreement with the experiment is observed.
\end{abstract}

\maketitle

%%%%%%%%%%%%%%%%%%%%%%%%%%%%%%%%%%5 Introduction
\section{Introduction}
The groundbreaking discovery of graphene and its unique properties has sparked further research on the development of other alternative layered and non-layered materials with various optical, electronic, and chemical properties~\cite{mas20112d,bhimanapati2015recent}. 
Recently, transition metal dichalcogenides (TMDs)~\cite{jariwala2014emerging, wang2012electronics} and black phosphorus (BP)~\cite{wu2021large, molas2021photoluminescence} have been extensively researched for their high carrier mobility~\cite{liu2014phosphorene, qiao2014high, podzorov2004high} and strong bound excitons~\cite{molas2021photoluminescence, ugeda2014giant, zhang2018determination,henriques2020excitons}. 
However, the instability of BP under air conditions limits its practical use~\cite{kim2019intrinsic,li2019recent,sadki2019oxidation}. 
An alternative class of materials with puckered honeycomb lattice, group-IV monochalcogenides MX (where M = Ge, Sn, or Pb and X = S, Se, or Te)~\cite{PhysRevB.92.085406,Lv2017TwodimensionalGM,wu2018few,yang2021plane} has emerged as a promising new class of layered van der Waals (vdW) semiconductors due to their benefits such as low toxicity~\cite{latiff2015toxicity}, high thermal stability~\cite{wiedemeier1977thermal, Liu_2018}, earth abundant~\cite{yang2021plane,feng2021interfacial}, excellent absorption energies observed for the visible frequency range~\cite{zawadzka2021anisotropic,ulaganathan2016high,wang2020sub, lan2015synthesis}, in addition to low thermal conductivities and significant anisotropic physical properties~\cite{zawadzka2021anisotropic, taniguchi1990core,tolloczko2020anisotropic,zhao2014ultralow,yang2021plane,guo2017thermoelectric}. 
Among group-IV monochalcogenides, germanium sulfide (GeS) is considered a promising material in optoelectronics applications due to its high optical absorption and optical band gap (BG) in the visible range, which can be effectively tuned by applying an external strain.
This allows for modulation of its emission wavelength~\cite{senske1978luminescence, wiley1980absorption,el2021layers,tan2017anisotropic, postorino2020interlayer,le2019strain}. 
GeS also exhibits high photosensitivity, broad spectral response, and giant piezoelectric because of its characteristic “puckered” symmetry~\cite{postorino2020interlayer,ulaganathan2016high,lan2015synthesis, fei2015giant}. 
Analogously to BP, the low symmetry orthorhombic crystal structure of GeS results in unique anisotropic optical, electronic, and vibrational properties along the zigzag (ZZ) and armchair (AC) axis~\cite{qin2016diverse, wu2016intrinsic,ho2017polarized,tan2017polarization,tan2017anisotropic}, which makes it suitable for large-scale applications in photovoltaic thermoelectric~\cite{shafique2017thermoelectric}, and optoelectronics~\cite{yang2021plane}. 
Polarization-resolved photoluminescence (PL), reflectance contrast (RC), and Raman scattering (RS) measurements performed over a wide temperature range in GeS confirmed its anisotropies characteristics~\cite{tolloczko2020anisotropic, zawadzka2021anisotropic, oliva2020valley}. 
This makes it a highly promising material for the development of polarization-sensitive photodetectors. 
It is important to examine the many-body interactions present in this material, particularly with regard to the excitonic effect~\cite{lan2015synthesis, postorino2020interlayer,yang2020intrinsic,ho2017polarized, tan2017polarization,tan2017anisotropic,li2016germanium}. Despite tremendous progress in the study of GeS, many of its fundamental optical anisotropies properties remain unknown~\cite{zawadzka2021anisotropic}.  

In this work,  we investigate the optical response of anisotropic excitons in GeS encapsulated by hexagonal BN ($h$-BN) flakes using a comprehensive approach that combines polarization- and temperature-dependent PL techniques, \textit{ab initio} calculations, and effective mass approximation (EMA). 
We employ a state-of-the-art approach that utilizing the density functional theory (DFT) and the Quasiparticle (QP) correction GW to determine the electronic band structure (BS) of bulk GeS, and the anisotropic effective mass of electrons and holes in different crystal directions.
Furthermore, using both the independent particle approximation (IPA) and the GW+Bethe Salpeter Equation (BSE)~\cite{PhysRev.84.1232, PhysRevB.62.4927}, we study the dielectric functions for light polarized along the in-plane ($x$-$y$) and out-of-plane ($z$) directions. 
Indeed, the obtained result from BSE modifies the optical absorption spectra in bulk GeS. 
Thus, it shows the signature of excitonic effects on the optical response.  
To evaluate the different optical selection rules, we calculate the direct interband optical transition matrix elements (OME) in different crystal directions, as well as the percentage of the atomic orbital contribution to the valence and conduction bands at specific $\mathbf{k}$-points of the Brillouin zone (BZ).  
We were also able to predict the quasi-hydrogenic exciton states and the oscillator strength (OS) of the exciton along both the ZZ and AC directions by integrating \textit{ab initio} calculations with the EMA. 
In fact, the anisotropic effect on the exciton characteristics in GeS (e.g., exciton binding energy (BE) and the spatial extension of exciton wavefunction) is controlled by its reduced mass and dielectric constant.
Unlike the isotropic hydrogenic model, the anisotropy lifts the degeneracy of the exciton states, which have the same principal quantum number but different radial and angular quantum numbers.
This increases in the number of allowed dipole transitions that can be probed by terahertz radiation, providing new ways of controlling device emissions.

This article is organized as follows: 
In Section~\ref{sec.exp}, we present the experimental results obtained for a GeS flake encapsulated by $h$-BN, which include its measured polarization- and temperature- dependent PL spectra. 
In Section~\ref{DFTGW}, to gain a deeper understanding of the excitons behavior in anisotropic materials, we used theoretical methods (DFT and GW+BSE) to calculate the electronic BS of GeS, taking into account the excitonic effect. 
In Section~\ref{efa}, we analyze the anisotropic behavior of GeS using the EMA. By varying the relevant parameters, we control the degree of anisotropy across multiple cases, ranging from isotropic to anisotropic. Our calculations of the excitonic BE and OS are performed for each case.
Overall, our study of GeS as a prototype system for anisotropic layered materials provides an understanding of the optical response of anisotropic excitons and can be tailored for other three-dimensional (3D) anisotropic materials.
%TC:endignore

%%%%%%%%%%%%%%%%%%%%%%%%%%%%%%%Section 2 Experimental results
\section{Experimental Results}\label{sec.exp}

\begin{figure*}[]
    \centering
    \includegraphics[width=0.9\textwidth]{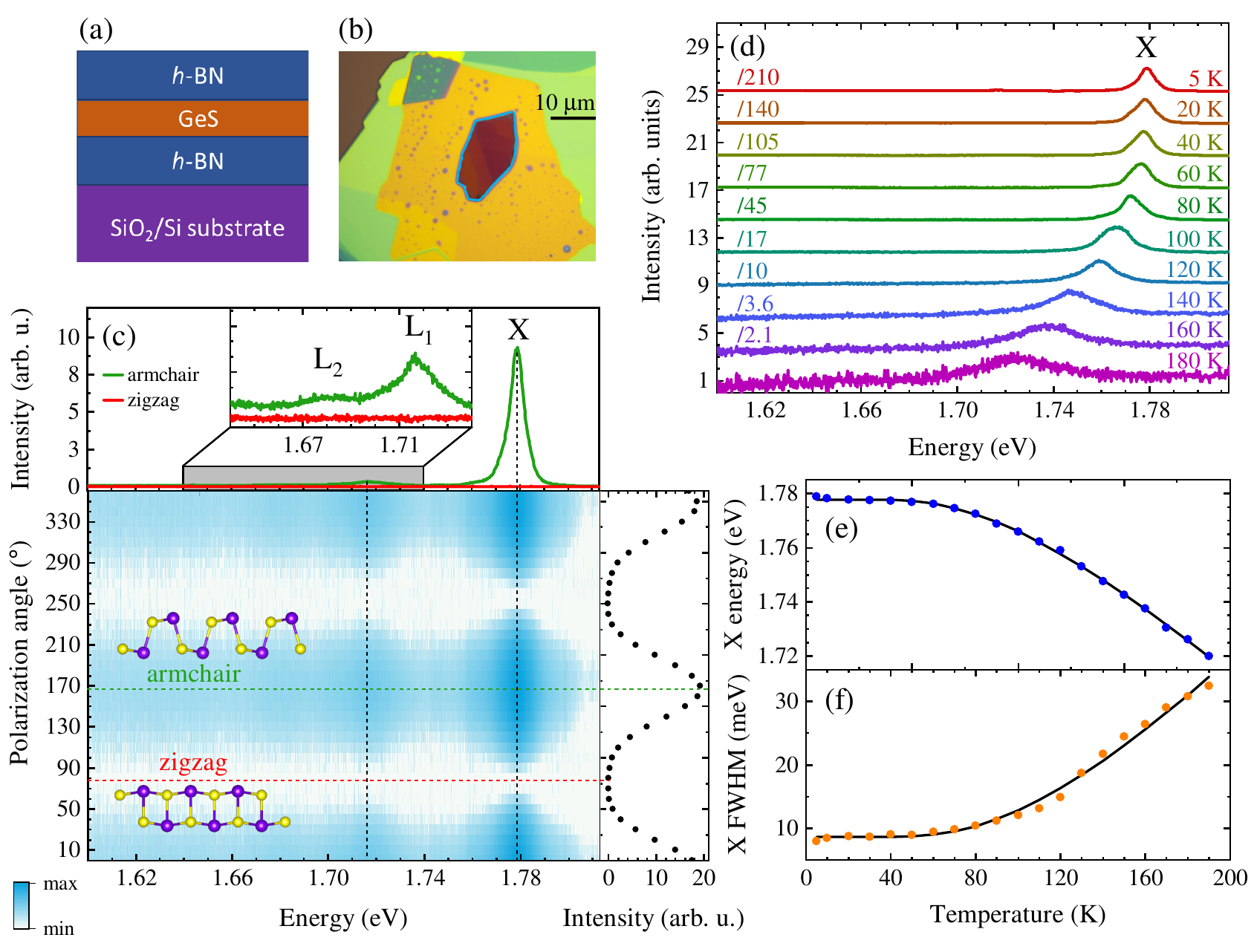}
    \caption{
    (a) Side-view scheme and (b) optical image of the investigated GeS encapsulated by $h$-BN flakes.  
    (c) False-colour map of the low temperature ($T$=5~K) polarization-resolved PL spectra measured on the GeS flake under excitation of 1.88~eV.
    Note that the intensity scale is logarithmic.
    The crystal structures along the ZZ and AC directions are drawn on top of the map.
    Top panel shows the corresponding PL spectra detected at AC and ZZ directions.
    Right panel demonstrates the integrated PL intensity as a function of detection angle. 
    (d) The corresponding temperature-dependent PL spectra measured on the GeS flake with 1.88 eV laser light excitation.
     The spectra are vertically shifted and are divided by scaling factors for clarity.
    The determined (e) energy and (f) full width at half maximum (FWHM) of the neutral exciton (X) line.
    The circles represent the experimental results while the curves are fits to the data obtained using Eq.~\ref{eq:odonnell} and \ref{eq:rudin}.     }
    \label{fig:res1}
\end{figure*}

To prevent degradation of the GeS flake, we encapsulated the thick GeS flake with a thickness of about $50$~nm in $h$-BN flakes, see Methods for details.
The side-view scheme and optical image of the investigated GeS are presented in Figure~\ref{fig:res1}(a) and (b).
First, we examine the polarization evolution of the PL spectra measured on the studied sample under the 1.88~eV excitation, see Figure~\ref{fig:res1}(c).
It is observed that the GeS emission is linearly polarized along the AC direction, whereas the PL signal is absent in the ZZ direction.
It is a hallmark of the anisotropic optical response of the GeS. 
The upper panel of Figure~\ref{fig:res1}(c) presents the low-temperature (T=$5$~K) PL spectra of GeS measured in two polarizations corresponding to the AC and ZZ orientations. 
The PL spectrum consists of three emission lines, denoted X, L$_1$, and L$_2$. 
In contrast, the corresponding polarization-resolved RC spectra, shown in Section~A of the Supporting Information (SI), consist of a single resonance, which energy coincides with the X emission line and is also polarized along the AC direction.
We can certainly attribute the X line to the free neutral exciton~\cite{zawadzka2021anisotropic}, while the assignment of the L$_1$ and L$_2$ peaks is more questionable, and hence we denoted them as localized excitons, see Section~A and B of SI for more details.
The measured shape of the PL spectrum with a rather low intensity of the L$_1$ and L$_2$ compared to that of X is different from those reported in Ref.~\cite{zawadzka2021anisotropic}.
It may suggest that the hBN encapsulation plays a similar role as for MoS$_2$ MLs, leading to complete quenching of the defect-related emission measured at liquid helium temperature~\cite{cadiz2017}.
Interestingly, the relative intensity of the localized and free excitons strongly depends on the excitation energy, see Section~A of the SI for details.
Figure~\ref{fig:res1}(d), presents the temperature evolution of the PL spectra measured from 5~K to 190~K. 
Our measurements indicate that with increasing temperature, the PL intensity of the exciton decreases until it completely vanishes at 200~K. 
This result suggests that the BE of the exciton, associated with its activation energy, is on the order of several meV.
As the L$_1$ and L$_2$ intensities are extremely small, we used different excitation energy, i.e., 2.41~eV, to investigate their temperature dependence (see Section~A in SI for details).
We found that at temperatures of approximately 70~K, the measured PL spectrum is contributed only by the neutral exciton line. 
To investigate in detail the temperature evolution of the X line, we deconvolute it using Gaussian.
The X energy experiences a redshift as the temperature increases from 5~K to 190~K, see Figure~\ref{fig:res1}(e). 
This type of evolution, characteristic of many semiconductors, can be expressed by the relation proposed by O'Donnell $et$ $al.$,~\cite{odonnell1991} which describes the temperature dependence of the BG in terms of an average phonon energy $<\hbar \omega>$ and reads
\begin{equation}
\label{eq:odonnell}
E_{bg} (T) = E_g - S <\hbar \omega> [\coth(<\hbar \omega>/2k_{\textrm{B}}T)-1],
\end{equation}
where $E_g$ stands for the BG at absolute zero temperature, $S$ is the coupling constant, and $k_{\textrm{B}}$ denotes the Boltzmann constant. 
As we can see that the used relation correctly reproduces the temperature evolution of the X energy, we conclude that its BE does not depend on the temperature.
The determined $E_g$ and $S$ values are of about 1.778~eV and 4.3, respectively.
The average phonon energy $<\hbar \omega>$ is found to be around 26~meV, which is close to the high density of phonon states around 28~meV.\cite{zawadzka2021anisotropic}
Panel (f) of Figure~\ref{fig:res1} presents the temperature evolution of the X linewidth.
As the temperature increases, the carriers have more thermal energy and move more rapidly, leading to a greater distribution of carrier velocities.
In semiconductors, such an evolution can be described by so-called Rudin's relation,~\cite{rudin1990} which is given by
\begin{equation}
\label{eq:rudin}
\gamma (T) = \gamma_0 + \sigma T + \gamma' \frac{1}{exp^{\hbar \omega / k T}- 1},
\end{equation}
where $\gamma_0$ denotes the broadening of a given spectral line at 0 K, the term linear in temperature ($\sigma$) quantifies the interaction of excitons with acoustic phonons (of negligible meaning for the present work), $\gamma' $ arises from the interaction of excitons with LO phonons, and $\hbar \omega$ is the LO phonon energy.
As can be appreciated in Figure~\ref{fig:res1}(f), Rudin's relation reproduces the experimental data quite well.
Due to the anisotropic structure of GeS, we consider $\hbar \omega$ as a free-fitting parameter in our analysis.
The determined values of $\gamma_0$ and $\gamma' $ correspond to about 8.5~meV and 0.13.
The fitted $\hbar \omega$ parameter is around 30~meV, which is close to the energy of the A$_\textrm{g}$ mode (31~meV at $T$=5~K).\cite{zawadzka2021anisotropic}

\begin{figure*}[!t]
    \centering
    \includegraphics[width=0.9\textwidth]{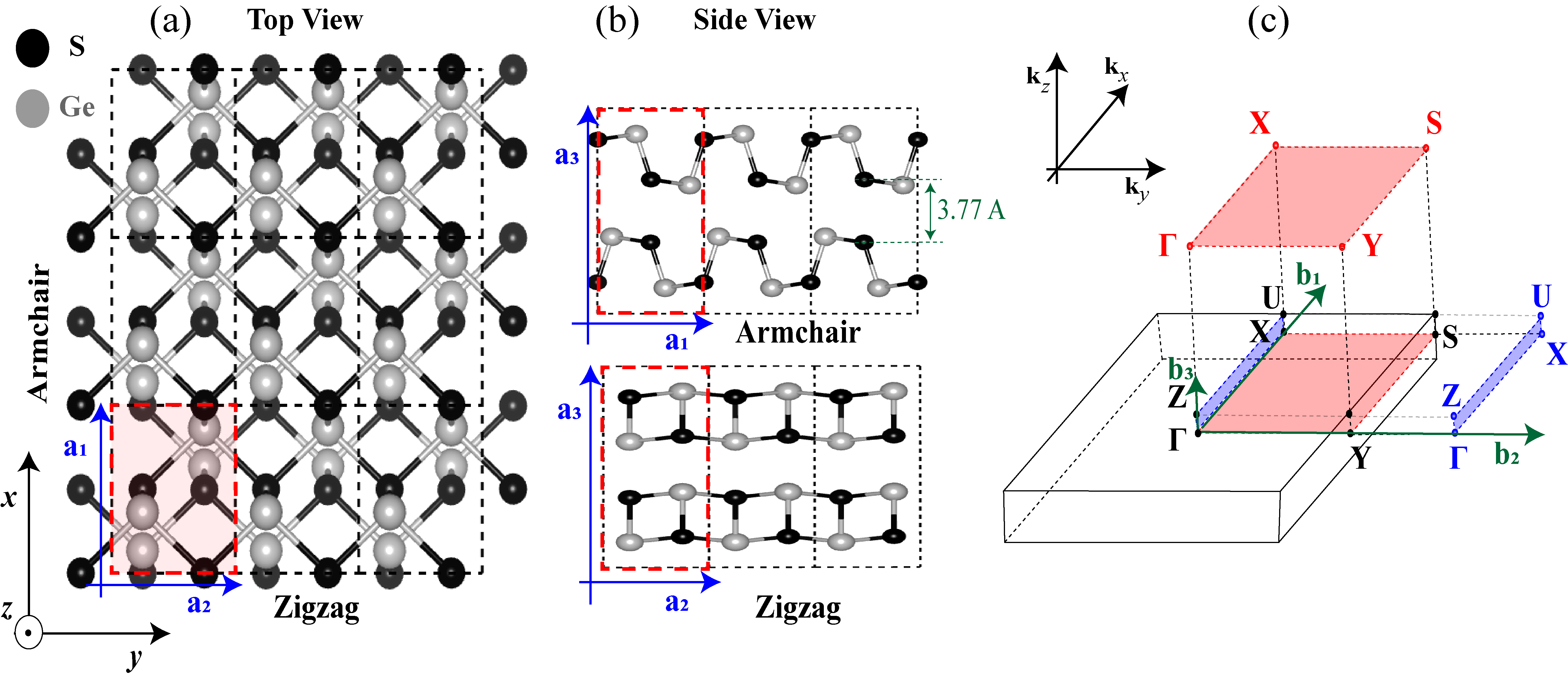}
    \caption{\textbf{Illustration of atomic structure for bulk GeS}. (a) Top view of the atomic structure of Bulk GeS in a $3$×$3$×$1$ supercell. (b) Side views of bulk GeS along the AC and ZZ directions, respectively. (c) The bulk first BZ of orthorhombic Bravais lattice and their projected surfaces. The gray and black spheres stand for the Ge and S atoms, respectively. The unit cell is indicated by the dashed red rectangle in (a, b).
    }
    \label{fig:structure}
\end{figure*}

The main objective of the present work is to construct a theoretical framework to analyze the aforementioned optical characteristics of free neutral excitons in GeS.
The free exciton is identified by its distinct spectral peak located around  $1.78$~eV in the PL spectra of the GeS flake with a thickness of $60$~nm at $T$=5~K.
The low-temperature conditions stabilize and clearly define the excitonic states, yielding an accurate assessment of their spectral properties. 
The reduced thermal excitation also enhances the analysis of GeS's intrinsic optical properties.

%%%%%%%%%%%%%%%%%%%%%%%%%%%%%%%%%%%%%% Section 3 First-Principles Calculations
\section{First-Principles Calculations of the Quasi-Particle and Excitonic Effect in GeS}\label{DFTGW}

GeS is a layered material that has an orthorhombic structure and belongs to the \textit{Pnma} ($D^{16}_{2h}$) space group~\cite{wiedemeier1977thermal}. 
This material crystallizes in double layers. 
The unit cell contains eight atoms organized in two adjacent double layers. 
The puckered honeycomb lattice of GeS has an anisotropic crystal structure characterized by the two orthogonal AC(x) and ZZ (y) directions, as indicated in Figure~\ref{fig:structure}.
The orthorhombic Bravais lattice of GeS can be specified by giving three primitive lattice vectors: $\mathbf{a}_1= a \hat{x}$, $\mathbf{a}_2= b \hat{y}$ and $\mathbf{a}_3= c \hat{z}$ and the reciprocal primitive lattice vectors are spanned by: $\mathbf{b}_1= \frac{2\pi}{a} \hat{k}_x$, $\mathbf{b}_2= \frac{2\pi}{b} \hat{k}_y$ and $\mathbf{b}_3= \frac{2\pi}{c} \hat{k}_z$.
$\hat{x}$, $\hat{y}$ and $\hat{z}$ are the unit vectors in the directions of the $x$-axis, the $y$-axis, and the $z$-axis, respectively.

In Table~\ref{tab:relaxed_parameter}, we present the final relaxed lattice parameters and compare them to previously reported experimental and theoretical data. All calculations are performed with the optimized lattice parameters.
\begin{table}[h]
    \centering
    \caption{Lattice parameters estimated at the DFT level while accounting for the vdW interaction are compared with experiment and recent theoretical results.}
    \renewcommand{\arraystretch}{1.3}
    \begin{tabular}{lccc}
        \hline
        References & a ($\AA$) & b ($\AA$) & c ($\AA$)\\
        \hline
        G.Ding \textit{et al.}~\cite{ding2015high} & 4.74 & 3.67 & 10.64 \\
        S.Hao \textit{et al.}~\cite{hao2016computational} & 4.44 & 3.67 & 10.76 \\
        T.Grandke \textit{et al.}~\cite{PhysRevB.16.832} & 4.30 & 3.64 & 10.47 \\
        Our work & 4.45 & 3.76 & 10.77 \\
        \hline
        \end{tabular}
    \label{tab:relaxed_parameter}
\end{table}

\begin{figure*}[]
    \centering
    \includegraphics[width=\textwidth]{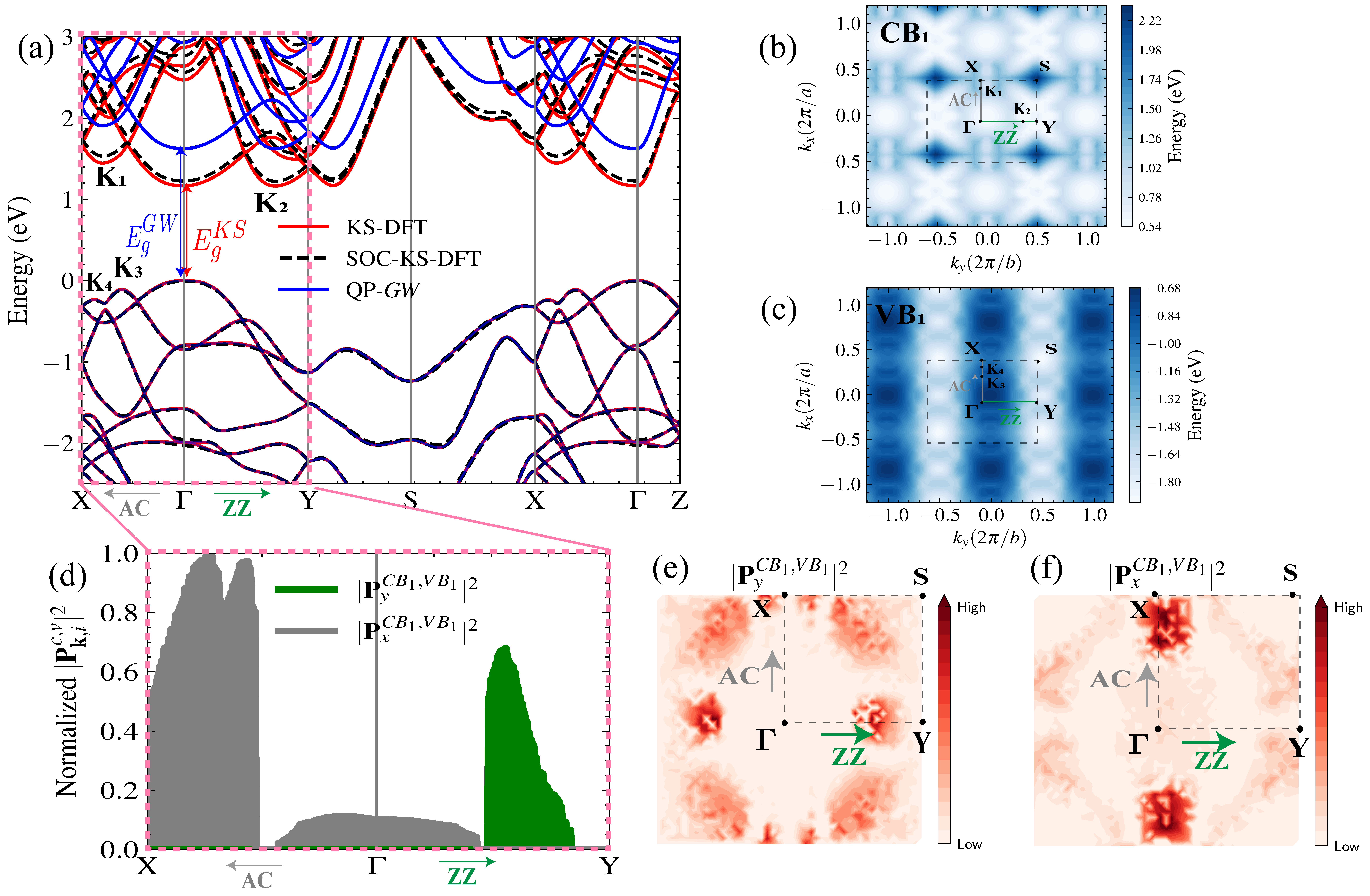}
    \caption{\textbf{Electronic properties of GeS from \textit{ab initio} calculations}. (a) Electronic BS of bulk GeS calculated by generalized gradient approximation (GGA) without spin-orbit coupling (red), with spin-orbit coupling (black dashed) and evGW (blue) methods. (b) and (c) are the $2$D BSs of the topmost valence band and bottom most conduction band, respectively. The global maximum (minimum) of VBM (CBM) is located at the AC direction $\Gamma X$ path of the BZ.
    (d) The interband transitions of OME between  $\boldsymbol{v}\rightarrow \boldsymbol{c}$ bands. (e) Color contour of the dipole OS distribution in the first BZ for light polarized in AC (e) and ZZ (f) directions, derived using DFT. Darker color indicates a larger dipole strength.}
    \label{fig:bs}
\end{figure*}

In Figure~\ref{fig:bs}~(a), we plot the BS of the bulk GeS using DFT with (black dashed curve) and without (red curve) spin-orbit interaction. 
The BS of GeS is not significantly impacted by the relativistic correction effect. 
Therefore, we disregard this in our calculation of many-body simulations. 
We found that the bulk GeS is a semiconductor with a BG of $1.23$~eV at the $\Gamma$ point. 
The indirect gap is only $3$~meV in energy higher than the direct one. 
Our results are consistent with previous DFT calculations~\cite{hao2016computational,ding2015high}. 
As we can see, the DFT direct BG (red curve) has been significantly underestimated because of the BG problem with the DFT Kohn-Sham (KS) approach.  
To overcome this issue, we calculate the relative QP BS using the perturbative method GW (blue curve). 
In fact, G and W are constructed from the KS-wavefunctions, and the perturbatively corrected KS-eigenvalues. 
The single-shot $G_0W_0$ gives an insufficiently small BG when compared to the experimental one. 
To solve this, we use the self-consistency of GW on eigenvalues only (evGW), see Section~D in the SI for details. 
At this theoretical level, the BG increases to $1.78$~eV with the QP correction of $0.55$~eV at $\Gamma$. 
Our results are in excellent agreement with the experimental bulk GeS BG energies extrapolated to 0~K given in Ref.~\cite{WILEY1975355, JPSJ.13.559}, as well as with previous theoretical calculations with GW in Ref.~\cite{ma13163568, PhysRevB.87.245312}. 
In addition, the QP corrections are slightly dispersed with respect to the KS-eigenenergies. 
In fact, we found that by fitting the QP corrections data to a linear curve, the conduction and valence bands are slightly stretched by $7$\% and $1$\%, respectively. 
The band profiles for the lowest conduction band and the highest valence band were also extracted from DFT and are shown in Figure~\ref{fig:bs}(b) and  Figure~\ref{fig:bs}(c), respectively. 
The valence band has three maximum locations in the AC direction: $\mathbf{K}_4$, $\mathbf{K}_3$, and $\Gamma$, with the highest maximum located at the $\Gamma$-point. 
The conduction band has a minimum located at the $\Gamma$-point.

\sloppy
In Figures~\ref{fig:bs}(d-f), we calculate the normalized squared OME $\mathbf{M}_{i}^{c,v}(\mathbf{k}) = \big\vert \mathbf{P}_{i}^{c,v} \big\vert^{2}=\left|\left\langle u_{v,\mathbf{k}}\middle| \mathbf{p}_i\middle|u_{c,\mathbf{k}}\right\rangle\right|^2$ between the top valence band and the bottom conduction band along a specific direction $i$. 
Here $u_{v (c),\mathbf{k}}$ is the single particle Bloch function of the valence (v) and conduction band (c) obtained by the DFT-KS calculation for the wavevector $\mathbf{k}$. ${\mathbf{p}}_i$ is the momentum operator along the $i$ direction. 
$\mathbf{M}_{i}^{c,v}(\mathbf{k})$ measures the $\mathbf{k}$-dependent optical strength of the c-v interband transition. 
Furthermore, the OME contains all the symmetry-imposed selection rules. 
Figures~\ref{fig:bs}(e) and (f) show that the $\mathbf{k}$-resolved OME for light polarized in the AC and ZZ directions exhibits distinct responses, which reveal the anisotropic nature of GeS. 
Of particular interest are the valleys located in $\mathbf{K}_1$, $\mathbf{K}_2$, $\mathbf{K}_3$, $\mathbf{K}_4$, and $\Gamma$ in the $\Gamma X$ and $\Gamma Y$  directions, where the allowed OME is significantly higher in $\Gamma X$ for light polarized in the AC-direction. 
Additionally, the OME are not allowed in the vicinity of $\Gamma$-point for light polarized in the ZZ-direction. 
By calculating the direct interband OME, we calculate the linear optical characteristics that can be derived from the complex dielectric tensor $\varepsilon_{i,j}(\omega, \mathbf{q})=\varepsilon_{1_{i,j}}(\omega,\mathbf{q})+i \varepsilon_{2_{i,j}}(\omega,\mathbf{q})$ for $\mathbf{q}\rightarrow 0$,  where $\varepsilon_{1_{i,j}}(\omega, \mathbf{q})$ and $\varepsilon_{2_{i,j}}(\omega, \mathbf{q})$ are the real and imaginary parts of the dielectric tensor, see Section~C and D of the SI. 
$\hbar\omega$ is the photon energy, $\mathbf{q}$ is the photon wavevector and $i,j$= $x$, $y$, or $z$ are the subscripts that correspond to the Cartesian directions.

\begin{figure}[]
    \centering  
    \includegraphics[width=\linewidth]{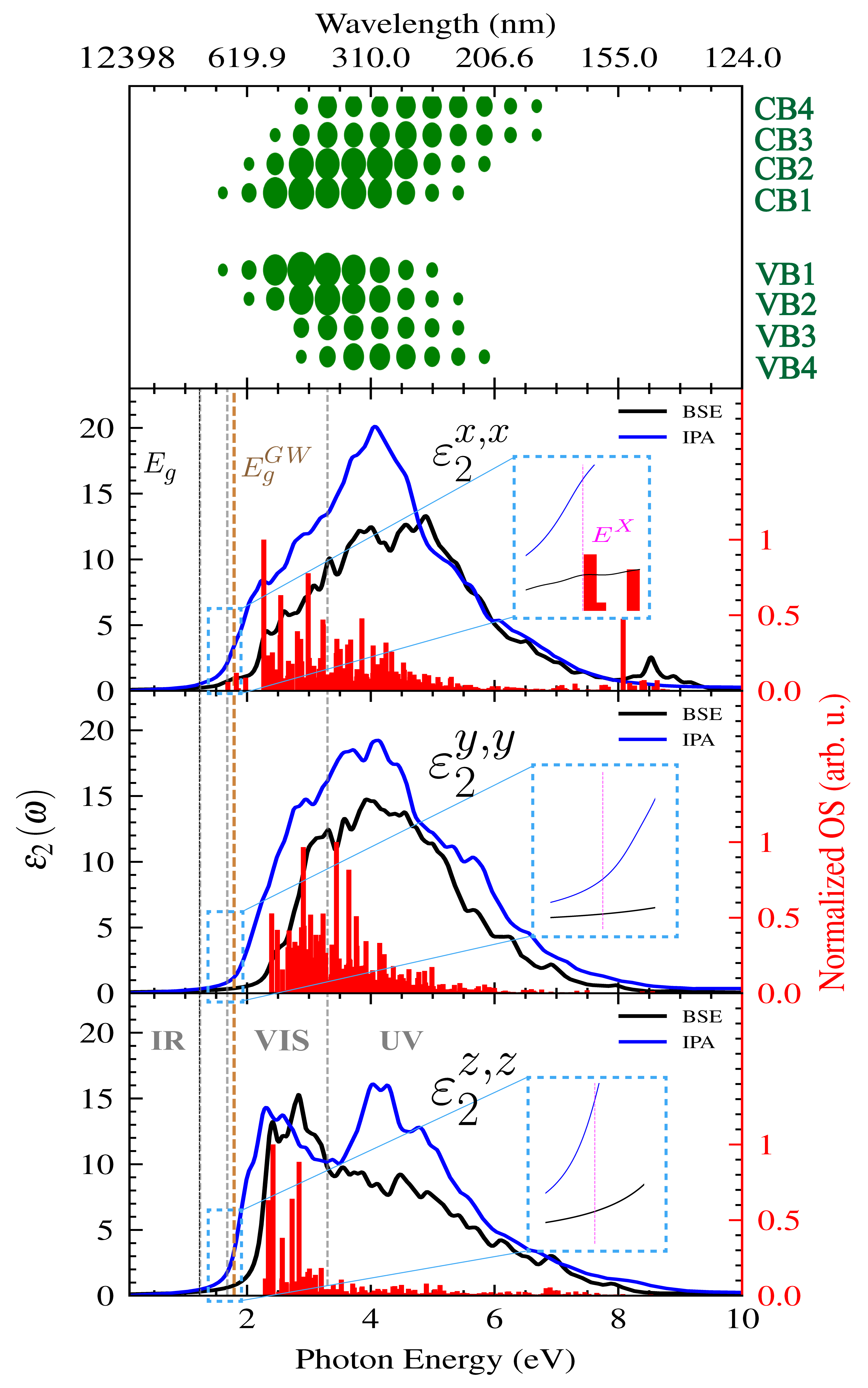}
    \caption{\textbf{Optical properties of GeS from \textit{ab initio} calculations}. Imaginary part of the dielectric functions calculated using IPA and GW+BSE methods, with light polarized in the $x$, $y$, and $z$ directions, respectively. 
    The red vertical bars represent the normalized OS (arb. units). The DFT and GW BGs are denoted by the black and orange dashed vertical lines, respectively. $E^X=1.77$ eV denotes the first bright exciton energy. The size of each green circle in the upper panel corresponds to the value of $\varepsilon_2 (\omega)$.}
    \label{fig:dielectric}
\end{figure}

In Figure~\ref{fig:dielectric}, we plot the three components of the imaginary part of the dielectric function for linear light polarized along the axis of the AC ($x$), ZZ ($y$) and perpendicular ($z$) directions to the atomic planes for bulk GeS. In fact, Figure~\ref{fig:dielectric} shows that $\varepsilon_2 (\omega)$ strongly depends on the direction of polarization. 
This behavior reveals the anisotropic character of bulk GeS. 
The light absorption of GeS has a wide range from near-infrared to near-ultraviolet light ($1\sim9$~eV). 
However, compared to the experimental results, the exciton peak does not appear in Figure~\ref{fig:dielectric} with the IPA method (blue line). 
This result is expected since the IPA method does not include the electron-hole (\textit{e-h}) interaction. 
To account for the \textit{e-h} interaction, we plot, in Figure~\ref{fig:dielectric} (black line), the dielectric function using the GW + BSE method. 
A detailed description of the GW + BSE method can be found in Section~D of the SI. 
In fact, compared with the IPA results, the Coulomb interaction included in the BSE method modifies the optical absorption spectra in bulk GeS as well as a redistribution of OS, implying that excitonic effects have an important influence on optical properties. 
The detected low-energy peak in the BSE spectrum is attributed to the presence of a bound exciton. 
In comparison to the experimental results, the first bright exciton peak is well reproduced by our calculation. 
In addition, our results show that GeS has remarkable optical anisotropy, as evidenced by the presence of an excitonic peak only for light polarized along the AC direction. 
This intriguing observation aligns very well with our experimental results in Figure~\ref{fig:res1}. 
The first bright exciton BE is $\sim 10$~meV. 
Our experimental measurements of temperature-dependent PL spectra also support this result. In Section~A of the SI, we estimated the exciton BE to be $\sim11$~meV under $2.41$~eV excitation, and this value can slightly increase ($\sim16$~meV) under resonant excitation at $1.88$~eV.
This BE is also comparable to those of bulk materials such as BP ($30$~meV)~\cite{PhysRevB.89.235319}, MoS$_2$ ($25$~meV)~\cite{PhysRevB.85.205302}, wurtzite GaN ($21$~meV)~\cite{PhysRevB.54.16369} and GaAs quantum well ($\sim$10 meV)~\cite{belov2017binding}. 
Moreover, the results in the upper panel of Figure~\ref{fig:dielectric}, show the contributions of the occupied and unoccupied bands to $\varepsilon_2$, where each band is labeled with a distinct index, $i.e.$ $VB_{1}\rightarrow VB_{4}$ for the occupied bands and $CB_{1}\rightarrow CB_{4}$ for the unoccupied bands. 
In particular, the size of each green circle is proportional to the value of $\varepsilon_2 (\omega)$. 
Indeed, the analysis reveals that the first peak in $\varepsilon_2 (\omega)$ for the AC direction is predominantly formed by the direct optical transition between the occupied band $VB_1$ and the unoccupied band $CB_1$. 
Note that we neglect the phonon-assisted related optical absorption, which requires more computational resources and is beyond the scope of this work. 

\begin{figure}[]
    \centering
    \includegraphics[width=\linewidth]{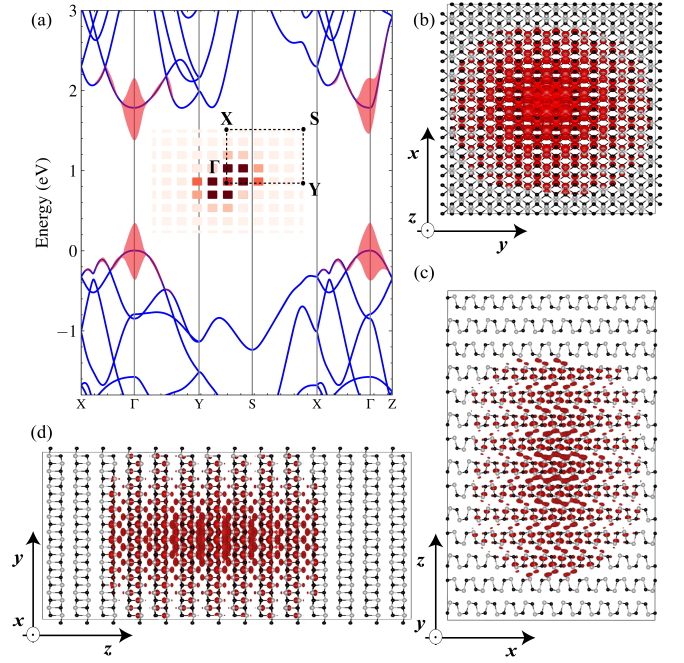}
    \caption{(a) Excitonic weights showing the most important electronic transitions along the high symmetry points of the BZ of GeS. The electronic transitions representing exciton weights are projected onto the ground-state electronic dispersion, which are depicted by red shade. The  colored  inset  shows  the  $2$D  projected  exciton wavefunction distribution in $\mathbf{k}$-space. Both the excitonic weights and $\mathbf{k}$-space wavefunction show that the first bright exciton originates from direct optical transitions in the vicinity of  $\mathbf{\Gamma}$ valley along $\mathbf{\Gamma} \mathbf{X}$, $\mathbf{\Gamma} \mathbf{Y}$ and $\mathbf{\Gamma}\mathbf{Z}$. (b, c, d) Normalized squared exciton wavefunction of first bright exciton in GeS using BSE for incident light polarized along the AC direction. We fix the position of the hole near the S atom. Side view on the bottom, top view on the top.}
    \label{fig:weight}
\end{figure}

The red shades in Figure~\ref{fig:weight}(a) show the first bright exciton weight originating from the vertical interband transition, which contributes to exciton formation and, as a result, to absorption spectra. 
The inset characterizes the transitions in the first reciprocal BZ. 
In GeS, it is mainly contributed by transitions in the nearby $\mathbf{\Gamma}$ valley along $\mathbf{\Gamma} \mathbf{X}$, $\mathbf{\Gamma} \mathbf{Y}$ and $\mathbf{\Gamma}\mathbf{Z}$. 
The valence band minimum (VBM) along these directions is occupied mainly by the $p$- and $s$- orbitals of the Ge and S atoms.
The conduction band maximum (CBM) is populated by $p$- and $s$- orbitals of the Ge atom and the S atom, see Section~E of the SI.
In order to explain the anisotropic behavior shown in Figure~\ref{fig:dielectric} and Figure~\ref{fig:weight}, we evaluate the different optical selection rules. 
In fact, in one-photon spectroscopy, the transition dipole selection rules have to satisfy two conditions: i) the change in angular momentum between the valence and conduction states should satisfy ($\Delta \ell= \ell-\ell^{\prime}= \pm 1$) and ii) since the parity of the momentum operator is odd, the conduction and valence bands should have opposite parity in the $i$ direction and the same parity in other directions. 
For incident light polarized along the AC direction, the dipole transitions are allowed for $p_{x} \longleftrightarrow s$, $p_{x}\longleftrightarrow d_{x^{2}-y^{2}}$. $p_{x}\longleftrightarrow d_{z^{2}}$, and $p_{y(z)}\longleftrightarrow d_{xy(zx)}$, 
Furthermore, the allowed transitions for the light polarized along the ZZ direction are $p_{y}\longleftrightarrow s$; $p_{y}\longleftrightarrow d_{x^{2}-y^{2}}$; $p_{y}\longleftrightarrow d_{z^{2}}$ and $p_{x(z)}\longleftrightarrow d_{xy(zy)}$.
For bulk GeS, we calculate in Table~\ref{tab:dsr}, the percentage of the contribution of the atomic orbital to the valence and the conduction wavefunctions at a particular $\mathbf{k}$-point. 
We can clearly see that the $d$-orbitals do not contribute to the interband transition between CBM and VBM. 
Thus, the optical absorption for polarization in the AC (ZZ) direction, which is related to the interband transition from VBM to CBM along the $\Gamma X$ ($\Gamma Y$) direction, occurs only from $p_{x(y)} \longleftrightarrow s$ transitions. 
The projected OME in the AC and ZZ directions (see Figure~\ref{fig:bs}(d)), shows that the allowed interband transition in the $\Gamma-X$ direction is more significant than the transition in $\Gamma-Y$. 
Indeed, in the vicinity of the high-symmetry $\Gamma$ point, $\mathbf{M}_{y}^{c,v}(\mathbf{k})$ vanish and only $\mathbf{M}_{x}^{c,v}(\mathbf{k})$ contribute to this interband transition, which may explain the strong signal of the exciton peak in the AC direction. 
Thus, bulk GeS strongly absorbs AC-polarized light with an energy $E^X$, and it is almost transparent along the ZZ direction for the same energy.
This phenomenon is the result of selection rules associated with the symmetries of this anisotropic material.

\begin{table}[]
    \centering
    %\small\setlength\extrarowheight{1ex}
    \caption{PDOS of the VBM and CBM band wavefunctions at the special $\mathbf{k}$-points. The percent contribution from each atomic orbital to these wavefunctions are listed.}
    \resizebox{\columnwidth}{!}{%
    \renewcommand{\arraystretch}{1.2}
    \begin{tabular}{cccccccccccccccc}
 \hline
 \hline
 $\text{Direction}$&$\mathbf{k}$&$ \text { State } $&  $\text{S atom} $&  &  & & $\text{Ge atom} $&  & &  & &&&& \\

& &   & s & $p_z$ & $p_x $& $p_y$ & s & $p_z$ & $p_x$ & $p_y$ & $d_{z^2}$ & $d_{x z}$ & $d_{y z}$ & $d_{x^2-y^2}$ & $d_{x y}$ \\
\hline
  &$\mathbf{\Gamma}$ & $\text { CBM }$ & 7 & 0.9 & 8.4 & 0 & 7.9 & 64 & 6.8 & 0 & 0&0&0&0&0 \\
& &$\text { VBM }$ & 0.9 & 59.1 & 5 & 0 & 20.9 & 12.6 & 0.3 & 0 & 0&0&0&0&0 \\
\hline
$\text{AC}$&$\mathbf{K}_1$ & $\text { CBM }$ & 6.2 & 3 & 7.5 & 0.0 & 8.4 & 60.2 & 9.2 & 0 & 0&0&0&0&0 \\
&& $\text { VBM }$ & 0.97 & 60 & 4.7 & 0.0 & 20 & 11.8 & 0.86 & 0.0 & 0&0&0&0&0 \\
\hline
$\text{ZZ}$&$\mathbf{K}_2$ & $\text { CBM }$ & 6.7 & 1.65 & 8 & 0.0 & 8.1 & 62 & 7.8 & 0.0 & 0&0&0&0&0 \\
&& $\text { VBM }$ & 0.9 & 59.8 & 4.8 & 0 & 20 & 12 & 0.58 & 0.0 & 0&0&0&0&0 \\
\hline
$\text{AC}$&$\mathbf{K}_3$ & $\text { CBM }$  & 4.8 & 8 & 6.2 & 0.0 & 9 & 54.7 & 12 & 0.0 & 0&0&0&0&0\\
&& $\text { VBM }$ & 1.2 & 57.3 & 5.1 & 0.0 & 20.6 & 13 & 1.6 & 0.0 & 0&0&0&0&0 \\
\hline
$\text{AC}$&$\mathbf{K}_4$ & $\text { CBM }$  & 15.5 & 7 & 9.1 & 1.3 & 3.7 & 11.5 & 3 & 45.5 & 0&0&0&0&0\\
&& $\text { VBM }$ & 2.2 & 13 & 26.4 & 2 & 31.4 & 12.2 & 6.14 & 6.1 & 0&0&0&0&0 \\
\hline
\hline
\end{tabular}
}
    \label{tab:dsr}
\end{table}

To investigate the exciton spatial extension, we plotted the relative \textit{e-h} wavefunction in real-space, see Figure~\ref{fig:weight}(b-d).
This shows how these excitonic wavefunctions unfold over the real-space lattice. 
Indeed, we fixed the position of the hole on the top of the S atom (which contributes mainly to the top of the valence band, see Section~E of the SI for more information on the PDOS) separated by about $1$ $\AA$ within the unit cell.
Because the exciton wavefunction spreads over many unit lattices (more delocalized), they are more like Wannier-type exciton.

\begin{table}[]
    \centering
    \caption{Values of effective $m_{e f f}^{e, i}$($m_{e f f}^{h, i}$) and reduced masses $\mu_i$ in unit of free electron mass ($m_{0}$), and dielectric constant $\varepsilon_i$ along $\Gamma X$, $\Gamma Y$ and $\Gamma Z$ direction. The subscripts $i$ refer to {x, y, z} crystallography direction.}
    \renewcommand{\arraystretch}{1.3}
    \begin{tabular}{ccccc}
    \hline
    \hline & $m_{e f f}^{e, i}$  ($m_{0}$) & $m_{e f f}^{h, i}$  ($m_{0}$) & $\mu_i$ ($m_{0}$) &  $\varepsilon_i $ \\
    \hline $\Gamma-X$ & 0.845 & 1.011 & 0.46  & 10.94 \\
    \hline $\Gamma-Y$ & 1.519 & 1.599 & 0.78  & 11.29 \\
    \hline $\Gamma-Z$ & 0.019 & 0.101 & 0.016 & 10.72 \\
    \hline
    \hline
    \end{tabular}
    \label{tab:parameters_1}
\end{table}

In the following section, we use a semi-analytical theoretical model based on the EMA. 
In Table~\ref{tab:parameters_1}, we determine the effective masses ($m_{e f f}^{\nu, i}$) of electrons ($\nu=e$) and holes ($\nu =h$) as well as the static dielectric constant ($\varepsilon_i$), in GeS for different crystal directions ($i$). 
These parameters can be inserted directly into the Schrödinger equation describing the interaction between \textit{e-h} pairs. 
$m_{e f f}^{e, i}$ and $m_{e f f}^{h, i}$ are calculated in terms of effective free electron mass units ($m_{0}$) by means of a parabolic fitting of the valence and conduction band curvature near the $\Gamma$ $\mathbf{k}$-point, see Figure~\ref{fig:bs}(a). $m_{e f f}^{\nu, i}$ and $\varepsilon_i$ are strongly dependent on the crystal direction, revealing the influence of the crystal anisotropy on the optical properties of the excitonic states. 
The exciton reduced mass ($\mu_{i} = m_{eff}^{e,i}m_{eff}^{h,i}/(m_{eff}^{e,i}+m_{eff}^{e,i})$) in the in-plane direction is larger compared to the out-of-plane direction, leading to a strong compression of the exciton Bohr radius ($a_b^{i} \propto  \varepsilon_i \hbar^{2} / e^{2} \mu_i$) in the in-plane direction. 
As a result, our models will consider the exciton as an unconfined \textit{e-h} pair, since the confinement potential is negligible compared to the Coulomb potential.

%%%%%%%%%%%%%%%%%%%%%%%%%%%%%%%%%%%% Section 4 Anisotropic Wannier exciton
\section{Anisotropic Wannier exciton theory within EMA}\label{efa}

BSE demonstrates that the dominant contribution to the first bright exciton in GeS originates from the band states that lie in the vicinity of the $\mathbf{\Gamma}$ valley along the $\mathbf{\Gamma} \mathbf{X}$, $\mathbf{\Gamma} \mathbf{Y}$, and $\mathbf{\Gamma}\mathbf{Z}$ directions (see Figure~\ref{fig:weight}). 
We use this result to analyze the Schrödinger equation that describes the unconfined exciton in a pristine sample, taking into account the interaction between \textit{e-h} pair with anisotropic $m_{e f f}^{\nu, i}$  and $\varepsilon_i$. 
The Hamiltonian for modeling excitons in bulk semiconductors, along with the transformation to exciton coordinates, is described in Section~F of the SI. 
Using the relative $\mathbf{r}=(x, y, z)$ and center-of-mass (COM) $\mathbf{R_{CM}}= (X_{CM}, Y_{CM}, Z_{CM})$ coordinates, the resulting Schrödinger equation can be written as~\cite{kuzuba1976nearly,baldereschi1970anisotropy,dresselhaus1956effective}
\begin{widetext} 
\begin{align}
 \Bigg[ -\dfrac{\hbar^{2}}{2} \Big(\dfrac{1}{M_{X,x}} \frac{\partial^2}{\partial X_{CM}^2} +\dfrac{1}{M_{X,y}} \frac{\partial^2}{\partial Y_{CM}^2}+\dfrac{1}{M_{X,z}} \frac{\partial^2}{\partial Z_{CM}^2}\Big) -\dfrac{\hbar^{2}}{2} \Big(\dfrac{1}{\mu_{x}} \frac{\partial^2}{\partial x^2} +\dfrac{1}{\mu_{y}} \frac{\partial^2}{\partial y^2}+\dfrac{1}{\mu_{z}} \frac{\partial^2}{\partial z^2}\Big)- & \notag\\  \dfrac{e^{2}}{\sqrt{\varepsilon_{y} \varepsilon_{z} x^{2}+\varepsilon_{x} \varepsilon_{z} y^{2}+\varepsilon_{x} \varepsilon_{y} z^{2}}}\Bigg] 
\mathbf{\Upsilon}_{j}^{X}(\mathbf{R}_{CM}, \mathbf{r})  =  E_{j}^{X} \mathbf{\Upsilon}_{j}^{X} (\mathbf{R}_{CM}, \mathbf{r}),   
\end{align}
\end{widetext} 
where the first terms describe the kinetic energy of the free COM motion of the exciton in direction $i$, with exciton masses $M_{X,i} = m^{e,i}_{eff} + m^{h,i}_{eff}$. 
The last two terms describe the relative motion of the \textit{e-h} pair, bound via the attractive Coulomb interaction. $E_{j}^{X}(\mathbf{K}) = E_g^{GW} + E_{j}^{\,rel} + \sum_{i={x,y,z}} \hbar^2 \mathbf{K}_i^2/ (2 M_{X,i})$ and $\mathbf{\Upsilon}_{j}^{X} (\mathbf{R}_{CM}, \mathbf{r})=\sqrt{1/V}\exp^{i\mathbf{R}_{CM} \mathbf{K}} \mathbf{\Psi}^{rel}_{j}(\mathbf{r})$ are the eigenenergy and eigenfunction solutions of the Schrödinger equation, respectively. 
Here, $E_{j}^{\,rel}$ and $\mathbf{\Psi}^{rel}_{j}(\mathbf{r})$ represent the eigenenergy and the eigenvector of the relative motion of \textit{e-h} pair. 
The COM wavevector is denoted by $\mathbf{K}=(K_x, K_y, K_z)$. The volume of a bulk semiconductor, $V=N\Omega$, depends on the number of primitive cells ($N$) and the volume of the unit cell ($\Omega$).
To obtain the equation that describes Wannier excitons in an anisotropic medium, it is more convenient to transform from anisotropic masses to anisotropic potential via the change of variables, $\xi=\sqrt{\mu_{x}/\bar{\mu}} \,x,\,  \eta=\sqrt{\mu_{y}/\bar{\mu}}\,y,\,  \zeta=\sqrt{\mu_{z}/\bar{\mu}}\,z$. 
Thus, the equation for the relative motion of \textit{e-h} pair with $\mathbf{K}=0$ reads
\begin{equation}\label{ham_2}
 \mathbf{H}_{X}^{rel}=-\dfrac{\hbar^{2}}{2\bar{\mu}} \Big(\frac{\partial^2}{\partial \xi^2} + \frac{\partial^2}{\partial \eta^2}+ \frac{\partial^2}{\partial \zeta^2}\Big)- \dfrac{e^{2}}{ \bar{\varepsilon}\sqrt{A \xi^{2}+B \eta^{2}+C \zeta^{2}}},
\end{equation}
where the kinetic term retains its conventional form but with a modified mass. 
The anisotropic parameters $A = \dfrac{\bar{\mu}}{\mu_{x}} \dfrac{\varepsilon_{y}\varepsilon_{z}}{\bar{\varepsilon}^{2}}$, $B = \dfrac{\bar{\mu}}{\mu_{y}} \dfrac{\varepsilon_{x}\varepsilon_{z}}{\bar{\varepsilon}^{2}}$ and $C = \dfrac{\bar{\mu}}{\mu_{z}} \dfrac{\varepsilon_{x}\varepsilon_{y}}{\bar{\varepsilon}^{2}}$, assumed to be real and positive, estimate the degree of anisotropy. 
$\bar{\mu}=(\frac{\bar{\varepsilon}}{3} (\frac{1}{\varepsilon_{x}\mu_{x}} + \frac{1}{\varepsilon_{y}\mu_{y}}+ \frac{1}{\varepsilon_{z}\mu_{z}}))^{-1}$ and $\bar{\varepsilon}=\sqrt[3]{\varepsilon_x \varepsilon_y \varepsilon_z}$ represent the average reduced mass and the dielectric constant, respectively. 
The problem is treated in spherical coordinates, where $\rho =\sqrt{\xi^{2}+\eta^{2}+\zeta^{2}}$. 
This transformation integrates the anisotropy into the potential while maintaining the conventional form of an isotropic COM system with a reduced mass equal to $\bar{\mu}$. 
We utilize the new coordinate system for our calculations and only switch back to Cartesian coordinates $x$, $y$, and $z$ when presenting the contour plots of the density wavefunction for clarity and interpretation. 
Expanding the relative wavefunction in a basis of $3$D-hydrogenic wavefunction, $\mathbf{\Phi}_{n,\ell,m}(\boldsymbol{\rho}, \theta, \phi)$, is convenient to solve Eq.~\ref{ham_2}. 
We express the relative wavefunction $\mathbf{\Psi}^{rel}_{j=(\Tilde{n},\Tilde{\ell},\Tilde{m})}(\boldsymbol{\rho}, \theta, \phi)=\sum_{n,\ell,m}C_{n,\ell,m} \mathbf{\Phi}_{n,\ell,m}(\boldsymbol{\rho}, \theta, \phi)$. Here, $C_{n,\ell,m}$ are expansion coefficients and $\mathbf{\Phi}_{n,\ell,m}(\boldsymbol{\rho}, \theta, \phi)$ are the basis wavefunctions, where 
$n$, $\ell$, and $m$ represent the primary, azimuthal, and magnetic quantum numbers, respectively. 
The numerical diagonalization method was adopted for the resolution of Eq.~\ref{ham_2}.
In fact, the indices $\Tilde{n}$, $\Tilde{\ell}$, and $\Tilde{m}$ refer to the dominant contributions of the coefficients $C_{n,\ell,m}$ to the relative excitonic wavefunction $\mathbf{\Psi}^{rel}_{j=(\Tilde{n},\Tilde{\ell},\Tilde{m})}(\boldsymbol{\rho}, \theta, \phi)$, corresponding to the coefficient of the highest weight. 
The matrix elements of the relative Hamiltonian can be written as
 \begin{align}\label{energybind}
 \begin{split}
\bra{\Phi_{n,\ell,m}}  \mathbf{H}_{X}^{rel} \ket{\Phi_{n^{'},\ell^{'},m^{'}}}  &=-\frac{\bar{R}_{y}}{n^{2}} \delta_{n,n^{'}} \delta_{\ell,\ell^{'}} \delta_{m,m^{'}}\\
&+\bra{\Phi_{n,\ell,m}}  \mathbf{H}_{per} \ket{\Phi_{n^{'},\ell^{'},m^{'}}} 
 \end{split}
 \end{align}
with $\bar{R_{y}}=e^{2} /(2\, \bar{\varepsilon} \bar{a}_{b} )$ is the $3$D-effective Rydberg energy. $\bar{a}_{b}=\bar{\varepsilon} \hbar^{2}/(\bar{\mu} e^{2})$ is the $3$D-exciton effective Bohr radius and the perturbed Hamiltonian
\begin{align}
\begin{split}
     &\mathbf{H}_{per}= \frac{e^2}{\bar{\varepsilon}\boldsymbol{\rho}}\times\\
     &\Bigg(1-\frac{1}{\sqrt{\Big(A \cos(\phi)^2+B \sin (\phi)^2\Big)\sin(\theta)^2+ C \cos(\theta)^2 }}\Bigg).
     \end{split}
\end{align}
The perturbation reduces the symmetry of the system and, therefore, breaks the degeneracy of the excitonic state compared to the isotropic cases. 
For further information on the matrix elements and the excitonic Hamiltonian, see Section~F of the SI.
\begin{figure*}[]
    \centering
    \includegraphics[width=\textwidth]{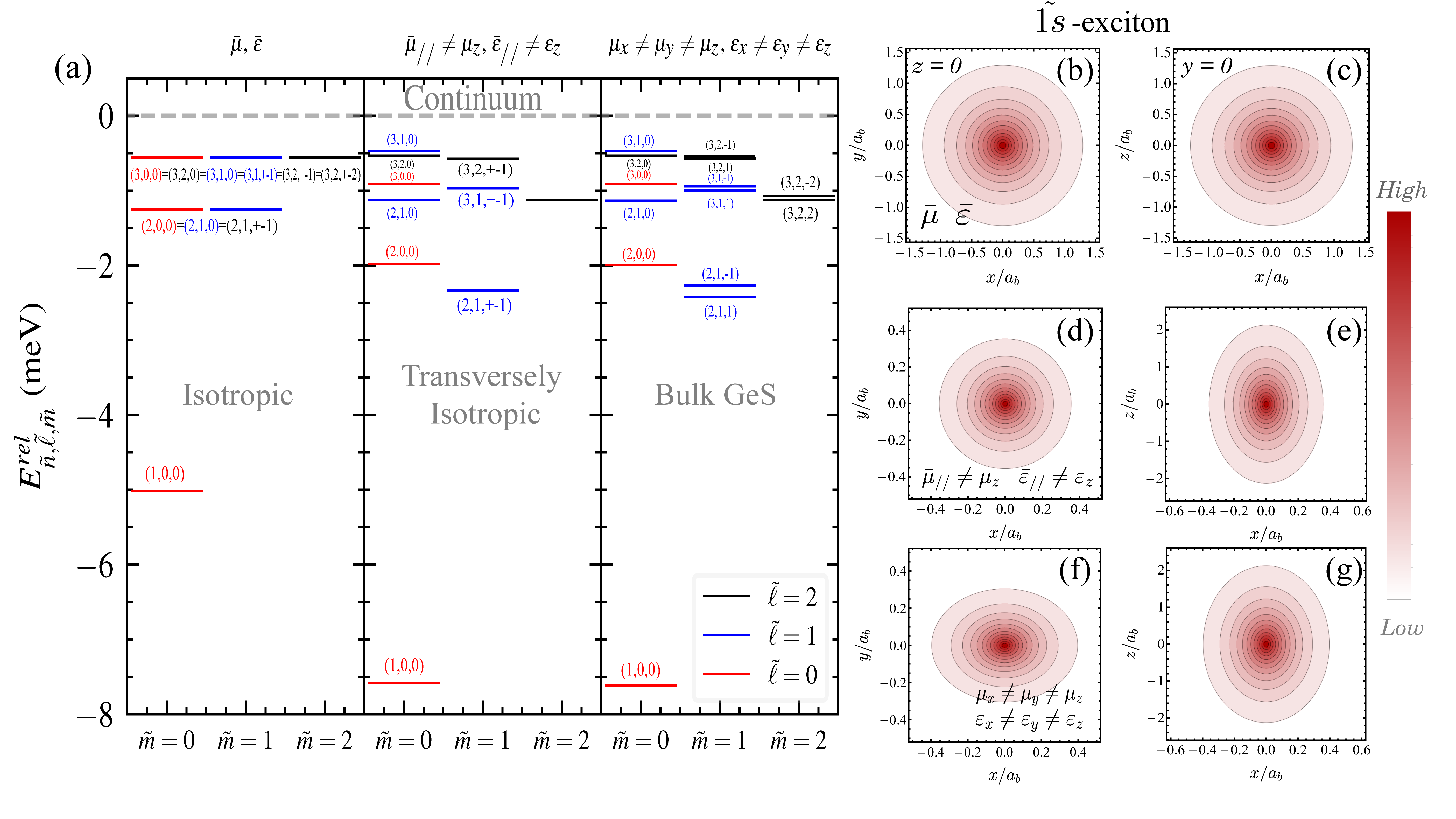}
    \caption{(a) BE of exciton states, showcasing the transition from isotropic to anisotropic cases across three panels. Each panel is further subdivided into columns, with each column denoting a magnetic quantum number $m$ (ranging from $0$, $\pm 1$, $\pm 2$, from left to right), while the different line colors represent $\ell$ states. The left panel illustrates the isotropic case with average values of the reduced exciton mass $\bar{\mu}$ and the dielectric constant $\bar{\varepsilon}$. In the middle panel, we consider the transversely isotropic scenario, where $\mu_{\parallel} \neq \mu_z$ and $\varepsilon_{\parallel} \neq \varepsilon_z$. The right panel showcases the BE of exciton states in the bulk GeS. (b-g) The comparison of probability density, which represents the squared modulus of the corrected wavefunctions for the ground excitonic state projected in real space, is presented in contour plots. These plots are shown in the $(x,y)$ plane for figures (b, d, f) and in the $(x,z)$ plane for figures (c, e, g). The unit of Bohr radius is used for all three cases. The figures correspond to the isotropic, transversely isotropic, and anisotropic cases, from top to bottom respectively.}
    \label{fig:diag_bi}
\end{figure*}

By the variation of the appropriate parameters ($\mu_{i}$, $\varepsilon_{i}$ and $A$, $B$, $C$), we can control the degree of anisotropy. 
To visualize the effects of anisotropy, Figure~\ref{fig:diag_bi}~(a) reports the first low-lying BEs of the relative excitonic states, $E_{\Tilde{n},\Tilde{\ell},\Tilde{m}}^{B} = -E_{\Tilde{n},\Tilde{\ell},\Tilde{m}}^{rel}$, obtained by numerical diagonalization of the matrix resulting from the projection of the Hamiltonian $\mathbf{H}_{X}^{rel}$ as given in Eq.~\ref{ham_2}. 
To assign the exciton state according to its primary orbital character, we label them according to the $(n,\ell,m)$ component that produces the highest probability density $ \lvert \mathbf{\Psi}^{rel}_{\Tilde{n},\Tilde{\ell},\Tilde{m}}(\boldsymbol{\rho}, \theta, \phi) \rvert^{2}$. 
For instance, the $\Tilde{1s}$-exciton is mainly dominated by the ($1,0,0$) component, while the ($2,1,0$) component largely contributes to $\Tilde{2p_{0}}$ exciton, and the ($2,1,\pm 1$) component contributes significantly to the $\Tilde{2p_{\pm 1}}$ exciton, etc.  
In bulk GeS case, as depicted in the right panel of Figure~\ref{fig:diag_bi}, $\mu_{i}$ had a strong anisotropy along the direction $i$, while $\varepsilon_i$ shows slight variations. 
In Figure~\ref{fig:diag_bi}, we also studied two other different cases: (i) First, the middle panel represents a transversely isotropic (such as uniaxial system). 
We found that the in-plane reduced mass $\mu_{\parallel}$=$(\frac{\varepsilon_{\parallel}}{2}(\frac{1}{\varepsilon_x \mu_x}+\frac{1}{\varepsilon_y \mu_y}))^{-1}=0.58$  $m_0 \neq \mu_{z}$. 
Similarly, the in-plane ($\varepsilon_{\parallel}=\sqrt{\varepsilon_{x} \varepsilon_{y}}=11.11$) and out-off-plane ($\varepsilon_{z}$) dielectric constants are also slightly different. (ii) Second, the left panel shows an isotropic system with a reduced mass of $\bar{\mu}=0.044$ $m_0$ and a dielectric constant of $\bar{\varepsilon}=10.98$. 
The comparison between these two cases and the fully anisotropic one (GeS) shows that the anisotropy strongly affects the degeneracy of the excitonic states. 
In fact, for an isotropic system, the results of the symmetric Coulomb potential are found to follow $-\bar{R}_{y}/n^{2}$, where the states are $(2n+1)$-fold degenerate.
In contrast to 3D hydrogenic-like models, the anisotropy clearly lifts the degeneracy of the different excited excitonic states. 
Additionally, the excitonic states are apparent in an anomalous energy level order of the azimuthal $\tilde{\ell}$ and the magnetic quantum number $\tilde{m}$.  
For example, the $\tilde{2p_{\pm 1}}$ states lay energetically below the $\tilde{2s}$ state, due to the asymmetry caused by the effective mass and the dielectric constant anisotropy. 
This means that the radial and angular dependencies in the Wannier equation can be separated. 
However, for bulk GeS, the anisotropy reduces the symmetry of the system.
In this case, the radial and angular degrees of freedom are coupled, which lift the degeneracy. 
Interestingly, a two-fold degeneracy is found for transverse isotropy, so that for the same $\tilde{\ell}$, positive and negative values of the magnetic quantum number $\tilde{m}$ lead to the same energy. 
This degeneracy is removed in GeS, producing, for example, a $0.2$~meV separation between $\tilde{2p_{+ 1}}$ and $\tilde{2p_{- 1}}$ states. 
In the transverse anisotropy case, the exciton Hamiltonian exhibits uniaxial symmetry. 
Consequently, it is possible to diagonalize the matrices separately for even and odd $\ell$ and for different $m$. In the angular momentum space, this special characteristic results in a block-diagonal eigenvalue problem. 
In contrast, for GeS, the anisotropy mixes states with the same $\ell$ but different $m$. 

In addition to lifting the degeneracy, we can clearly notice that the exciton BE strongly depends on the reduced mass and the dielectric constant and hence on the anisotropic parameters A, B, and C.
Strong anisotropy ($A \neq B \neq C$) leads to the increasing ($E^{B}/\bar{R}_{y}=1.5$) of the exciton BE compared to the isotropic case ($A=B=C$). In the EMA, we found that the $\tilde{1s}$-exciton BE and Bohr radius of the bulk GeS were $\sim 7.6$~meV and $13$~nm, respectively. 
This is consistent with the BSE method, which indicates that the exciton BE is $\sim10$~meV.
In fact, the small BE (largest Bohr radius) is due to the highest effective dielectric function $\bar{\varepsilon}$ and the smaller reduced mass $\bar{\mu}$, where the reduced mass out-of-plane is significantly smaller than those in-plane. 
These values account for the swift disappearance of exciton emission at high temperatures, since their magnitudes are lower than the thermal energy ($k_B T$). 
In fact, in Section A of the SI using temperature-dependent PL, we estimate the quenching of the exciton emission at temperatures 130~K under 2.41~eV excitation, compared to 190~K under 1.88~eV resonant excitation.

To better understand the impact of anisotropy on exciton wavefunctions, we compare in Figure~\ref{fig:diag_bi}(b-g) the probability density (squared modulus of the excitonic wavefunctions) for the ground-state excitonic wavefunction $\tilde{1s}$ for the three previous cases. 
The isotropic case produces the well-known hydrogen-like wavefunctions. 
For the transversal isotropic case, as expected, the excitonic wavefunction for the in-plane compound has a spherical symmetry, while the orbital becomes a disk stretched along the $z$-direction, since the reduced mass in the $z$-direction is much smaller than the average in-plane reduced mass. 
For GeS, the anisotropy modifies the ground state from a spherically symmetric $s$ (isotropic) to a squeezed wavefunction with a slightly peanut shape. 
In fact, we found that the $\tilde{1s}$ state stretched in the three directions, however, the distortions (squeezing) are more significant in the out-of-plane direction due to the heavier mass of this direction. 
The smaller area across which the probability density extends in the in-plane direction for the anisotropic case as compared to the isotopic one is due to the higher BE, which leads to more localized wavefunctions.

\begin{figure}[]
    \centering
    \includegraphics[width=\linewidth,height=8.5cm]{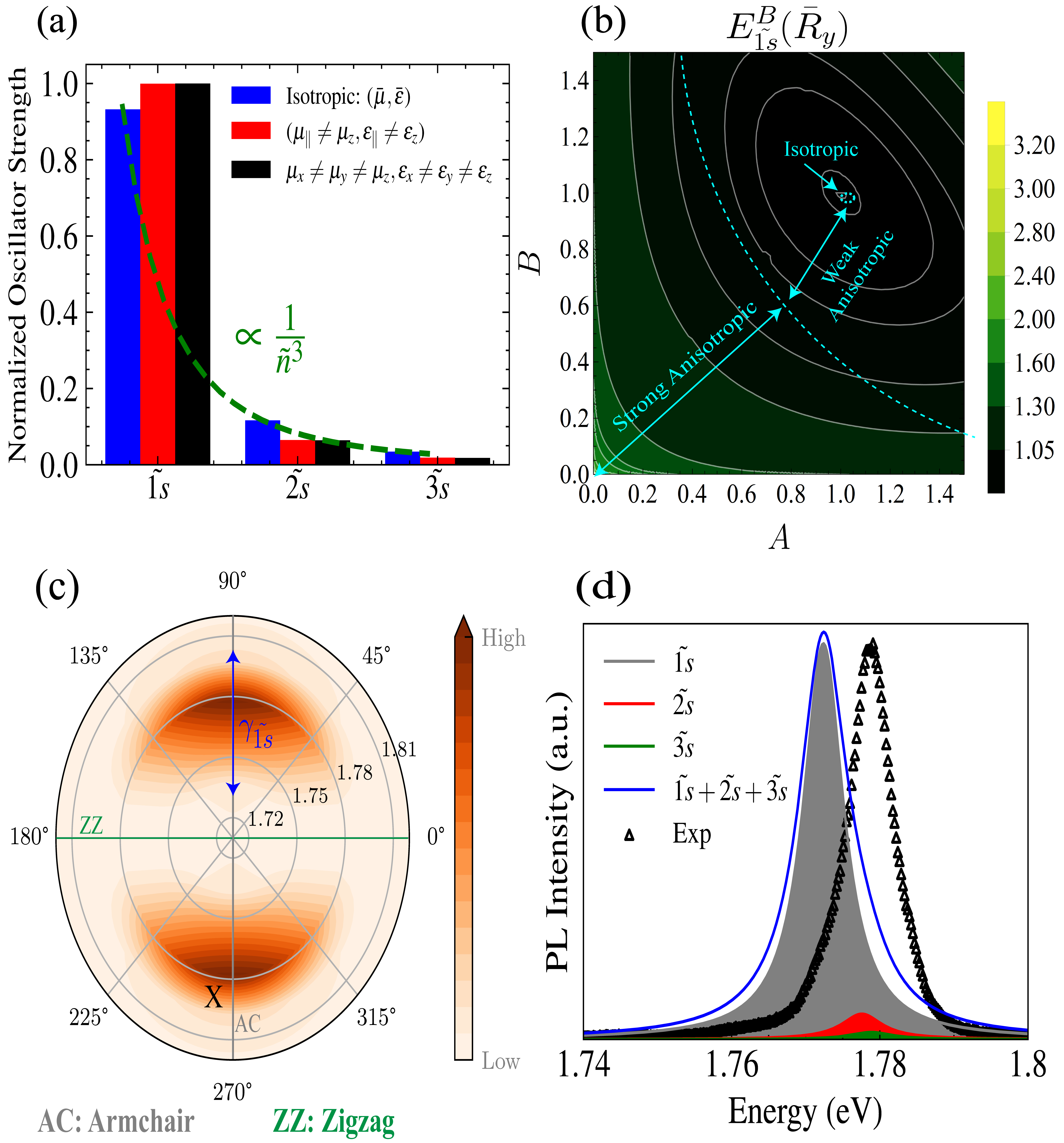}
    \caption{\textbf{Anisotropy effect on OS for low-lying exciton states in GeS}. (a) Normalized OS for the first low-lying $\tilde{ns}$-state for different cases. (b) $\tilde{1s}$-Exciton BE as a function of anisotropy parameters. The plot present three critical regions (isotropic, weak anisotropic, and strong anisotropic), which are separated by a dashed cyan line. (c) Theoretical prediction of the angle-resolved PL spectrum for GeS. (d) Direct comparison between experimental and theoretical PL for low-lying exciton states.}
    \label{fig:pl}
\end{figure}

Figure~\ref{fig:pl} (b) illustrates the impact of anisotropic parameters (A, B, C) on the ground state ($\tilde{1s}$) and highlights the strong dependence of the exciton energy on the degree of anisotropy. 
By adjusting those parameters, we can shift from the strong anisotropy case to the isotropy one. 
In fact, depending on the values of $A$ and $B$, we found three limiting regimes. 
If $A = B = C = 1$, Eq.~\ref{ham_2} and Eq.~\ref{energybind} becomes the well-known equation for the isotropic case, whose eigenvalues are $-\bar{R}_{y}/n^{2}$, with the well-known degree of degeneracy $n^{2}$ so that $E_{1s}^{B}=\bar{R}_{y}$, as clearly shown in Figure~\ref{fig:pl}(c). 
Then, with decreasing of A and B, the degree of perturbation induced by the anisotropy increases, leading to an increase in the exciton BE.
Indeed, for $A=B \ge 0.1$, the excitonic ground state tends to the well-known 2D-hydrogenic energy $E_{\tilde{1s}}^{B}=4\bar{R}_{y}$ due to the asymmetry in the Coulomb potential caused by the strong perturbation induced by the anisotropy in this regime.
Interestingly, for $A=B \ge 1$, we found that the $\tilde{1s}$-exciton energy is lower than the Rydberg energy $E_{\tilde{1s}}^{B} < \bar{R}_{y}$. 

From the knowledge of the exciton energies and wavefunctions as functions of the anisotropic parameters, we can compute the OS and hence the PL signal for these exciton states. 
Indeed, the OS is a dimensionless quantity that gives the relative strength of a particular optical transition. 
Here, we consider only the most common case of direct allowed optical transitions between valence and conduction bands. 
For instance, the OS of the optical interband transition for exciton states is
   $f_{ \tilde{n},\tilde{\ell},\tilde{m}}^{\alpha_{\mathbf{q}}} =
    \frac{2}{m_0 \hbar \omega_{0}} \times \left|\bar{F}_{\mathbf{q},c,v}^{\tilde{n},\tilde{\ell},\tilde{m}}\right|^{2}$, 
    where $\omega_{0}$ is the angular frequency of the optical transition with energy $\hbar\omega_{0}= E^X_{\tilde{n},\tilde{\ell},\tilde{m}}$.  The quantity $\bar{F}_{\mathbf{q},c,v}^{\tilde{n},\tilde{\ell},\tilde{m}}=\boldsymbol{\alpha}_{\mathbf{q}} \cdot \left\langle \emptyset \left|e^{i \mathbf{q} \cdot \mathbf{r}} \mathbf{p}\right| \zeta^{X}_{\tilde{n},\tilde{\ell},\tilde{m}} \right \rangle$ is the OME between the crystal ground state $\ket{\emptyset}$  and the excited states $\ket{\zeta_X^{\tilde{n},\tilde{\ell},\tilde{m}}}$ corresponding to the direct exciton in bulk GeS. Here,   
$\boldsymbol{\zeta}^{X}_{\tilde{n},\tilde{\ell},\tilde{m}}\left(\mathbf{r_{e}}, \mathbf{r_{h}}\right)=\mathbf{\Upsilon}_{\tilde{n},\tilde{\ell},\tilde{m}}(\mathbf{R_{CM}}, \boldsymbol{\rho})  u_{c, \mathbf{k}_{e}}\left(\mathbf{r}_e\right) u_{v, \mathbf{k}_{h}}^*\left(\mathbf{r}_h\right)
$, $ \boldsymbol{\alpha}_{\mathbf{q}}$ denotes the photon polarization unit vector. 
For the scenario of unconfined exciton COM motion in the bulk GeS, the OS is expressed as follows (see Section~F in the SI for more details)~\cite{Shinada_}
\begin{equation}
f_{ \tilde{n},\tilde{\ell},\tilde{m}}^{\alpha_{\mathbf{q}}}= \frac{2 V}{m_0 E^X_{\tilde{n},\tilde{\ell},\tilde{m}}} \left|  \sum_{n,\ell,m}C_{n,\ell,m} \mathbf{\Phi}_{n,\ell,m}(\boldsymbol{\rho}=0) \right|^2 
\times \mathbf{M}_{i, i}^{c,v}(\mathbf{k}).
\end{equation}
The OME and the relative wavefunction give rise to different types of selection rules. 
Indeed, only allowed transitions occur within the excitonic state wherein the relative wavefunction $\mathbf{\Psi}_{\tilde{n},\tilde{\ell},\tilde{m}}^{rel}(\rho=0) \neq 0$.
For the OME, the selection rules come mainly from the interband coupling term, which depends on the nature of the Bloch function.

Figure~\ref{fig:pl}(a) presents the calculated normalized OS of the three low-lying s-exciton states $\tilde{1s}$, $\tilde{2s}$, and $\tilde{3s}$ in bulk GeS (black line). 
For comparison, we also plot the OS for the isotropic (blue line) and transversal isotropic (red line) cases. 
It is seen that the OS decay ratio in the anisotropic case shows an anomalous behavior compared to that in the isotopic case. 
In GeS, the intensity ratio $f_{\tilde{1s}}^{AC}/f_{ \tilde{ns}}^{AC}$ exhibits different values compared to a bare isotropic hydrogen-like Coulomb potential, where the exciton OS of states $\tilde{ns}$ decay as $f_{\tilde{ns}}^{AC}= f_{\tilde{1s}}^{AC}/n^{3}$. 
Specifically, the ratio is approximately 15 for $\tilde{1s}/ \tilde{2s}$ and 55 for $\tilde{1s} / \tilde{3s}$, while it is equal to 8 and 27, respectively, in the isotropic case. 
This finding is consistent with the results reported by T. Shubina et al. for InSe~\cite{Shubina2019InSeAA}.  
Indeed, this difference is due to the fact that, in contrast to the isotropic case, the perturbation induced by anisotropy generates linear combinations of basis functions, so different orbitals with different weights, $C_{n,\ell,m}$, contribute in each anisotropic excitonic transition. 
Therefore, the $s$, $p$ and even the $d$-state shell excitons are mixed due to the reduction of symmetry induced by the anisotropy of GeS. 
This mixing makes the $p$-shell states optically active. 
Hence, both $s$-shell and $p$-shell excitons are active in both one- and two-photon processes, providing an efficient mechanism of second-harmonic generation. 
The discernible characteristic may be observed through the perturbed matrix elements. 
Moreover, this illustrates that states with different angular momentum components are indistinguishable in GeS, which well describes the mixing of excitonic states with different parities, particularly the $s$-shell and $p$-shell states. 
These non-linear contributions are similar to those related to the magneto-Stark effect or the electric-field-induced mixing of excitons in ZnO~\cite{PhysRevLett.110.116402} and GaAs~\cite{PhysRevB.92.085202}. 
The non-linear contributions to the exciton emission are beyond the scope of this work, as the one-photon emission is investigated. 
In addition, we found that the $\tilde{ns}$-state BE for the transversal isotropic and full anisotropic cases is almost similar. 
This is also shown in Figure~\ref{fig:diag_bi}, in which the exciton BE of the $\tilde{ns}$-state for these two cases is almost equal. 
In fact, due to the effective light mass along the $z$ direction compared to the in-plane effective mass, the total effective mass in the anisotropic case ($\bar{\mu}$) is almost equal to the transversal isotopic case ($\mu_{\parallel}$), which produces a comparable Rydberg energy ($\bar{R}_{y}$) and Bohr radius ($\bar{a}_{b}$) in both cases.

The OS per unit volume can be related to the probability of emission of a photon over all the photon modes by the following formula
 \begin{equation}\label{final_pl_mesurement}
    \mathbf{P}_{\alpha_\mathbf{q}}^{\tilde{ns}}(\omega_\mathbf{q}) = \int d\boldsymbol{\mathrm{q}} \,f_{\alpha_{\mathbf{q}}}^{\tilde{ns}}  \mathcal{L}\left(\hbar \omega_{\mathbf{q}}-E^X_{\tilde{ns}}\right)
\end{equation}
where the Lorentzian $\mathcal{L}(\hbar \omega_{\mathbf{q}}-E^X_{\tilde{ns}})=\gamma_{\tilde{ns}}/\pi\big[(\hbar \omega_{\mathbf{q}}-E^X_{\tilde{ns}})^2+\gamma_{\tilde{ns}}^2\big]$ expresses the energy conservation taking into account the state of the linewidth broadening extracted from our experiment, $\gamma_{\tilde{1s}}=8$~meV. 
Notably, the OS of exciton states is generally different between one- and two-photon processes. 
In Figure~\ref{fig:pl}(d), we plot the PL spectra for different excitonic states, and in the same plot we show the experimental PL spectrum of the free exciton at low temperature. 
We found that the exciton ground state $\tilde{1s}$, located around $1.772$~eV dominates the PL spectrum, which is expected since it has the largest OS. 
When comparing our experimental results, it is evident that the excited excitonic states, such as $\tilde{2s}$ and $\tilde{3s}$, are not clearly visible in our measurements. 
This is due to different factors: (i) the largest linewidth broadening ($\gamma_{\tilde{1s}}$), which is significantly greater than the energy separation between different excitonic $\tilde{s}$-states ($\delta E_{\tilde{1s},\tilde{ns}}= E_{\tilde{ns}}^{X}- E_{\tilde{1s}}^{X}$). (ii) The value of the OS (which is directly proportional to the PL intensity) that corresponds to the excited exciton $\tilde{ns}$-states is lower than the $\tilde{1s}$-state. 
Thus, the nearly energetic $\tilde{ns}$-states overlap and assemble into a single peak in the PL spectra, which appears to be broadened and asymmetric with a lower slope on the lower-energy side. 
It is important to note that this peak's asymmetry may also be due to phonon-assisted processes.~\cite{PhysRevResearch.3.L042019}

Using Eq.~\eqref{final_pl_mesurement}, we calculate in Figure~\ref{fig:pl}(d) the angle-resolved PL at low temperature. 
The PL signal is strongly dependent on the polarization angle. 
In fact, the exciton signal is maximum along the AC direction, then the signal intensity decreases until it disappears in the ZZ direction. 
This demonstrates the anisotropic optical signature in GeS, and agrees well with the experimental results in Figure~\ref{fig:res1}. 

Our experimental results reveal the existence of multiple emission peaks at low temperature in GeS encapsulated with $h$-BN, in addition to the free neutral exciton peak, these peaks are located in the energy range of approximately $60 - 100$~meV below the neutral exciton. 
Despite thorough investigations, the nature of these low-energy emission remains unknown, both from theoretical and experimental perspectives. 
During this study, to identify these peaks, we qualitatively studied different scenarios, including biexciton, optical phonon-assisted exciton recombination, emission of charged excitons (trions) and their fine structure, and exciton plus a localized exciton.  
More details about the nature of these peaks can be found in Section~A and B of the SI, however, a deep quantitative study of these low-energy emission is beyond the scope of the present paper.

%TC:ignore
\section{Conclusions}
Anisotropy of the low-symmetry orthorhombic crystal structure of GeS offers possibilities for the manipulation of physical properties along different crystal directions. 
We presented a detailed study of the anisotropic properties of GeS using temperature- and polarization-dependent PL techniques supported by \textit{ab initio} calculations. 
Temperature-dependent PL experiments enabled us to investigate the excitonic impact on the optical properties of GeS, whilst polarization-dependent PL measurements revealed the anisotropic nature of the material. 
The \textit{ab initio} calculations were used to theoretically predict the electronic BS and optical properties of GeS and to interpret the experimental results. 
By applying both the IPA and the GW+BSE methods, we calculated the dielectric functions and showed that including exciton effects through the BSE method modifies the optical absorption spectra in bulk GeS. 
Using the EMA, we showed that compared to an isotropic system, the anisotropy in GeS breaks the degeneracy, causes mixing of states with the same quantum number but different orbital and angular quantum numbers, and increases the BE.
We also found a relatively low BE, on the order of $\sim10$~meV and a quite large exciton Bohr radius, on the order of $13$~nm, which explains the observed rapid decrease in the exciton peak as temperature increases. 
This result provides essential understanding into the complexities of anisotropic materials and may have ramifications for the design of certain technological applications.

%%%%%%%%%%%%%%%%%%%%%%%%%%%%%%%%%%%%%%%%%%% Methods
\section*{Methods}
\subsection*{Sample}
The studied sample is composed of a thin layer of GeS, characterized by thickness of about 60~nm, encapsulated in $h$-BN flakes and placed on a Si/SiO$_2$ substrate. 
The GeS crystal, which was used for the preparation of the investigated sample, was purchased from HQ graphene. 
Thin GeS flakes were directly exfoliated on 285~nm SiO$_2$/Si substrates in an inert gas glovebox (O$_2$$<1$~ppm, H$_2$O$<1$~ppm). 
Then we used a poly (bisphenol A carbonate)/polydimethylsiloxane stamp on a glass slide to pick up top $h$-BN, thin flake of GeS, bottom $h$-BN at 80$^{\circ}$C with the assistance of the transfer stage in the glove box. 
Finally, the stack was released on a Si/SiO$_2$ substrate. 
The thicknesses of the GeS flakes were first identified by optical contrast and then measured more precisely with an atomic force microscope.

\subsection*{Experimental techniques}
The PL spectra were measured using different illuminations with a series of continuous wave (CW) laser diodes: $\lambda$=488~nm (2.54~eV), $\lambda$=515~nm (2.41~eV), $\lambda$=561~nm (2.21~eV), $\lambda$=633~nm (1.96~eV), and $\lambda$=660~nm (1.88~eV).
The PLE experiment was carried out using a supercontinuum light source coupled with a monochromator as an excitation source.
For the RC studies, the only difference in the experimental setup with respect to the one used to record the PL and PLE signals concerned the excitation source, which was replaced by a tungsten halogen lamp.
The studied samples were placed on a cold finger in a continuous-flow cryostat mounted on $x$-$y$ motorized positioners. 
The excitation light was focused by means of a 100x long working distance objective with a 0.55 numerical aperture that produced a spot of about 1/4 $\mu$m diameter in the PL/RC measurements.
The signal was collected via the same microscope objective, sent through a \mbox{0.75 m} monochromator, and then detected using a cooled charge-coupled device camera cooled with liquid nitrogen. 
The excitation power focused on the sample was kept at $100$~$\mu$W during all measurements to avoid local heating.
The polarization-resolved PL and RC spectra were measured using a motorized half-wave plate mounted on top of the microscope objective and a fixed linear polarizer placed in the detection path, which provides simultaneous rotation of the co-linearly polarized excitation light and detected signal.

\subsection*{Computational Details}
\subsubsection*{Ground-state calculations from first principles}
\sloppy
Structural relaxation and electronic properties were performed using the \textit{QUANTUM ESPRESSO}~\cite{Giannozzi_2009, Giannozzi_2017} (QE) package with DFT~\cite{PhysRev.140.A1133} based on the GGA of the Perdew-Burke-Ernzerhof (PBE) exchange-correlation functional~\cite{PhysRevLett.77.3865} within the plane-wave expansion. The vdW correction of the PBE functional was performed using the Grimme DFT-D2 method~\cite{grimme}. The optimized norm-conserving pseudopotentials~\cite{VANSETTEN201839} from the QE repository are used to describe the core–valence interaction. To evaluate the role of spin–orbit interaction, the calculations are performed using the fully relativistic version of the same pseudopotentials from the QE repository. The atomic positions and lattice constants were fully relaxed by DFT. It is assumed that the relaxation of the structure has reached convergence when the maximum component of the residual ionic forces is less than $10^{-10}$ Ry/Bohr. The Broyden–Fletcher–Goldfarb–Shanno~\cite{fischer1992general}  method was used for structural optimization. After convergence tests, the energy cutoff for the plane-wave expansion of the wavefunction is set to $816$ eV for all calculations and an appropriate Monkhorst-Pack $\mathbf{k}$-point sampling~\cite{PhysRevB.13.5188} in the BZ is centered with $18 \times 16 \times 4$, $18\times 16\times 4$, and $24\times 22\times 6$ meshes for geometry optimizations, self-consistent calculation, and projected density of states, respectively. The cutoff energy and $\mathbf{k}$-point sample were tested with a PBE-GGA calculation in the convergence study to ensure numerical stability. The criteria for the convergence of forces and total energy in optimization were set to  $10^{-4}$ eV/$\AA$ and $10^{-5}$~eV, respectively. 

\subsubsection*{Excitation energies and exciton wavefunctions from first principles}
Self and non-self consistent DFT calculations are performed to obtain KS eigenvalues and eigenfunctions to be used in the many-body perturbation theory~\cite{RevModPhys.74.601, PhysRevB.34.5390, strinati1988application}, by using the YAMBO code~\cite{MARINI20091392, sangalli2019many}. YAMBO interfaced with QE, which allows for the calculation of the optical response starting from the previously generated KS wavefunctions and energies in a plane-wave basis set. The YAMBO code was used to calculate QP adjustments at G$_0$W$_0$ and evGW, which was then used to calculate the optical excitation energies and optical spectra by solving the BSE. For GW simulations, the inverse of the microscopic dynamic dielectric function, $\varepsilon_{\mathbf{G}, \mathbf{G}^{\prime}}^{-1}$, is obtained within the plasmon-pole approximation~\cite{PhysRevB.88.125205, sangalli2019many}. After we calculated the convergence test of the parameters, we set the following parameters as the starting point for our calculation. We have used 200 bands, a $45$~Ry energy cutoff for the self-energy exchange component (the number of $\mathbf{G}$-vectors in the exchange) and a $18$~Ry cutoff for the correlation part (energy cutoff in the screening) or response block size. To speed up convergence with respect to empty states, we adopt the technique described in Ref.~\cite{PhysRevB.78.085125} as implemented in the YAMBO code. The QP BSs are then used to build up the excitonic Hamiltonian and to solve the BSE. We obtained converged excitation energies considering, respectively, six empty states and six occupied states in the excitonic Hamiltonian, the irreducible BZ being sampled up to a $24$ ×$ 22$ ×$ 4$ $\mathbf{k}$-point mesh. We used the Tamm-Dancoff approximation~\cite{PhysRev.78.382, RevModPhys.74.601} for the Bethe-Salpeter Hamiltonian and took into account the local field effects. 

\section*{Notes}
The authors declare no competing financial interest.
\section*{acknowledgement}
The work was supported by the National Science Centre, Poland (Grant No. 2017/27/B/ST3/00205 and 2018/31/B/ST3/02111), the Ministry of Education (Singapore) through the Research Centre of Excellence program (grant EDUN C-33-18-279-V12, I-FIM) and under its Academic Research Fund Tier 2 (MOE-T2EP50122-0012), and the Air Force Office of Scientific Research and the Office of Naval Research Global under award number FA8655-21-1-7026.
K.W. and T.T. acknowledge support from the JSPS KAKENHI (Grant Numbers 20H00354 and 23H02052) and World Premier International Research Center Initiative (WPI), MEXT, Japan.

\section*{Supporting Information Available}
The SI for this article includes additional experimental results on polarization-resolved RC spectra, the influence of excitation energy on the PL spectra, and the polarization- and temperature-resolved PL spectra of GeS. It also contains information about the identification of low-energy peaks, IPA for linear optical response, the GW method, BSE, PDOS, and EMA.

\bibliographystyle{apsrev4-2}
\bibliography{revtex}

\begin{thebibliography}{122}%
\makeatletter
\providecommand \@ifxundefined [1]{%
 \@ifx{#1\undefined}
}%
\providecommand \@ifnum [1]{%
 \ifnum #1\expandafter \@firstoftwo
 \else \expandafter \@secondoftwo
 \fi
}%
\providecommand \@ifx [1]{%
 \ifx #1\expandafter \@firstoftwo
 \else \expandafter \@secondoftwo
 \fi
}%
\providecommand \natexlab [1]{#1}%
\providecommand \enquote  [1]{``#1''}%
\providecommand \bibnamefont  [1]{#1}%
\providecommand \bibfnamefont [1]{#1}%
\providecommand \citenamefont [1]{#1}%
\providecommand \href@noop [0]{\@secondoftwo}%
\providecommand \href [0]{\begingroup \@sanitize@url \@href}%
\providecommand \@href[1]{\@@startlink{#1}\@@href}%
\providecommand \@@href[1]{\endgroup#1\@@endlink}%
\providecommand \@sanitize@url [0]{\catcode `\\12\catcode `\$12\catcode
  `\&12\catcode `\#12\catcode `\^12\catcode `\_12\catcode `\%12\relax}%
\providecommand \@@startlink[1]{}%
\providecommand \@@endlink[0]{}%
\providecommand \url  [0]{\begingroup\@sanitize@url \@url }%
\providecommand \@url [1]{\endgroup\@href {#1}{\urlprefix }}%
\providecommand \urlprefix  [0]{URL }%
\providecommand \Eprint [0]{\href }%
\providecommand \doibase [0]{https://doi.org/}%
\providecommand \selectlanguage [0]{\@gobble}%
\providecommand \bibinfo  [0]{\@secondoftwo}%
\providecommand \bibfield  [0]{\@secondoftwo}%
\providecommand \translation [1]{[#1]}%
\providecommand \BibitemOpen [0]{}%
\providecommand \bibitemStop [0]{}%
\providecommand \bibitemNoStop [0]{.\EOS\space}%
\providecommand \EOS [0]{\spacefactor3000\relax}%
\providecommand \BibitemShut  [1]{\csname bibitem#1\endcsname}%
\let\auto@bib@innerbib\@empty
%</preamble>
\bibitem [{\citenamefont {Mas-Balleste}\ \emph {et~al.}(2011)\citenamefont
  {Mas-Balleste}, \citenamefont {Gomez-Navarro}, \citenamefont
  {Gomez-Herrero},\ and\ \citenamefont {Zamora}}]{mas20112d}%
  \BibitemOpen
  \bibfield  {author} {\bibinfo {author} {\bibfnamefont {R.}~\bibnamefont
  {Mas-Balleste}}, \bibinfo {author} {\bibfnamefont {C.}~\bibnamefont
  {Gomez-Navarro}}, \bibinfo {author} {\bibfnamefont {J.}~\bibnamefont
  {Gomez-Herrero}},\ and\ \bibinfo {author} {\bibfnamefont {F.}~\bibnamefont
  {Zamora}},\ }\href {https://doi.org/https://doi.org/10.1039/C0NR00323A}
  {\bibfield  {journal} {\bibinfo  {journal} {Nanoscale}\ }\textbf {\bibinfo
  {volume} {3}},\ \bibinfo {pages} {20} (\bibinfo {year} {2011})}\BibitemShut
  {NoStop}%
\bibitem [{\citenamefont {Bhimanapati}\ \emph {et~al.}(2015)\citenamefont
  {Bhimanapati}, \citenamefont {Lin}, \citenamefont {Meunier}, \citenamefont
  {Jung}, \citenamefont {Cha}, \citenamefont {Das}, \citenamefont {Xiao},
  \citenamefont {Son}, \citenamefont {Strano}, \citenamefont {Cooper} \emph
  {et~al.}}]{bhimanapati2015recent}%
  \BibitemOpen
  \bibfield  {author} {\bibinfo {author} {\bibfnamefont {G.~R.}\ \bibnamefont
  {Bhimanapati}}, \bibinfo {author} {\bibfnamefont {Z.}~\bibnamefont {Lin}},
  \bibinfo {author} {\bibfnamefont {V.}~\bibnamefont {Meunier}}, \bibinfo
  {author} {\bibfnamefont {Y.}~\bibnamefont {Jung}}, \bibinfo {author}
  {\bibfnamefont {J.}~\bibnamefont {Cha}}, \bibinfo {author} {\bibfnamefont
  {S.}~\bibnamefont {Das}}, \bibinfo {author} {\bibfnamefont {D.}~\bibnamefont
  {Xiao}}, \bibinfo {author} {\bibfnamefont {Y.}~\bibnamefont {Son}}, \bibinfo
  {author} {\bibfnamefont {M.~S.}\ \bibnamefont {Strano}}, \bibinfo {author}
  {\bibfnamefont {V.~R.}\ \bibnamefont {Cooper}}, \emph {et~al.},\ }\href
  {https://doi.org/https://doi.org/10.1021/acsnano.5b05556} {\bibfield
  {journal} {\bibinfo  {journal} {ACS nano}\ }\textbf {\bibinfo {volume} {9}},\
  \bibinfo {pages} {11509} (\bibinfo {year} {2015})}\BibitemShut {NoStop}%
\bibitem [{\citenamefont {Jariwala}\ \emph {et~al.}(2014)\citenamefont
  {Jariwala}, \citenamefont {Sangwan}, \citenamefont {Lauhon}, \citenamefont
  {Marks},\ and\ \citenamefont {Hersam}}]{jariwala2014emerging}%
  \BibitemOpen
  \bibfield  {author} {\bibinfo {author} {\bibfnamefont {D.}~\bibnamefont
  {Jariwala}}, \bibinfo {author} {\bibfnamefont {V.~K.}\ \bibnamefont
  {Sangwan}}, \bibinfo {author} {\bibfnamefont {L.~J.}\ \bibnamefont {Lauhon}},
  \bibinfo {author} {\bibfnamefont {T.~J.}\ \bibnamefont {Marks}},\ and\
  \bibinfo {author} {\bibfnamefont {M.~C.}\ \bibnamefont {Hersam}},\ }\href
  {https://doi.org/https://doi.org/10.1021/nn500064s} {\bibfield  {journal}
  {\bibinfo  {journal} {ACS nano}\ }\textbf {\bibinfo {volume} {8}},\ \bibinfo
  {pages} {1102} (\bibinfo {year} {2014})}\BibitemShut {NoStop}%
\bibitem [{\citenamefont {Wang}\ \emph {et~al.}(2012)\citenamefont {Wang},
  \citenamefont {Kalantar-Zadeh}, \citenamefont {Kis}, \citenamefont
  {Coleman},\ and\ \citenamefont {Strano}}]{wang2012electronics}%
  \BibitemOpen
  \bibfield  {author} {\bibinfo {author} {\bibfnamefont {Q.~H.}\ \bibnamefont
  {Wang}}, \bibinfo {author} {\bibfnamefont {K.}~\bibnamefont
  {Kalantar-Zadeh}}, \bibinfo {author} {\bibfnamefont {A.}~\bibnamefont {Kis}},
  \bibinfo {author} {\bibfnamefont {J.~N.}\ \bibnamefont {Coleman}},\ and\
  \bibinfo {author} {\bibfnamefont {M.~S.}\ \bibnamefont {Strano}},\ }\href
  {https://doi.org/https://doi.org/10.1038/nnano.2012.193} {\bibfield
  {journal} {\bibinfo  {journal} {Nature nanotechnology}\ }\textbf {\bibinfo
  {volume} {7}},\ \bibinfo {pages} {699} (\bibinfo {year} {2012})}\BibitemShut
  {NoStop}%
\bibitem [{\citenamefont {Wu}\ \emph {et~al.}(2021)\citenamefont {Wu},
  \citenamefont {Lyu}, \citenamefont {Zhang}, \citenamefont {Ding},
  \citenamefont {Zheng}, \citenamefont {Yang}, \citenamefont {Lau},
  \citenamefont {Chen},\ and\ \citenamefont {Hao}}]{wu2021large}%
  \BibitemOpen
  \bibfield  {author} {\bibinfo {author} {\bibfnamefont {Z.}~\bibnamefont
  {Wu}}, \bibinfo {author} {\bibfnamefont {Y.}~\bibnamefont {Lyu}}, \bibinfo
  {author} {\bibfnamefont {Y.}~\bibnamefont {Zhang}}, \bibinfo {author}
  {\bibfnamefont {R.}~\bibnamefont {Ding}}, \bibinfo {author} {\bibfnamefont
  {B.}~\bibnamefont {Zheng}}, \bibinfo {author} {\bibfnamefont
  {Z.}~\bibnamefont {Yang}}, \bibinfo {author} {\bibfnamefont {S.~P.}\
  \bibnamefont {Lau}}, \bibinfo {author} {\bibfnamefont {X.~H.}\ \bibnamefont
  {Chen}},\ and\ \bibinfo {author} {\bibfnamefont {J.}~\bibnamefont {Hao}},\
  }\href {https://doi.org/https://doi.org/10.1038/s41563-021-01001-7}
  {\bibfield  {journal} {\bibinfo  {journal} {Nature materials}\ }\textbf
  {\bibinfo {volume} {20}},\ \bibinfo {pages} {1203} (\bibinfo {year}
  {2021})}\BibitemShut {NoStop}%
\bibitem [{\citenamefont {Molas}\ \emph {et~al.}(2021)\citenamefont {Molas},
  \citenamefont {Macewicz}, \citenamefont {Wieloszy{\'n}ska}, \citenamefont
  {Jak{\'o}bczyk}, \citenamefont {Wysmo{\l}ek}, \citenamefont {Bogdanowicz},\
  and\ \citenamefont {Jasinski}}]{molas2021photoluminescence}%
  \BibitemOpen
  \bibfield  {author} {\bibinfo {author} {\bibfnamefont {M.~R.}\ \bibnamefont
  {Molas}}, \bibinfo {author} {\bibfnamefont {{\L}.}~\bibnamefont {Macewicz}},
  \bibinfo {author} {\bibfnamefont {A.}~\bibnamefont {Wieloszy{\'n}ska}},
  \bibinfo {author} {\bibfnamefont {P.}~\bibnamefont {Jak{\'o}bczyk}}, \bibinfo
  {author} {\bibfnamefont {A.}~\bibnamefont {Wysmo{\l}ek}}, \bibinfo {author}
  {\bibfnamefont {R.}~\bibnamefont {Bogdanowicz}},\ and\ \bibinfo {author}
  {\bibfnamefont {J.~B.}\ \bibnamefont {Jasinski}},\ }\href
  {https://doi.org/https://doi.org/10.1038/s41699-021-00263-8} {\bibfield
  {journal} {\bibinfo  {journal} {npj 2D Materials and Applications}\ }\textbf
  {\bibinfo {volume} {5}},\ \bibinfo {pages} {83} (\bibinfo {year}
  {2021})}\BibitemShut {NoStop}%
\bibitem [{\citenamefont {Liu}\ \emph {et~al.}(2014)\citenamefont {Liu},
  \citenamefont {Neal}, \citenamefont {Zhu}, \citenamefont {Luo}, \citenamefont
  {Xu}, \citenamefont {Tom{\'a}nek},\ and\ \citenamefont
  {Ye}}]{liu2014phosphorene}%
  \BibitemOpen
  \bibfield  {author} {\bibinfo {author} {\bibfnamefont {H.}~\bibnamefont
  {Liu}}, \bibinfo {author} {\bibfnamefont {A.~T.}\ \bibnamefont {Neal}},
  \bibinfo {author} {\bibfnamefont {Z.}~\bibnamefont {Zhu}}, \bibinfo {author}
  {\bibfnamefont {Z.}~\bibnamefont {Luo}}, \bibinfo {author} {\bibfnamefont
  {X.}~\bibnamefont {Xu}}, \bibinfo {author} {\bibfnamefont {D.}~\bibnamefont
  {Tom{\'a}nek}},\ and\ \bibinfo {author} {\bibfnamefont {P.~D.}\ \bibnamefont
  {Ye}},\ }\href {https://doi.org/https://doi.org/10.1021/nn501226z} {\bibfield
   {journal} {\bibinfo  {journal} {ACS nano}\ }\textbf {\bibinfo {volume}
  {8}},\ \bibinfo {pages} {4033} (\bibinfo {year} {2014})}\BibitemShut
  {NoStop}%
\bibitem [{\citenamefont {Qiao}\ \emph {et~al.}(2014)\citenamefont {Qiao},
  \citenamefont {Kong}, \citenamefont {Hu}, \citenamefont {Yang},\ and\
  \citenamefont {Ji}}]{qiao2014high}%
  \BibitemOpen
  \bibfield  {author} {\bibinfo {author} {\bibfnamefont {J.}~\bibnamefont
  {Qiao}}, \bibinfo {author} {\bibfnamefont {X.}~\bibnamefont {Kong}}, \bibinfo
  {author} {\bibfnamefont {Z.-X.}\ \bibnamefont {Hu}}, \bibinfo {author}
  {\bibfnamefont {F.}~\bibnamefont {Yang}},\ and\ \bibinfo {author}
  {\bibfnamefont {W.}~\bibnamefont {Ji}},\ }\href
  {https://doi.org/https://doi.org/10.1038/ncomms5475} {\bibfield  {journal}
  {\bibinfo  {journal} {Nature communications}\ }\textbf {\bibinfo {volume}
  {5}},\ \bibinfo {pages} {4475} (\bibinfo {year} {2014})}\BibitemShut
  {NoStop}%
\bibitem [{\citenamefont {Podzorov}\ \emph {et~al.}(2004)\citenamefont
  {Podzorov}, \citenamefont {Gershenson}, \citenamefont {Kloc}, \citenamefont
  {Zeis},\ and\ \citenamefont {Bucher}}]{podzorov2004high}%
  \BibitemOpen
  \bibfield  {author} {\bibinfo {author} {\bibfnamefont {V.}~\bibnamefont
  {Podzorov}}, \bibinfo {author} {\bibfnamefont {M.}~\bibnamefont
  {Gershenson}}, \bibinfo {author} {\bibfnamefont {C.}~\bibnamefont {Kloc}},
  \bibinfo {author} {\bibfnamefont {R.}~\bibnamefont {Zeis}},\ and\ \bibinfo
  {author} {\bibfnamefont {E.}~\bibnamefont {Bucher}},\ }\href
  {https://doi.org/https://doi.org/10.1063/1.1723695} {\bibfield  {journal}
  {\bibinfo  {journal} {Applied Physics Letters}\ }\textbf {\bibinfo {volume}
  {84}},\ \bibinfo {pages} {3301} (\bibinfo {year} {2004})}\BibitemShut
  {NoStop}%
\bibitem [{\citenamefont {Ugeda}\ \emph {et~al.}(2014)\citenamefont {Ugeda},
  \citenamefont {Bradley}, \citenamefont {Shi}, \citenamefont {Da~Jornada},
  \citenamefont {Zhang}, \citenamefont {Qiu}, \citenamefont {Ruan},
  \citenamefont {Mo}, \citenamefont {Hussain}, \citenamefont {Shen} \emph
  {et~al.}}]{ugeda2014giant}%
  \BibitemOpen
  \bibfield  {author} {\bibinfo {author} {\bibfnamefont {M.~M.}\ \bibnamefont
  {Ugeda}}, \bibinfo {author} {\bibfnamefont {A.~J.}\ \bibnamefont {Bradley}},
  \bibinfo {author} {\bibfnamefont {S.-F.}\ \bibnamefont {Shi}}, \bibinfo
  {author} {\bibfnamefont {F.~H.}\ \bibnamefont {Da~Jornada}}, \bibinfo
  {author} {\bibfnamefont {Y.}~\bibnamefont {Zhang}}, \bibinfo {author}
  {\bibfnamefont {D.~Y.}\ \bibnamefont {Qiu}}, \bibinfo {author} {\bibfnamefont
  {W.}~\bibnamefont {Ruan}}, \bibinfo {author} {\bibfnamefont {S.-K.}\
  \bibnamefont {Mo}}, \bibinfo {author} {\bibfnamefont {Z.}~\bibnamefont
  {Hussain}}, \bibinfo {author} {\bibfnamefont {Z.-X.}\ \bibnamefont {Shen}},
  \emph {et~al.},\ }\href {https://doi.org/https://doi.org/10.1038/nmat4061}
  {\bibfield  {journal} {\bibinfo  {journal} {Nature materials}\ }\textbf
  {\bibinfo {volume} {13}},\ \bibinfo {pages} {1091} (\bibinfo {year}
  {2014})}\BibitemShut {NoStop}%
\bibitem [{\citenamefont {Zhang}\ \emph {et~al.}(2018)\citenamefont {Zhang},
  \citenamefont {Chaves}, \citenamefont {Huang}, \citenamefont {Wang},
  \citenamefont {Xing}, \citenamefont {Low},\ and\ \citenamefont
  {Yan}}]{zhang2018determination}%
  \BibitemOpen
  \bibfield  {author} {\bibinfo {author} {\bibfnamefont {G.}~\bibnamefont
  {Zhang}}, \bibinfo {author} {\bibfnamefont {A.}~\bibnamefont {Chaves}},
  \bibinfo {author} {\bibfnamefont {S.}~\bibnamefont {Huang}}, \bibinfo
  {author} {\bibfnamefont {F.}~\bibnamefont {Wang}}, \bibinfo {author}
  {\bibfnamefont {Q.}~\bibnamefont {Xing}}, \bibinfo {author} {\bibfnamefont
  {T.}~\bibnamefont {Low}},\ and\ \bibinfo {author} {\bibfnamefont
  {H.}~\bibnamefont {Yan}},\ }\href {https://doi.org/h10.1126/sciadv.aap99}
  {\bibfield  {journal} {\bibinfo  {journal} {Science advances}\ }\textbf
  {\bibinfo {volume} {4}},\ \bibinfo {pages} {eaap9977} (\bibinfo {year}
  {2018})}\BibitemShut {NoStop}%
\bibitem [{\citenamefont {Henriques}\ and\ \citenamefont
  {Peres}(2020)}]{henriques2020excitons}%
  \BibitemOpen
  \bibfield  {author} {\bibinfo {author} {\bibfnamefont {J.}~\bibnamefont
  {Henriques}}\ and\ \bibinfo {author} {\bibfnamefont {N.}~\bibnamefont
  {Peres}},\ }\href
  {https://doi.org/https://doi.org/10.1103/PhysRevB.101.035406} {\bibfield
  {journal} {\bibinfo  {journal} {Physical Review B}\ }\textbf {\bibinfo
  {volume} {101}},\ \bibinfo {pages} {035406} (\bibinfo {year}
  {2020})}\BibitemShut {NoStop}%
\bibitem [{\citenamefont {Kim}\ \emph {et~al.}(2019)\citenamefont {Kim},
  \citenamefont {Kim}, \citenamefont {Park}, \citenamefont {Kim}, \citenamefont
  {Choi}, \citenamefont {Im}, \citenamefont {Lee}, \citenamefont {Kim},\ and\
  \citenamefont {Yi}}]{kim2019intrinsic}%
  \BibitemOpen
  \bibfield  {author} {\bibinfo {author} {\bibfnamefont {M.}~\bibnamefont
  {Kim}}, \bibinfo {author} {\bibfnamefont {H.-g.}\ \bibnamefont {Kim}},
  \bibinfo {author} {\bibfnamefont {S.}~\bibnamefont {Park}}, \bibinfo {author}
  {\bibfnamefont {J.~S.}\ \bibnamefont {Kim}}, \bibinfo {author} {\bibfnamefont
  {H.~J.}\ \bibnamefont {Choi}}, \bibinfo {author} {\bibfnamefont
  {S.}~\bibnamefont {Im}}, \bibinfo {author} {\bibfnamefont {H.}~\bibnamefont
  {Lee}}, \bibinfo {author} {\bibfnamefont {T.}~\bibnamefont {Kim}},\ and\
  \bibinfo {author} {\bibfnamefont {Y.}~\bibnamefont {Yi}},\ }\href
  {https://doi.org/https://doi.org/10.1002/anie.201811743} {\bibfield
  {journal} {\bibinfo  {journal} {Angewandte Chemie International Edition}\
  }\textbf {\bibinfo {volume} {58}},\ \bibinfo {pages} {3754} (\bibinfo {year}
  {2019})}\BibitemShut {NoStop}%
\bibitem [{\citenamefont {Li}\ \emph {et~al.}(2019)\citenamefont {Li},
  \citenamefont {Zhou}, \citenamefont {Shi}, \citenamefont {Chen},\ and\
  \citenamefont {Wang}}]{li2019recent}%
  \BibitemOpen
  \bibfield  {author} {\bibinfo {author} {\bibfnamefont {Q.}~\bibnamefont
  {Li}}, \bibinfo {author} {\bibfnamefont {Q.}~\bibnamefont {Zhou}}, \bibinfo
  {author} {\bibfnamefont {L.}~\bibnamefont {Shi}}, \bibinfo {author}
  {\bibfnamefont {Q.}~\bibnamefont {Chen}},\ and\ \bibinfo {author}
  {\bibfnamefont {J.}~\bibnamefont {Wang}},\ }\href
  {https://doi.org/https://doi.org/10.1039/C8TA10306B} {\bibfield  {journal}
  {\bibinfo  {journal} {Journal of Materials Chemistry A}\ }\textbf {\bibinfo
  {volume} {7}},\ \bibinfo {pages} {4291} (\bibinfo {year} {2019})}\BibitemShut
  {NoStop}%
\bibitem [{\citenamefont {Sadki}\ \emph {et~al.}(2019)\citenamefont {Sadki},
  \citenamefont {Sadki},\ and\ \citenamefont {Drissi}}]{sadki2019oxidation}%
  \BibitemOpen
  \bibfield  {author} {\bibinfo {author} {\bibfnamefont {K.}~\bibnamefont
  {Sadki}}, \bibinfo {author} {\bibfnamefont {S.}~\bibnamefont {Sadki}},\ and\
  \bibinfo {author} {\bibfnamefont {L.~B.}\ \bibnamefont {Drissi}},\ }\href
  {https://doi.org/https://doi.org/10.1016/j.jpcs.2018.10.008} {\bibfield
  {journal} {\bibinfo  {journal} {Journal of Physics and Chemistry of Solids}\
  }\textbf {\bibinfo {volume} {130}},\ \bibinfo {pages} {13} (\bibinfo {year}
  {2019})}\BibitemShut {NoStop}%
\bibitem [{\citenamefont {Gomes}\ and\ \citenamefont
  {Carvalho}(2015)}]{PhysRevB.92.085406}%
  \BibitemOpen
  \bibfield  {author} {\bibinfo {author} {\bibfnamefont {L.~C.}\ \bibnamefont
  {Gomes}}\ and\ \bibinfo {author} {\bibfnamefont {A.}~\bibnamefont
  {Carvalho}},\ }\href {https://doi.org/10.1103/PhysRevB.92.085406} {\bibfield
  {journal} {\bibinfo  {journal} {Phys. Rev. B}\ }\textbf {\bibinfo {volume}
  {92}},\ \bibinfo {pages} {085406} (\bibinfo {year} {2015})}\BibitemShut
  {NoStop}%
\bibitem [{\citenamefont {Lv}\ \emph {et~al.}(2017)\citenamefont {Lv},
  \citenamefont {Wei}, \citenamefont {Sun}, \citenamefont {Li}, \citenamefont
  {Huang},\ and\ \citenamefont {Dai}}]{Lv2017TwodimensionalGM}%
  \BibitemOpen
  \bibfield  {author} {\bibinfo {author} {\bibfnamefont {X.}~\bibnamefont
  {Lv}}, \bibinfo {author} {\bibfnamefont {W.}~\bibnamefont {Wei}}, \bibinfo
  {author} {\bibfnamefont {Q.}~\bibnamefont {Sun}}, \bibinfo {author}
  {\bibfnamefont {F.}~\bibnamefont {Li}}, \bibinfo {author} {\bibfnamefont
  {B.}~\bibnamefont {Huang}},\ and\ \bibinfo {author} {\bibfnamefont
  {Y.}~\bibnamefont {Dai}},\ }\href
  {https://doi.org/https://doi.org/10.1016/j.apcatb.2017.05.087} {\bibfield
  {journal} {\bibinfo  {journal} {Applied Catalysis B-environmental}\ }\textbf
  {\bibinfo {volume} {217}},\ \bibinfo {pages} {275} (\bibinfo {year}
  {2017})}\BibitemShut {NoStop}%
\bibitem [{\citenamefont {Wu}\ \emph {et~al.}(2018)\citenamefont {Wu},
  \citenamefont {Xie}, \citenamefont {Lu}, \citenamefont {Zhao}, \citenamefont
  {Wang}, \citenamefont {Jiang}, \citenamefont {Ge}, \citenamefont {Zhang},
  \citenamefont {Lu}, \citenamefont {Guo} \emph {et~al.}}]{wu2018few}%
  \BibitemOpen
  \bibfield  {author} {\bibinfo {author} {\bibfnamefont {L.}~\bibnamefont
  {Wu}}, \bibinfo {author} {\bibfnamefont {Z.}~\bibnamefont {Xie}}, \bibinfo
  {author} {\bibfnamefont {L.}~\bibnamefont {Lu}}, \bibinfo {author}
  {\bibfnamefont {J.}~\bibnamefont {Zhao}}, \bibinfo {author} {\bibfnamefont
  {Y.}~\bibnamefont {Wang}}, \bibinfo {author} {\bibfnamefont {X.}~\bibnamefont
  {Jiang}}, \bibinfo {author} {\bibfnamefont {Y.}~\bibnamefont {Ge}}, \bibinfo
  {author} {\bibfnamefont {F.}~\bibnamefont {Zhang}}, \bibinfo {author}
  {\bibfnamefont {S.}~\bibnamefont {Lu}}, \bibinfo {author} {\bibfnamefont
  {Z.}~\bibnamefont {Guo}}, \emph {et~al.},\ }\href
  {https://doi.org/https://doi.org/10.1002/adom.201700985} {\bibfield
  {journal} {\bibinfo  {journal} {Advanced Optical Materials}\ }\textbf
  {\bibinfo {volume} {6}},\ \bibinfo {pages} {1700985} (\bibinfo {year}
  {2018})}\BibitemShut {NoStop}%
\bibitem [{\citenamefont {Yang}\ \emph {et~al.}(2021)\citenamefont {Yang},
  \citenamefont {Liu}, \citenamefont {Li}, \citenamefont {Xue},\ and\
  \citenamefont {Hu}}]{yang2021plane}%
  \BibitemOpen
  \bibfield  {author} {\bibinfo {author} {\bibfnamefont {Y.}~\bibnamefont
  {Yang}}, \bibinfo {author} {\bibfnamefont {S.-C.}\ \bibnamefont {Liu}},
  \bibinfo {author} {\bibfnamefont {Z.}~\bibnamefont {Li}}, \bibinfo {author}
  {\bibfnamefont {D.-J.}\ \bibnamefont {Xue}},\ and\ \bibinfo {author}
  {\bibfnamefont {J.-S.}\ \bibnamefont {Hu}},\ }\href
  {https://doi.org/https://doi.org/10.1039/D0CC04476H} {\bibfield  {journal}
  {\bibinfo  {journal} {Chemical Communications}\ }\textbf {\bibinfo {volume}
  {57}},\ \bibinfo {pages} {565} (\bibinfo {year} {2021})}\BibitemShut
  {NoStop}%
\bibitem [{\citenamefont {Latiff}\ \emph {et~al.}(2015)\citenamefont {Latiff},
  \citenamefont {Teo}, \citenamefont {Sofer}, \citenamefont {Huber},
  \citenamefont {Fisher},\ and\ \citenamefont {Pumera}}]{latiff2015toxicity}%
  \BibitemOpen
  \bibfield  {author} {\bibinfo {author} {\bibfnamefont {N.}~\bibnamefont
  {Latiff}}, \bibinfo {author} {\bibfnamefont {W.~Z.}\ \bibnamefont {Teo}},
  \bibinfo {author} {\bibfnamefont {Z.}~\bibnamefont {Sofer}}, \bibinfo
  {author} {\bibfnamefont {{\v{S}}.}~\bibnamefont {Huber}}, \bibinfo {author}
  {\bibfnamefont {A.~C.}\ \bibnamefont {Fisher}},\ and\ \bibinfo {author}
  {\bibfnamefont {M.}~\bibnamefont {Pumera}},\ }\href
  {https://doi.org/https://doi.org/10.1039/C5RA09404F} {\bibfield  {journal}
  {\bibinfo  {journal} {RSC Advances}\ }\textbf {\bibinfo {volume} {5}},\
  \bibinfo {pages} {67485} (\bibinfo {year} {2015})}\BibitemShut {NoStop}%
\bibitem [{\citenamefont {Wiedemeier}\ and\ \citenamefont
  {Siemers}(1977)}]{wiedemeier1977thermal}%
  \BibitemOpen
  \bibfield  {author} {\bibinfo {author} {\bibfnamefont {H.}~\bibnamefont
  {Wiedemeier}}\ and\ \bibinfo {author} {\bibfnamefont {P.}~\bibnamefont
  {Siemers}},\ }\href
  {https://doi.org/https://doi.org/10.1002/zaac.19774310134} {\bibfield
  {journal} {\bibinfo  {journal} {Zeitschrift f{\"u}r anorganische und
  allgemeine Chemie}\ }\textbf {\bibinfo {volume} {431}},\ \bibinfo {pages}
  {299} (\bibinfo {year} {1977})}\BibitemShut {NoStop}%
\bibitem [{\citenamefont {Liu}\ and\ \citenamefont
  {Pantelides}(2018)}]{Liu_2018}%
  \BibitemOpen
  \bibfield  {author} {\bibinfo {author} {\bibfnamefont {J.}~\bibnamefont
  {Liu}}\ and\ \bibinfo {author} {\bibfnamefont {S.~T.}\ \bibnamefont
  {Pantelides}},\ }\href {https://doi.org/10.7567/APEX.11.101301} {\bibfield
  {journal} {\bibinfo  {journal} {Applied Physics Express}\ }\textbf {\bibinfo
  {volume} {11}},\ \bibinfo {pages} {101301} (\bibinfo {year}
  {2018})}\BibitemShut {NoStop}%
\bibitem [{\citenamefont {Feng}\ \emph {et~al.}(2021)\citenamefont {Feng},
  \citenamefont {Liu}, \citenamefont {Hu}, \citenamefont {Wu}, \citenamefont
  {Liu}, \citenamefont {Xue}, \citenamefont {Hu},\ and\ \citenamefont
  {Wan}}]{feng2021interfacial}%
  \BibitemOpen
  \bibfield  {author} {\bibinfo {author} {\bibfnamefont {M.}~\bibnamefont
  {Feng}}, \bibinfo {author} {\bibfnamefont {S.-C.}\ \bibnamefont {Liu}},
  \bibinfo {author} {\bibfnamefont {L.}~\bibnamefont {Hu}}, \bibinfo {author}
  {\bibfnamefont {J.}~\bibnamefont {Wu}}, \bibinfo {author} {\bibfnamefont
  {X.}~\bibnamefont {Liu}}, \bibinfo {author} {\bibfnamefont {D.-J.}\
  \bibnamefont {Xue}}, \bibinfo {author} {\bibfnamefont {J.-S.}\ \bibnamefont
  {Hu}},\ and\ \bibinfo {author} {\bibfnamefont {L.-J.}\ \bibnamefont {Wan}},\
  }\href {https://doi.org/https://doi.org/10.1021/jacs.1c04734} {\bibfield
  {journal} {\bibinfo  {journal} {Journal of the American Chemical Society}\
  }\textbf {\bibinfo {volume} {143}},\ \bibinfo {pages} {9664} (\bibinfo {year}
  {2021})}\BibitemShut {NoStop}%
\bibitem [{\citenamefont {Zawadzka}\ \emph {et~al.}(2021)\citenamefont
  {Zawadzka}, \citenamefont {Kipczak}, \citenamefont {Wo{\'z}niak},
  \citenamefont {Olkowska-Pucko}, \citenamefont {Grzeszczyk}, \citenamefont
  {Babi{\'n}ski},\ and\ \citenamefont {Molas}}]{zawadzka2021anisotropic}%
  \BibitemOpen
  \bibfield  {author} {\bibinfo {author} {\bibfnamefont {N.}~\bibnamefont
  {Zawadzka}}, \bibinfo {author} {\bibfnamefont {{\L}.}~\bibnamefont
  {Kipczak}}, \bibinfo {author} {\bibfnamefont {T.}~\bibnamefont
  {Wo{\'z}niak}}, \bibinfo {author} {\bibfnamefont {K.}~\bibnamefont
  {Olkowska-Pucko}}, \bibinfo {author} {\bibfnamefont {M.}~\bibnamefont
  {Grzeszczyk}}, \bibinfo {author} {\bibfnamefont {A.}~\bibnamefont
  {Babi{\'n}ski}},\ and\ \bibinfo {author} {\bibfnamefont {M.~R.}\ \bibnamefont
  {Molas}},\ }\href {https://doi.org/https://doi.org/10.3390/nano11113109}
  {\bibfield  {journal} {\bibinfo  {journal} {Nanomaterials}\ }\textbf
  {\bibinfo {volume} {11}},\ \bibinfo {pages} {3109} (\bibinfo {year}
  {2021})}\BibitemShut {NoStop}%
\bibitem [{\citenamefont {Ulaganathan}\ \emph {et~al.}(2016)\citenamefont
  {Ulaganathan}, \citenamefont {Lu}, \citenamefont {Kuo}, \citenamefont
  {Tamalampudi}, \citenamefont {Sankar}, \citenamefont {Boopathi},
  \citenamefont {Anand}, \citenamefont {Yadav}, \citenamefont {Mathew},
  \citenamefont {Liu} \emph {et~al.}}]{ulaganathan2016high}%
  \BibitemOpen
  \bibfield  {author} {\bibinfo {author} {\bibfnamefont {R.~K.}\ \bibnamefont
  {Ulaganathan}}, \bibinfo {author} {\bibfnamefont {Y.-Y.}\ \bibnamefont {Lu}},
  \bibinfo {author} {\bibfnamefont {C.-J.}\ \bibnamefont {Kuo}}, \bibinfo
  {author} {\bibfnamefont {S.~R.}\ \bibnamefont {Tamalampudi}}, \bibinfo
  {author} {\bibfnamefont {R.}~\bibnamefont {Sankar}}, \bibinfo {author}
  {\bibfnamefont {K.~M.}\ \bibnamefont {Boopathi}}, \bibinfo {author}
  {\bibfnamefont {A.}~\bibnamefont {Anand}}, \bibinfo {author} {\bibfnamefont
  {K.}~\bibnamefont {Yadav}}, \bibinfo {author} {\bibfnamefont {R.~J.}\
  \bibnamefont {Mathew}}, \bibinfo {author} {\bibfnamefont {C.-R.}\
  \bibnamefont {Liu}}, \emph {et~al.},\ }\href
  {https://doi.org/https://doi.org/10.1039/C5NR05988G} {\bibfield  {journal}
  {\bibinfo  {journal} {Nanoscale}\ }\textbf {\bibinfo {volume} {8}},\ \bibinfo
  {pages} {2284} (\bibinfo {year} {2016})}\BibitemShut {NoStop}%
\bibitem [{\citenamefont {Wang}\ \emph {et~al.}(2020)\citenamefont {Wang},
  \citenamefont {Tan}, \citenamefont {Han}, \citenamefont {Wang}, \citenamefont
  {Lu}, \citenamefont {Du}, \citenamefont {Jia}, \citenamefont {Deringer},
  \citenamefont {Zhou},\ and\ \citenamefont {Zhang}}]{wang2020sub}%
  \BibitemOpen
  \bibfield  {author} {\bibinfo {author} {\bibfnamefont {X.}~\bibnamefont
  {Wang}}, \bibinfo {author} {\bibfnamefont {J.}~\bibnamefont {Tan}}, \bibinfo
  {author} {\bibfnamefont {C.}~\bibnamefont {Han}}, \bibinfo {author}
  {\bibfnamefont {J.-J.}\ \bibnamefont {Wang}}, \bibinfo {author}
  {\bibfnamefont {L.}~\bibnamefont {Lu}}, \bibinfo {author} {\bibfnamefont
  {H.}~\bibnamefont {Du}}, \bibinfo {author} {\bibfnamefont {C.-L.}\
  \bibnamefont {Jia}}, \bibinfo {author} {\bibfnamefont {V.~L.}\ \bibnamefont
  {Deringer}}, \bibinfo {author} {\bibfnamefont {J.}~\bibnamefont {Zhou}},\
  and\ \bibinfo {author} {\bibfnamefont {W.}~\bibnamefont {Zhang}},\ }\href
  {https://doi.org/https://doi.org/10.1021/acsnano.9b10057} {\bibfield
  {journal} {\bibinfo  {journal} {ACS nano}\ }\textbf {\bibinfo {volume}
  {14}},\ \bibinfo {pages} {4456} (\bibinfo {year} {2020})}\BibitemShut
  {NoStop}%
\bibitem [{\citenamefont {Lan}\ \emph {et~al.}(2015)\citenamefont {Lan},
  \citenamefont {Li}, \citenamefont {Yin}, \citenamefont {Guo},\ and\
  \citenamefont {Wang}}]{lan2015synthesis}%
  \BibitemOpen
  \bibfield  {author} {\bibinfo {author} {\bibfnamefont {C.}~\bibnamefont
  {Lan}}, \bibinfo {author} {\bibfnamefont {C.}~\bibnamefont {Li}}, \bibinfo
  {author} {\bibfnamefont {Y.}~\bibnamefont {Yin}}, \bibinfo {author}
  {\bibfnamefont {H.}~\bibnamefont {Guo}},\ and\ \bibinfo {author}
  {\bibfnamefont {S.}~\bibnamefont {Wang}},\ }\href
  {https://doi.org/https://doi.org/10.1039/C5TC01435B} {\bibfield  {journal}
  {\bibinfo  {journal} {Journal of Materials Chemistry C}\ }\textbf {\bibinfo
  {volume} {3}},\ \bibinfo {pages} {8074} (\bibinfo {year} {2015})}\BibitemShut
  {NoStop}%
\bibitem [{\citenamefont {Taniguchi}\ \emph {et~al.}(1990)\citenamefont
  {Taniguchi}, \citenamefont {Johnson}, \citenamefont {Ghijsen},\ and\
  \citenamefont {Cardona}}]{taniguchi1990core}%
  \BibitemOpen
  \bibfield  {author} {\bibinfo {author} {\bibfnamefont {M.}~\bibnamefont
  {Taniguchi}}, \bibinfo {author} {\bibfnamefont {R.}~\bibnamefont {Johnson}},
  \bibinfo {author} {\bibfnamefont {J.}~\bibnamefont {Ghijsen}},\ and\ \bibinfo
  {author} {\bibfnamefont {M.}~\bibnamefont {Cardona}},\ }\href
  {https://doi.org/10.1103/physrevb.42.3634} {\bibfield  {journal} {\bibinfo
  {journal} {Physical Review B}\ }\textbf {\bibinfo {volume} {42}},\ \bibinfo
  {pages} {3634} (\bibinfo {year} {1990})}\BibitemShut {NoStop}%
\bibitem [{\citenamefont {To{\l}{\l}oczko}\ \emph {et~al.}(2020)\citenamefont
  {To{\l}{\l}oczko}, \citenamefont {Oliva}, \citenamefont {Wo{\'z}niak},
  \citenamefont {Kopaczek}, \citenamefont {Scharoch},\ and\ \citenamefont
  {Kudrawiec}}]{tolloczko2020anisotropic}%
  \BibitemOpen
  \bibfield  {author} {\bibinfo {author} {\bibfnamefont {A.}~\bibnamefont
  {To{\l}{\l}oczko}}, \bibinfo {author} {\bibfnamefont {R.}~\bibnamefont
  {Oliva}}, \bibinfo {author} {\bibfnamefont {T.}~\bibnamefont {Wo{\'z}niak}},
  \bibinfo {author} {\bibfnamefont {J.}~\bibnamefont {Kopaczek}}, \bibinfo
  {author} {\bibfnamefont {P.}~\bibnamefont {Scharoch}},\ and\ \bibinfo
  {author} {\bibfnamefont {R.}~\bibnamefont {Kudrawiec}},\ }\href
  {https://doi.org/https://doi.org/10.1039/D0MA00146E} {\bibfield  {journal}
  {\bibinfo  {journal} {Materials Advances}\ }\textbf {\bibinfo {volume} {1}},\
  \bibinfo {pages} {1886} (\bibinfo {year} {2020})}\BibitemShut {NoStop}%
\bibitem [{\citenamefont {Zhao}\ \emph {et~al.}(2014)\citenamefont {Zhao},
  \citenamefont {Lo}, \citenamefont {Zhang}, \citenamefont {Sun}, \citenamefont
  {Tan}, \citenamefont {Uher}, \citenamefont {Wolverton}, \citenamefont
  {Dravid},\ and\ \citenamefont {Kanatzidis}}]{zhao2014ultralow}%
  \BibitemOpen
  \bibfield  {author} {\bibinfo {author} {\bibfnamefont {L.-D.}\ \bibnamefont
  {Zhao}}, \bibinfo {author} {\bibfnamefont {S.-H.}\ \bibnamefont {Lo}},
  \bibinfo {author} {\bibfnamefont {Y.}~\bibnamefont {Zhang}}, \bibinfo
  {author} {\bibfnamefont {H.}~\bibnamefont {Sun}}, \bibinfo {author}
  {\bibfnamefont {G.}~\bibnamefont {Tan}}, \bibinfo {author} {\bibfnamefont
  {C.}~\bibnamefont {Uher}}, \bibinfo {author} {\bibfnamefont {C.}~\bibnamefont
  {Wolverton}}, \bibinfo {author} {\bibfnamefont {V.~P.}\ \bibnamefont
  {Dravid}},\ and\ \bibinfo {author} {\bibfnamefont {M.~G.}\ \bibnamefont
  {Kanatzidis}},\ }\href {https://doi.org/https://doi.org/10.1038/nature13184}
  {\bibfield  {journal} {\bibinfo  {journal} {nature}\ }\textbf {\bibinfo
  {volume} {508}},\ \bibinfo {pages} {373} (\bibinfo {year}
  {2014})}\BibitemShut {NoStop}%
\bibitem [{\citenamefont {Guo}\ and\ \citenamefont
  {Wang}(2017)}]{guo2017thermoelectric}%
  \BibitemOpen
  \bibfield  {author} {\bibinfo {author} {\bibfnamefont {S.-D.}\ \bibnamefont
  {Guo}}\ and\ \bibinfo {author} {\bibfnamefont {Y.-H.}\ \bibnamefont {Wang}},\
  }\href {https://doi.org/https://doi.org/10.1063/1.4974200} {\bibfield
  {journal} {\bibinfo  {journal} {Journal of Applied Physics}\ }\textbf
  {\bibinfo {volume} {121}},\ \bibinfo {pages} {034302} (\bibinfo {year}
  {2017})}\BibitemShut {NoStop}%
\bibitem [{\citenamefont {Senske}\ \emph {et~al.}(1978)\citenamefont {Senske},
  \citenamefont {Street}, \citenamefont {Nowitzki},\ and\ \citenamefont
  {Wiesner}}]{senske1978luminescence}%
  \BibitemOpen
  \bibfield  {author} {\bibinfo {author} {\bibfnamefont {W.}~\bibnamefont
  {Senske}}, \bibinfo {author} {\bibfnamefont {R.}~\bibnamefont {Street}},
  \bibinfo {author} {\bibfnamefont {A.}~\bibnamefont {Nowitzki}},\ and\
  \bibinfo {author} {\bibfnamefont {P.}~\bibnamefont {Wiesner}},\ }\href
  {https://doi.org/https://doi.org/10.1016/0022-2313(78)90079-0} {\bibfield
  {journal} {\bibinfo  {journal} {Journal of Luminescence}\ }\textbf {\bibinfo
  {volume} {16}},\ \bibinfo {pages} {343} (\bibinfo {year} {1978})}\BibitemShut
  {NoStop}%
\bibitem [{\citenamefont {Wiley}\ \emph {et~al.}(1980)\citenamefont {Wiley},
  \citenamefont {Thomas}, \citenamefont {Sch{\"o}nherr},\ and\ \citenamefont
  {Breitschwerdt}}]{wiley1980absorption}%
  \BibitemOpen
  \bibfield  {author} {\bibinfo {author} {\bibfnamefont {J.}~\bibnamefont
  {Wiley}}, \bibinfo {author} {\bibfnamefont {D.}~\bibnamefont {Thomas}},
  \bibinfo {author} {\bibfnamefont {E.}~\bibnamefont {Sch{\"o}nherr}},\ and\
  \bibinfo {author} {\bibfnamefont {A.}~\bibnamefont {Breitschwerdt}},\ }\href
  {https://doi.org/https://doi.org/10.1016/0022-3697(80)90091-8} {\bibfield
  {journal} {\bibinfo  {journal} {Journal of Physics and Chemistry of Solids}\
  }\textbf {\bibinfo {volume} {41}},\ \bibinfo {pages} {801} (\bibinfo {year}
  {1980})}\BibitemShut {NoStop}%
\bibitem [{\citenamefont {El-Bakkali}\ \emph {et~al.}(2021)\citenamefont
  {El-Bakkali}, \citenamefont {Sadki}, \citenamefont {Drissi},\ and\
  \citenamefont {Djeffal}}]{el2021layers}%
  \BibitemOpen
  \bibfield  {author} {\bibinfo {author} {\bibfnamefont {A.}~\bibnamefont
  {El-Bakkali}}, \bibinfo {author} {\bibfnamefont {S.}~\bibnamefont {Sadki}},
  \bibinfo {author} {\bibfnamefont {L.~B.}\ \bibnamefont {Drissi}},\ and\
  \bibinfo {author} {\bibfnamefont {F.}~\bibnamefont {Djeffal}},\ }\href
  {https://doi.org/https://doi.org/10.1016/j.physe.2021.114791} {\bibfield
  {journal} {\bibinfo  {journal} {Physica E: Low-dimensional Systems and
  Nanostructures}\ }\textbf {\bibinfo {volume} {133}},\ \bibinfo {pages}
  {114791} (\bibinfo {year} {2021})}\BibitemShut {NoStop}%
\bibitem [{\citenamefont {Tan}\ \emph {et~al.}(2017{\natexlab{a}})\citenamefont
  {Tan}, \citenamefont {Lim}, \citenamefont {Wang}, \citenamefont {Mohamed},
  \citenamefont {Mouri}, \citenamefont {Zhang}, \citenamefont {Miyauchi},
  \citenamefont {Ohfuchi},\ and\ \citenamefont {Matsuda}}]{tan2017anisotropic}%
  \BibitemOpen
  \bibfield  {author} {\bibinfo {author} {\bibfnamefont {D.}~\bibnamefont
  {Tan}}, \bibinfo {author} {\bibfnamefont {H.~E.}\ \bibnamefont {Lim}},
  \bibinfo {author} {\bibfnamefont {F.}~\bibnamefont {Wang}}, \bibinfo {author}
  {\bibfnamefont {N.~B.}\ \bibnamefont {Mohamed}}, \bibinfo {author}
  {\bibfnamefont {S.}~\bibnamefont {Mouri}}, \bibinfo {author} {\bibfnamefont
  {W.}~\bibnamefont {Zhang}}, \bibinfo {author} {\bibfnamefont
  {Y.}~\bibnamefont {Miyauchi}}, \bibinfo {author} {\bibfnamefont
  {M.}~\bibnamefont {Ohfuchi}},\ and\ \bibinfo {author} {\bibfnamefont
  {K.}~\bibnamefont {Matsuda}},\ }\href
  {https://doi.org/https://doi.org/10.1007/s12274-016-1312-6} {\bibfield
  {journal} {\bibinfo  {journal} {Nano Research}\ }\textbf {\bibinfo {volume}
  {10}},\ \bibinfo {pages} {546} (\bibinfo {year}
  {2017}{\natexlab{a}})}\BibitemShut {NoStop}%
\bibitem [{\citenamefont {Postorino}\ \emph
  {et~al.}(2020{\natexlab{a}})\citenamefont {Postorino}, \citenamefont {Sun},
  \citenamefont {Fiedler}, \citenamefont {Lee Cheong~Lem}, \citenamefont
  {Palummo},\ and\ \citenamefont {Camilli}}]{postorino2020interlayer}%
  \BibitemOpen
  \bibfield  {author} {\bibinfo {author} {\bibfnamefont {S.}~\bibnamefont
  {Postorino}}, \bibinfo {author} {\bibfnamefont {J.}~\bibnamefont {Sun}},
  \bibinfo {author} {\bibfnamefont {S.}~\bibnamefont {Fiedler}}, \bibinfo
  {author} {\bibfnamefont {L.~O.}\ \bibnamefont {Lee Cheong~Lem}}, \bibinfo
  {author} {\bibfnamefont {M.}~\bibnamefont {Palummo}},\ and\ \bibinfo {author}
  {\bibfnamefont {L.}~\bibnamefont {Camilli}},\ }\href
  {https://doi.org/10.3390/ma13163568} {\bibfield  {journal} {\bibinfo
  {journal} {Materials}\ }\textbf {\bibinfo {volume} {13}},\ \bibinfo {pages}
  {3568} (\bibinfo {year} {2020}{\natexlab{a}})}\BibitemShut {NoStop}%
\bibitem [{\citenamefont {Le}\ \emph {et~al.}(2019)\citenamefont {Le},
  \citenamefont {Nguyen}, \citenamefont {Thuan}, \citenamefont {Vu},
  \citenamefont {Ilyasov}, \citenamefont {Poklonski}, \citenamefont {Phuc},
  \citenamefont {Ershov}, \citenamefont {Geguzina}, \citenamefont {Hieu} \emph
  {et~al.}}]{le2019strain}%
  \BibitemOpen
  \bibfield  {author} {\bibinfo {author} {\bibfnamefont {P.}~\bibnamefont
  {Le}}, \bibinfo {author} {\bibfnamefont {C.~V.}\ \bibnamefont {Nguyen}},
  \bibinfo {author} {\bibfnamefont {D.~V.}\ \bibnamefont {Thuan}}, \bibinfo
  {author} {\bibfnamefont {T.~V.}\ \bibnamefont {Vu}}, \bibinfo {author}
  {\bibfnamefont {V.}~\bibnamefont {Ilyasov}}, \bibinfo {author} {\bibfnamefont
  {N.}~\bibnamefont {Poklonski}}, \bibinfo {author} {\bibfnamefont {H.~V.}\
  \bibnamefont {Phuc}}, \bibinfo {author} {\bibfnamefont {I.}~\bibnamefont
  {Ershov}}, \bibinfo {author} {\bibfnamefont {G.}~\bibnamefont {Geguzina}},
  \bibinfo {author} {\bibfnamefont {N.~V.}\ \bibnamefont {Hieu}}, \emph
  {et~al.},\ }\href
  {https://doi.org/https://doi.org/10.1007/s11664-019-06980-7} {\bibfield
  {journal} {\bibinfo  {journal} {Journal of Electronic Materials}\ }\textbf
  {\bibinfo {volume} {48}},\ \bibinfo {pages} {2902} (\bibinfo {year}
  {2019})}\BibitemShut {NoStop}%
\bibitem [{\citenamefont {Fei}\ \emph {et~al.}(2015)\citenamefont {Fei},
  \citenamefont {Li}, \citenamefont {Li},\ and\ \citenamefont
  {Yang}}]{fei2015giant}%
  \BibitemOpen
  \bibfield  {author} {\bibinfo {author} {\bibfnamefont {R.}~\bibnamefont
  {Fei}}, \bibinfo {author} {\bibfnamefont {W.}~\bibnamefont {Li}}, \bibinfo
  {author} {\bibfnamefont {J.}~\bibnamefont {Li}},\ and\ \bibinfo {author}
  {\bibfnamefont {L.}~\bibnamefont {Yang}},\ }\href
  {https://doi.org/https://doi.org/10.1063/1.4934750} {\bibfield  {journal}
  {\bibinfo  {journal} {Applied Physics Letters}\ }\textbf {\bibinfo {volume}
  {107}},\ \bibinfo {pages} {173104} (\bibinfo {year} {2015})}\BibitemShut
  {NoStop}%
\bibitem [{\citenamefont {Qin}\ \emph {et~al.}(2016)\citenamefont {Qin},
  \citenamefont {Qin}, \citenamefont {Fang}, \citenamefont {Zhang},
  \citenamefont {Yue}, \citenamefont {Yan}, \citenamefont {Hu},\ and\
  \citenamefont {Su}}]{qin2016diverse}%
  \BibitemOpen
  \bibfield  {author} {\bibinfo {author} {\bibfnamefont {G.}~\bibnamefont
  {Qin}}, \bibinfo {author} {\bibfnamefont {Z.}~\bibnamefont {Qin}}, \bibinfo
  {author} {\bibfnamefont {W.-Z.}\ \bibnamefont {Fang}}, \bibinfo {author}
  {\bibfnamefont {L.-C.}\ \bibnamefont {Zhang}}, \bibinfo {author}
  {\bibfnamefont {S.-Y.}\ \bibnamefont {Yue}}, \bibinfo {author} {\bibfnamefont
  {Q.-B.}\ \bibnamefont {Yan}}, \bibinfo {author} {\bibfnamefont
  {M.}~\bibnamefont {Hu}},\ and\ \bibinfo {author} {\bibfnamefont
  {G.}~\bibnamefont {Su}},\ }\href
  {https://doi.org/https://doi.org/10.1039/C6NR01349J} {\bibfield  {journal}
  {\bibinfo  {journal} {Nanoscale}\ }\textbf {\bibinfo {volume} {8}},\ \bibinfo
  {pages} {11306} (\bibinfo {year} {2016})}\BibitemShut {NoStop}%
\bibitem [{\citenamefont {Wu}\ and\ \citenamefont
  {Zeng}(2016)}]{wu2016intrinsic}%
  \BibitemOpen
  \bibfield  {author} {\bibinfo {author} {\bibfnamefont {M.}~\bibnamefont
  {Wu}}\ and\ \bibinfo {author} {\bibfnamefont {X.~C.}\ \bibnamefont {Zeng}},\
  }\href {https://doi.org/https://doi.org/10.1021/acs.nanolett.6b00726}
  {\bibfield  {journal} {\bibinfo  {journal} {Nano letters}\ }\textbf {\bibinfo
  {volume} {16}},\ \bibinfo {pages} {3236} (\bibinfo {year}
  {2016})}\BibitemShut {NoStop}%
\bibitem [{\citenamefont {Ho}\ and\ \citenamefont
  {Li}(2017)}]{ho2017polarized}%
  \BibitemOpen
  \bibfield  {author} {\bibinfo {author} {\bibfnamefont {C.-H.}\ \bibnamefont
  {Ho}}\ and\ \bibinfo {author} {\bibfnamefont {J.-X.}\ \bibnamefont {Li}},\
  }\href {https://doi.org/https://doi.org/10.1002/adom.201600814} {\bibfield
  {journal} {\bibinfo  {journal} {Advanced Optical Materials}\ }\textbf
  {\bibinfo {volume} {5}},\ \bibinfo {pages} {1600814} (\bibinfo {year}
  {2017})}\BibitemShut {NoStop}%
\bibitem [{\citenamefont {Tan}\ \emph {et~al.}(2017{\natexlab{b}})\citenamefont
  {Tan}, \citenamefont {Zhang}, \citenamefont {Wang}, \citenamefont {Koirala},
  \citenamefont {Miyauchi},\ and\ \citenamefont
  {Matsuda}}]{tan2017polarization}%
  \BibitemOpen
  \bibfield  {author} {\bibinfo {author} {\bibfnamefont {D.}~\bibnamefont
  {Tan}}, \bibinfo {author} {\bibfnamefont {W.}~\bibnamefont {Zhang}}, \bibinfo
  {author} {\bibfnamefont {X.}~\bibnamefont {Wang}}, \bibinfo {author}
  {\bibfnamefont {S.}~\bibnamefont {Koirala}}, \bibinfo {author} {\bibfnamefont
  {Y.}~\bibnamefont {Miyauchi}},\ and\ \bibinfo {author} {\bibfnamefont
  {K.}~\bibnamefont {Matsuda}},\ }\href
  {https://doi.org/https://doi.org/10.1039/C7NR03040A} {\bibfield  {journal}
  {\bibinfo  {journal} {Nanoscale}\ }\textbf {\bibinfo {volume} {9}},\ \bibinfo
  {pages} {12425} (\bibinfo {year} {2017}{\natexlab{b}})}\BibitemShut {NoStop}%
\bibitem [{\citenamefont {Shafique}\ and\ \citenamefont
  {Shin}(2017)}]{shafique2017thermoelectric}%
  \BibitemOpen
  \bibfield  {author} {\bibinfo {author} {\bibfnamefont {A.}~\bibnamefont
  {Shafique}}\ and\ \bibinfo {author} {\bibfnamefont {Y.-H.}\ \bibnamefont
  {Shin}},\ }\href {https://doi.org/https://doi.org/10.1038/s41598-017-00598-7}
  {\bibfield  {journal} {\bibinfo  {journal} {Scientific reports}\ }\textbf
  {\bibinfo {volume} {7}},\ \bibinfo {pages} {506} (\bibinfo {year}
  {2017})}\BibitemShut {NoStop}%
\bibitem [{\citenamefont {Oliva}\ \emph
  {et~al.}(2020{\natexlab{a}})\citenamefont {Oliva}, \citenamefont
  {Wo{\'z}niak}, \citenamefont {Dybala}, \citenamefont {To{\l}{\l}oczko},
  \citenamefont {Kopaczek}, \citenamefont {Scharoch},\ and\ \citenamefont
  {Kudrawiec}}]{oliva2020valley}%
  \BibitemOpen
  \bibfield  {author} {\bibinfo {author} {\bibfnamefont {R.}~\bibnamefont
  {Oliva}}, \bibinfo {author} {\bibfnamefont {T.}~\bibnamefont {Wo{\'z}niak}},
  \bibinfo {author} {\bibfnamefont {F.}~\bibnamefont {Dybala}}, \bibinfo
  {author} {\bibfnamefont {A.}~\bibnamefont {To{\l}{\l}oczko}}, \bibinfo
  {author} {\bibfnamefont {J.}~\bibnamefont {Kopaczek}}, \bibinfo {author}
  {\bibfnamefont {P.}~\bibnamefont {Scharoch}},\ and\ \bibinfo {author}
  {\bibfnamefont {R.}~\bibnamefont {Kudrawiec}},\ }\href
  {https://doi.org/https://doi.org/10.1103/PhysRevB.101.235205} {\bibfield
  {journal} {\bibinfo  {journal} {Physical Review B}\ }\textbf {\bibinfo
  {volume} {101}},\ \bibinfo {pages} {235205} (\bibinfo {year}
  {2020}{\natexlab{a}})}\BibitemShut {NoStop}%
\bibitem [{\citenamefont {Yang}\ \emph {et~al.}(2020)\citenamefont {Yang},
  \citenamefont {Cao}, \citenamefont {You}, \citenamefont {Shi},\ and\
  \citenamefont {Qian}}]{yang2020intrinsic}%
  \BibitemOpen
  \bibfield  {author} {\bibinfo {author} {\bibfnamefont {M.}~\bibnamefont
  {Yang}}, \bibinfo {author} {\bibfnamefont {S.}~\bibnamefont {Cao}}, \bibinfo
  {author} {\bibfnamefont {Q.}~\bibnamefont {You}}, \bibinfo {author}
  {\bibfnamefont {L.-B.}\ \bibnamefont {Shi}},\ and\ \bibinfo {author}
  {\bibfnamefont {P.}~\bibnamefont {Qian}},\ }\href
  {https://doi.org/https://doi.org/10.1016/j.physe.2019.113877} {\bibfield
  {journal} {\bibinfo  {journal} {Physica E: Low-dimensional Systems and
  Nanostructures}\ }\textbf {\bibinfo {volume} {118}},\ \bibinfo {pages}
  {113877} (\bibinfo {year} {2020})}\BibitemShut {NoStop}%
\bibitem [{\citenamefont {Li}\ \emph {et~al.}(2016)\citenamefont {Li},
  \citenamefont {Liu}, \citenamefont {Wang},\ and\ \citenamefont
  {Li}}]{li2016germanium}%
  \BibitemOpen
  \bibfield  {author} {\bibinfo {author} {\bibfnamefont {F.}~\bibnamefont
  {Li}}, \bibinfo {author} {\bibfnamefont {X.}~\bibnamefont {Liu}}, \bibinfo
  {author} {\bibfnamefont {Y.}~\bibnamefont {Wang}},\ and\ \bibinfo {author}
  {\bibfnamefont {Y.}~\bibnamefont {Li}},\ }\href
  {https://doi.org/https://doi.org/10.1039/C6TC00454G} {\bibfield  {journal}
  {\bibinfo  {journal} {Journal of Materials Chemistry C}\ }\textbf {\bibinfo
  {volume} {4}},\ \bibinfo {pages} {2155} (\bibinfo {year} {2016})}\BibitemShut
  {NoStop}%
\bibitem [{\citenamefont {Salpeter}\ and\ \citenamefont
  {Bethe}(1951)}]{PhysRev.84.1232}%
  \BibitemOpen
  \bibfield  {author} {\bibinfo {author} {\bibfnamefont {E.~E.}\ \bibnamefont
  {Salpeter}}\ and\ \bibinfo {author} {\bibfnamefont {H.~A.}\ \bibnamefont
  {Bethe}},\ }\href {https://doi.org/10.1103/PhysRev.84.1232} {\bibfield
  {journal} {\bibinfo  {journal} {Phys. Rev.}\ }\textbf {\bibinfo {volume}
  {84}},\ \bibinfo {pages} {1232} (\bibinfo {year} {1951})}\BibitemShut
  {NoStop}%
\bibitem [{\citenamefont {Rohlfing}\ and\ \citenamefont
  {Louie}(2000)}]{PhysRevB.62.4927}%
  \BibitemOpen
  \bibfield  {author} {\bibinfo {author} {\bibfnamefont {M.}~\bibnamefont
  {Rohlfing}}\ and\ \bibinfo {author} {\bibfnamefont {S.~G.}\ \bibnamefont
  {Louie}},\ }\href {https://doi.org/10.1103/PhysRevB.62.4927} {\bibfield
  {journal} {\bibinfo  {journal} {Phys. Rev. B}\ }\textbf {\bibinfo {volume}
  {62}},\ \bibinfo {pages} {4927} (\bibinfo {year} {2000})}\BibitemShut
  {NoStop}%
\bibitem [{\citenamefont {Cadiz}\ \emph {et~al.}(2017)\citenamefont {Cadiz},
  \citenamefont {Courtade}, \citenamefont {Robert}, \citenamefont {Wang},
  \citenamefont {Shen}, \citenamefont {Cai}, \citenamefont {Taniguchi},
  \citenamefont {Watanabe}, \citenamefont {Carrere}, \citenamefont {Lagarde},
  \citenamefont {Manca}, \citenamefont {Amand}, \citenamefont {Renucci},
  \citenamefont {Tongay}, \citenamefont {Marie},\ and\ \citenamefont
  {Urbaszek}}]{cadiz2017}%
  \BibitemOpen
  \bibfield  {author} {\bibinfo {author} {\bibfnamefont {F.}~\bibnamefont
  {Cadiz}}, \bibinfo {author} {\bibfnamefont {E.}~\bibnamefont {Courtade}},
  \bibinfo {author} {\bibfnamefont {C.}~\bibnamefont {Robert}}, \bibinfo
  {author} {\bibfnamefont {G.}~\bibnamefont {Wang}}, \bibinfo {author}
  {\bibfnamefont {Y.}~\bibnamefont {Shen}}, \bibinfo {author} {\bibfnamefont
  {H.}~\bibnamefont {Cai}}, \bibinfo {author} {\bibfnamefont {T.}~\bibnamefont
  {Taniguchi}}, \bibinfo {author} {\bibfnamefont {K.}~\bibnamefont {Watanabe}},
  \bibinfo {author} {\bibfnamefont {H.}~\bibnamefont {Carrere}}, \bibinfo
  {author} {\bibfnamefont {D.}~\bibnamefont {Lagarde}}, \bibinfo {author}
  {\bibfnamefont {M.}~\bibnamefont {Manca}}, \bibinfo {author} {\bibfnamefont
  {T.}~\bibnamefont {Amand}}, \bibinfo {author} {\bibfnamefont
  {P.}~\bibnamefont {Renucci}}, \bibinfo {author} {\bibfnamefont
  {S.}~\bibnamefont {Tongay}}, \bibinfo {author} {\bibfnamefont
  {X.}~\bibnamefont {Marie}},\ and\ \bibinfo {author} {\bibfnamefont
  {B.}~\bibnamefont {Urbaszek}},\ }\href
  {https://doi.org/10.1103/PhysRevX.7.021026} {\bibfield  {journal} {\bibinfo
  {journal} {Phys. Rev. X}\ }\textbf {\bibinfo {volume} {7}},\ \bibinfo {pages}
  {021026} (\bibinfo {year} {2017})}\BibitemShut {NoStop}%
\bibitem [{\citenamefont {O’Donnell}\ and\ \citenamefont
  {Chen}(1991)}]{odonnell1991}%
  \BibitemOpen
  \bibfield  {author} {\bibinfo {author} {\bibfnamefont {K.~P.}\ \bibnamefont
  {O’Donnell}}\ and\ \bibinfo {author} {\bibfnamefont {X.}~\bibnamefont
  {Chen}},\ }\href {https://doi.org/10.1063/1.104723} {\bibfield  {journal}
  {\bibinfo  {journal} {Applied Physics Letters}\ }\textbf {\bibinfo {volume}
  {58}},\ \bibinfo {pages} {2924} (\bibinfo {year} {1991})},\ \Eprint
  {https://arxiv.org/abs/https://doi.org/10.1063/1.104723}
  {https://doi.org/10.1063/1.104723} \BibitemShut {NoStop}%
\bibitem [{\citenamefont {Rudin}\ \emph {et~al.}(1990)\citenamefont {Rudin},
  \citenamefont {Reinecke},\ and\ \citenamefont {Segall}}]{rudin1990}%
  \BibitemOpen
  \bibfield  {author} {\bibinfo {author} {\bibfnamefont {S.}~\bibnamefont
  {Rudin}}, \bibinfo {author} {\bibfnamefont {T.~L.}\ \bibnamefont
  {Reinecke}},\ and\ \bibinfo {author} {\bibfnamefont {B.}~\bibnamefont
  {Segall}},\ }\href {https://doi.org/10.1103/PhysRevB.42.11218} {\bibfield
  {journal} {\bibinfo  {journal} {Phys. Rev. B}\ }\textbf {\bibinfo {volume}
  {42}},\ \bibinfo {pages} {11218} (\bibinfo {year} {1990})}\BibitemShut
  {NoStop}%
\bibitem [{\citenamefont {Ding}\ \emph {et~al.}(2015)\citenamefont {Ding},
  \citenamefont {Gao},\ and\ \citenamefont {Yao}}]{ding2015high}%
  \BibitemOpen
  \bibfield  {author} {\bibinfo {author} {\bibfnamefont {G.}~\bibnamefont
  {Ding}}, \bibinfo {author} {\bibfnamefont {G.}~\bibnamefont {Gao}},\ and\
  \bibinfo {author} {\bibfnamefont {K.}~\bibnamefont {Yao}},\ }\href
  {https://doi.org/https://doi.org/10.1038/srep09567} {\bibfield  {journal}
  {\bibinfo  {journal} {Scientific reports}\ }\textbf {\bibinfo {volume} {5}},\
  \bibinfo {pages} {9567} (\bibinfo {year} {2015})}\BibitemShut {NoStop}%
\bibitem [{\citenamefont {Hao}\ \emph {et~al.}(2016)\citenamefont {Hao},
  \citenamefont {Shi}, \citenamefont {Dravid}, \citenamefont {Kanatzidis},\
  and\ \citenamefont {Wolverton}}]{hao2016computational}%
  \BibitemOpen
  \bibfield  {author} {\bibinfo {author} {\bibfnamefont {S.}~\bibnamefont
  {Hao}}, \bibinfo {author} {\bibfnamefont {F.}~\bibnamefont {Shi}}, \bibinfo
  {author} {\bibfnamefont {V.~P.}\ \bibnamefont {Dravid}}, \bibinfo {author}
  {\bibfnamefont {M.~G.}\ \bibnamefont {Kanatzidis}},\ and\ \bibinfo {author}
  {\bibfnamefont {C.}~\bibnamefont {Wolverton}},\ }\href
  {https://doi.org/https://doi.org/10.1021/acs.chemmater.6b01164} {\bibfield
  {journal} {\bibinfo  {journal} {Chemistry of Materials}\ }\textbf {\bibinfo
  {volume} {28}},\ \bibinfo {pages} {3218} (\bibinfo {year}
  {2016})}\BibitemShut {NoStop}%
\bibitem [{\citenamefont {Grandke}\ and\ \citenamefont
  {Ley}(1977)}]{PhysRevB.16.832}%
  \BibitemOpen
  \bibfield  {author} {\bibinfo {author} {\bibfnamefont {T.}~\bibnamefont
  {Grandke}}\ and\ \bibinfo {author} {\bibfnamefont {L.}~\bibnamefont {Ley}},\
  }\href {https://doi.org/10.1103/PhysRevB.16.832} {\bibfield  {journal}
  {\bibinfo  {journal} {Phys. Rev. B}\ }\textbf {\bibinfo {volume} {16}},\
  \bibinfo {pages} {832} (\bibinfo {year} {1977})}\BibitemShut {NoStop}%
\bibitem [{\citenamefont {Wiley}\ \emph {et~al.}(1975)\citenamefont {Wiley},
  \citenamefont {Breitschwerdt},\ and\ \citenamefont
  {Schönherr}}]{WILEY1975355}%
  \BibitemOpen
  \bibfield  {author} {\bibinfo {author} {\bibfnamefont {J.}~\bibnamefont
  {Wiley}}, \bibinfo {author} {\bibfnamefont {A.}~\bibnamefont
  {Breitschwerdt}},\ and\ \bibinfo {author} {\bibfnamefont {E.}~\bibnamefont
  {Schönherr}},\ }\href
  {https://doi.org/https://doi.org/10.1016/0038-1098(75)90311-7} {\bibfield
  {journal} {\bibinfo  {journal} {Solid State Communications}\ }\textbf
  {\bibinfo {volume} {17}},\ \bibinfo {pages} {355} (\bibinfo {year}
  {1975})}\BibitemShut {NoStop}%
\bibitem [{\citenamefont {Yabumoto}(1958)}]{JPSJ.13.559}%
  \BibitemOpen
  \bibfield  {author} {\bibinfo {author} {\bibfnamefont {T.}~\bibnamefont
  {Yabumoto}},\ }\href {https://doi.org/10.1143/JPSJ.13.559} {\bibfield
  {journal} {\bibinfo  {journal} {Journal of the Physical Society of Japan}\
  }\textbf {\bibinfo {volume} {13}},\ \bibinfo {pages} {559} (\bibinfo {year}
  {1958})}\BibitemShut {NoStop}%
\bibitem [{\citenamefont {Postorino}\ \emph
  {et~al.}(2020{\natexlab{b}})\citenamefont {Postorino}, \citenamefont {Sun},
  \citenamefont {Fiedler}, \citenamefont {Lee Cheong~Lem}, \citenamefont
  {Palummo},\ and\ \citenamefont {Camilli}}]{ma13163568}%
  \BibitemOpen
  \bibfield  {author} {\bibinfo {author} {\bibfnamefont {S.}~\bibnamefont
  {Postorino}}, \bibinfo {author} {\bibfnamefont {J.}~\bibnamefont {Sun}},
  \bibinfo {author} {\bibfnamefont {S.}~\bibnamefont {Fiedler}}, \bibinfo
  {author} {\bibfnamefont {L.~O.}\ \bibnamefont {Lee Cheong~Lem}}, \bibinfo
  {author} {\bibfnamefont {M.}~\bibnamefont {Palummo}},\ and\ \bibinfo {author}
  {\bibfnamefont {L.}~\bibnamefont {Camilli}},\ }\bibfield  {journal} {\bibinfo
   {journal} {Materials}\ }\textbf {\bibinfo {volume} {13}},\ \href
  {https://doi.org/10.3390/ma13163568} {10.3390/ma13163568} (\bibinfo {year}
  {2020}{\natexlab{b}})\BibitemShut {NoStop}%
\bibitem [{\citenamefont {Malone}\ and\ \citenamefont
  {Kaxiras}(2013)}]{PhysRevB.87.245312}%
  \BibitemOpen
  \bibfield  {author} {\bibinfo {author} {\bibfnamefont {B.~D.}\ \bibnamefont
  {Malone}}\ and\ \bibinfo {author} {\bibfnamefont {E.}~\bibnamefont
  {Kaxiras}},\ }\href {https://doi.org/10.1103/PhysRevB.87.245312} {\bibfield
  {journal} {\bibinfo  {journal} {Phys. Rev. B}\ }\textbf {\bibinfo {volume}
  {87}},\ \bibinfo {pages} {245312} (\bibinfo {year} {2013})}\BibitemShut
  {NoStop}%
\bibitem [{\citenamefont {Tran}\ \emph {et~al.}(2014)\citenamefont {Tran},
  \citenamefont {Soklaski}, \citenamefont {Liang},\ and\ \citenamefont
  {Yang}}]{PhysRevB.89.235319}%
  \BibitemOpen
  \bibfield  {author} {\bibinfo {author} {\bibfnamefont {V.}~\bibnamefont
  {Tran}}, \bibinfo {author} {\bibfnamefont {R.}~\bibnamefont {Soklaski}},
  \bibinfo {author} {\bibfnamefont {Y.}~\bibnamefont {Liang}},\ and\ \bibinfo
  {author} {\bibfnamefont {L.}~\bibnamefont {Yang}},\ }\href
  {https://doi.org/10.1103/PhysRevB.89.235319} {\bibfield  {journal} {\bibinfo
  {journal} {Phys. Rev. B}\ }\textbf {\bibinfo {volume} {89}},\ \bibinfo
  {pages} {235319} (\bibinfo {year} {2014})}\BibitemShut {NoStop}%
\bibitem [{\citenamefont {Cheiwchanchamnangij}\ and\ \citenamefont
  {Lambrecht}(2012)}]{PhysRevB.85.205302}%
  \BibitemOpen
  \bibfield  {author} {\bibinfo {author} {\bibfnamefont {T.}~\bibnamefont
  {Cheiwchanchamnangij}}\ and\ \bibinfo {author} {\bibfnamefont {W.~R.~L.}\
  \bibnamefont {Lambrecht}},\ }\href
  {https://doi.org/10.1103/PhysRevB.85.205302} {\bibfield  {journal} {\bibinfo
  {journal} {Phys. Rev. B}\ }\textbf {\bibinfo {volume} {85}},\ \bibinfo
  {pages} {205302} (\bibinfo {year} {2012})}\BibitemShut {NoStop}%
\bibitem [{\citenamefont {Shan}\ \emph {et~al.}(1996)\citenamefont {Shan},
  \citenamefont {Little}, \citenamefont {Fischer}, \citenamefont {Song},
  \citenamefont {Goldenberg}, \citenamefont {Perry}, \citenamefont {Bremser},\
  and\ \citenamefont {Davis}}]{PhysRevB.54.16369}%
  \BibitemOpen
  \bibfield  {author} {\bibinfo {author} {\bibfnamefont {W.}~\bibnamefont
  {Shan}}, \bibinfo {author} {\bibfnamefont {B.~D.}\ \bibnamefont {Little}},
  \bibinfo {author} {\bibfnamefont {A.~J.}\ \bibnamefont {Fischer}}, \bibinfo
  {author} {\bibfnamefont {J.~J.}\ \bibnamefont {Song}}, \bibinfo {author}
  {\bibfnamefont {B.}~\bibnamefont {Goldenberg}}, \bibinfo {author}
  {\bibfnamefont {W.~G.}\ \bibnamefont {Perry}}, \bibinfo {author}
  {\bibfnamefont {M.~D.}\ \bibnamefont {Bremser}},\ and\ \bibinfo {author}
  {\bibfnamefont {R.~F.}\ \bibnamefont {Davis}},\ }\href
  {https://doi.org/10.1103/PhysRevB.54.16369} {\bibfield  {journal} {\bibinfo
  {journal} {Phys. Rev. B}\ }\textbf {\bibinfo {volume} {54}},\ \bibinfo
  {pages} {16369} (\bibinfo {year} {1996})}\BibitemShut {NoStop}%
\bibitem [{\citenamefont {Belov}\ and\ \citenamefont
  {Khramtsov}(2017)}]{belov2017binding}%
  \BibitemOpen
  \bibfield  {author} {\bibinfo {author} {\bibfnamefont {P.}~\bibnamefont
  {Belov}}\ and\ \bibinfo {author} {\bibfnamefont {E.}~\bibnamefont
  {Khramtsov}},\ }in\ \href {https://doi.org/10.1088/1742-6596/816/1/012018}
  {\emph {\bibinfo {booktitle} {Journal of Physics: Conference Series}}},\
  Vol.\ \bibinfo {volume} {816}\ (\bibinfo {organization} {IOP Publishing},\
  \bibinfo {year} {2017})\ p.\ \bibinfo {pages} {012018}\BibitemShut {NoStop}%
\bibitem [{\citenamefont {Kuzuba}\ and\ \citenamefont
  {Era}(1976)}]{kuzuba1976nearly}%
  \BibitemOpen
  \bibfield  {author} {\bibinfo {author} {\bibfnamefont {T.}~\bibnamefont
  {Kuzuba}}\ and\ \bibinfo {author} {\bibfnamefont {K.}~\bibnamefont {Era}},\
  }\href {https://doi.org/https://doi.org/10.1143/JPSJ.40.134} {\bibfield
  {journal} {\bibinfo  {journal} {Journal of the Physical Society of Japan}\
  }\textbf {\bibinfo {volume} {40}},\ \bibinfo {pages} {134} (\bibinfo {year}
  {1976})}\BibitemShut {NoStop}%
\bibitem [{\citenamefont {Baldereschi}\ and\ \citenamefont
  {Diaz}(1970)}]{baldereschi1970anisotropy}%
  \BibitemOpen
  \bibfield  {author} {\bibinfo {author} {\bibfnamefont {A.}~\bibnamefont
  {Baldereschi}}\ and\ \bibinfo {author} {\bibfnamefont {M.}~\bibnamefont
  {Diaz}},\ }\href {https://doi.org/https://doi.org/10.1007/BF02710415}
  {\bibfield  {journal} {\bibinfo  {journal} {Il Nuovo Cimento B (1965-1970)}\
  }\textbf {\bibinfo {volume} {68}},\ \bibinfo {pages} {217} (\bibinfo {year}
  {1970})}\BibitemShut {NoStop}%
\bibitem [{\citenamefont {Dresselhaus}(1956)}]{dresselhaus1956effective}%
  \BibitemOpen
  \bibfield  {author} {\bibinfo {author} {\bibfnamefont {G.}~\bibnamefont
  {Dresselhaus}},\ }\href
  {https://doi.org/https://doi.org/10.1016/0022-3697(56)90004-X} {\bibfield
  {journal} {\bibinfo  {journal} {Journal of Physics and Chemistry of Solids}\
  }\textbf {\bibinfo {volume} {1}},\ \bibinfo {pages} {14} (\bibinfo {year}
  {1956})}\BibitemShut {NoStop}%
\bibitem [{\citenamefont {Shinada}\ and\ \citenamefont
  {Sugano}(1966)}]{Shinada_}%
  \BibitemOpen
  \bibfield  {author} {\bibinfo {author} {\bibfnamefont {M.}~\bibnamefont
  {Shinada}}\ and\ \bibinfo {author} {\bibfnamefont {S.}~\bibnamefont
  {Sugano}},\ }\href {https://doi.org/10.1143/JPSJ.21.1936} {\bibfield
  {journal} {\bibinfo  {journal} {Journal of the Physical Society of Japan}\
  }\textbf {\bibinfo {volume} {21}},\ \bibinfo {pages} {1936} (\bibinfo {year}
  {1966})}\BibitemShut {NoStop}%
\bibitem [{\citenamefont {Shubina}\ \emph {et~al.}(2019)\citenamefont
  {Shubina}, \citenamefont {Desrat}, \citenamefont {Moret}, \citenamefont
  {Tiberj}, \citenamefont {Briot}, \citenamefont {Davydov}, \citenamefont
  {Platonov}, \citenamefont {Semina},\ and\ \citenamefont
  {Gil}}]{Shubina2019InSeAA}%
  \BibitemOpen
  \bibfield  {author} {\bibinfo {author} {\bibfnamefont {T.~V.}\ \bibnamefont
  {Shubina}}, \bibinfo {author} {\bibfnamefont {W.}~\bibnamefont {Desrat}},
  \bibinfo {author} {\bibfnamefont {M.}~\bibnamefont {Moret}}, \bibinfo
  {author} {\bibfnamefont {A.}~\bibnamefont {Tiberj}}, \bibinfo {author}
  {\bibfnamefont {O.}~\bibnamefont {Briot}}, \bibinfo {author} {\bibfnamefont
  {V.~Y.}\ \bibnamefont {Davydov}}, \bibinfo {author} {\bibfnamefont {A.~V.}\
  \bibnamefont {Platonov}}, \bibinfo {author} {\bibfnamefont {M.~A.}\
  \bibnamefont {Semina}},\ and\ \bibinfo {author} {\bibfnamefont
  {B.}~\bibnamefont {Gil}},\ }\bibfield  {journal} {\bibinfo  {journal} {Nature
  Communications}\ }\textbf {\bibinfo {volume} {10}},\ \href
  {https://doi.org/10.1038/s41467-019-11487-0} {10.1038/s41467-019-11487-0}
  (\bibinfo {year} {2019})\BibitemShut {NoStop}%
\bibitem [{\citenamefont {Lafrentz}\ \emph {et~al.}(2013)\citenamefont
  {Lafrentz}, \citenamefont {Brunne}, \citenamefont {Kaminski}, \citenamefont
  {Pavlov}, \citenamefont {Rodina}, \citenamefont {Pisarev}, \citenamefont
  {Yakovlev}, \citenamefont {Bakin},\ and\ \citenamefont
  {Bayer}}]{PhysRevLett.110.116402}%
  \BibitemOpen
  \bibfield  {author} {\bibinfo {author} {\bibfnamefont {M.}~\bibnamefont
  {Lafrentz}}, \bibinfo {author} {\bibfnamefont {D.}~\bibnamefont {Brunne}},
  \bibinfo {author} {\bibfnamefont {B.}~\bibnamefont {Kaminski}}, \bibinfo
  {author} {\bibfnamefont {V.~V.}\ \bibnamefont {Pavlov}}, \bibinfo {author}
  {\bibfnamefont {A.~V.}\ \bibnamefont {Rodina}}, \bibinfo {author}
  {\bibfnamefont {R.~V.}\ \bibnamefont {Pisarev}}, \bibinfo {author}
  {\bibfnamefont {D.~R.}\ \bibnamefont {Yakovlev}}, \bibinfo {author}
  {\bibfnamefont {A.}~\bibnamefont {Bakin}},\ and\ \bibinfo {author}
  {\bibfnamefont {M.}~\bibnamefont {Bayer}},\ }\href
  {https://doi.org/10.1103/PhysRevLett.110.116402} {\bibfield  {journal}
  {\bibinfo  {journal} {Phys. Rev. Lett.}\ }\textbf {\bibinfo {volume} {110}},\
  \bibinfo {pages} {116402} (\bibinfo {year} {2013})}\BibitemShut {NoStop}%
\bibitem [{\citenamefont {Brunne}\ \emph {et~al.}(2015)\citenamefont {Brunne},
  \citenamefont {Lafrentz}, \citenamefont {Pavlov}, \citenamefont {Pisarev},
  \citenamefont {Rodina}, \citenamefont {Yakovlev},\ and\ \citenamefont
  {Bayer}}]{PhysRevB.92.085202}%
  \BibitemOpen
  \bibfield  {author} {\bibinfo {author} {\bibfnamefont {D.}~\bibnamefont
  {Brunne}}, \bibinfo {author} {\bibfnamefont {M.}~\bibnamefont {Lafrentz}},
  \bibinfo {author} {\bibfnamefont {V.~V.}\ \bibnamefont {Pavlov}}, \bibinfo
  {author} {\bibfnamefont {R.~V.}\ \bibnamefont {Pisarev}}, \bibinfo {author}
  {\bibfnamefont {A.~V.}\ \bibnamefont {Rodina}}, \bibinfo {author}
  {\bibfnamefont {D.~R.}\ \bibnamefont {Yakovlev}},\ and\ \bibinfo {author}
  {\bibfnamefont {M.}~\bibnamefont {Bayer}},\ }\href
  {https://doi.org/10.1103/PhysRevB.92.085202} {\bibfield  {journal} {\bibinfo
  {journal} {Phys. Rev. B}\ }\textbf {\bibinfo {volume} {92}},\ \bibinfo
  {pages} {085202} (\bibinfo {year} {2015})}\BibitemShut {NoStop}%
\bibitem [{\citenamefont {Funk}\ \emph {et~al.}(2021)\citenamefont {Funk},
  \citenamefont {Wagner}, \citenamefont {Wietek}, \citenamefont {Ziegler},
  \citenamefont {F\"orste}, \citenamefont {Lindlau}, \citenamefont {F\"org},
  \citenamefont {Watanabe}, \citenamefont {Taniguchi}, \citenamefont
  {Chernikov},\ and\ \citenamefont {H\"ogele}}]{PhysRevResearch.3.L042019}%
  \BibitemOpen
  \bibfield  {author} {\bibinfo {author} {\bibfnamefont {V.}~\bibnamefont
  {Funk}}, \bibinfo {author} {\bibfnamefont {K.}~\bibnamefont {Wagner}},
  \bibinfo {author} {\bibfnamefont {E.}~\bibnamefont {Wietek}}, \bibinfo
  {author} {\bibfnamefont {J.~D.}\ \bibnamefont {Ziegler}}, \bibinfo {author}
  {\bibfnamefont {J.}~\bibnamefont {F\"orste}}, \bibinfo {author}
  {\bibfnamefont {J.}~\bibnamefont {Lindlau}}, \bibinfo {author} {\bibfnamefont
  {M.}~\bibnamefont {F\"org}}, \bibinfo {author} {\bibfnamefont
  {K.}~\bibnamefont {Watanabe}}, \bibinfo {author} {\bibfnamefont
  {T.}~\bibnamefont {Taniguchi}}, \bibinfo {author} {\bibfnamefont
  {A.}~\bibnamefont {Chernikov}},\ and\ \bibinfo {author} {\bibfnamefont
  {A.}~\bibnamefont {H\"ogele}},\ }\href
  {https://doi.org/10.1103/PhysRevResearch.3.L042019} {\bibfield  {journal}
  {\bibinfo  {journal} {Phys. Rev. Res.}\ }\textbf {\bibinfo {volume} {3}},\
  \bibinfo {pages} {L042019} (\bibinfo {year} {2021})}\BibitemShut {NoStop}%
\bibitem [{\citenamefont {Giannozzi}\ \emph {et~al.}(2009)\citenamefont
  {Giannozzi}, \citenamefont {Baroni}, \citenamefont {Bonini}, \citenamefont
  {Calandra}, \citenamefont {Car}, \citenamefont {Cavazzoni}, \citenamefont
  {Ceresoli}, \citenamefont {Chiarotti}, \citenamefont {Cococcioni},
  \citenamefont {Dabo}, \citenamefont {Corso}, \citenamefont {de~Gironcoli},
  \citenamefont {Fabris}, \citenamefont {Fratesi}, \citenamefont {Gebauer},
  \citenamefont {Gerstmann}, \citenamefont {Gougoussis}, \citenamefont
  {Kokalj}, \citenamefont {Lazzeri}, \citenamefont {Martin-Samos},
  \citenamefont {Marzari}, \citenamefont {Mauri}, \citenamefont {Mazzarello},
  \citenamefont {Paolini}, \citenamefont {Pasquarello}, \citenamefont
  {Paulatto}, \citenamefont {Sbraccia}, \citenamefont {Scandolo}, \citenamefont
  {Sclauzero}, \citenamefont {Seitsonen}, \citenamefont {Smogunov},
  \citenamefont {Umari},\ and\ \citenamefont {Wentzcovitch}}]{Giannozzi_2009}%
  \BibitemOpen
  \bibfield  {author} {\bibinfo {author} {\bibfnamefont {P.}~\bibnamefont
  {Giannozzi}}, \bibinfo {author} {\bibfnamefont {S.}~\bibnamefont {Baroni}},
  \bibinfo {author} {\bibfnamefont {N.}~\bibnamefont {Bonini}}, \bibinfo
  {author} {\bibfnamefont {M.}~\bibnamefont {Calandra}}, \bibinfo {author}
  {\bibfnamefont {R.}~\bibnamefont {Car}}, \bibinfo {author} {\bibfnamefont
  {C.}~\bibnamefont {Cavazzoni}}, \bibinfo {author} {\bibfnamefont
  {D.}~\bibnamefont {Ceresoli}}, \bibinfo {author} {\bibfnamefont {G.~L.}\
  \bibnamefont {Chiarotti}}, \bibinfo {author} {\bibfnamefont {M.}~\bibnamefont
  {Cococcioni}}, \bibinfo {author} {\bibfnamefont {I.}~\bibnamefont {Dabo}},
  \bibinfo {author} {\bibfnamefont {A.~D.}\ \bibnamefont {Corso}}, \bibinfo
  {author} {\bibfnamefont {S.}~\bibnamefont {de~Gironcoli}}, \bibinfo {author}
  {\bibfnamefont {S.}~\bibnamefont {Fabris}}, \bibinfo {author} {\bibfnamefont
  {G.}~\bibnamefont {Fratesi}}, \bibinfo {author} {\bibfnamefont
  {R.}~\bibnamefont {Gebauer}}, \bibinfo {author} {\bibfnamefont
  {U.}~\bibnamefont {Gerstmann}}, \bibinfo {author} {\bibfnamefont
  {C.}~\bibnamefont {Gougoussis}}, \bibinfo {author} {\bibfnamefont
  {A.}~\bibnamefont {Kokalj}}, \bibinfo {author} {\bibfnamefont
  {M.}~\bibnamefont {Lazzeri}}, \bibinfo {author} {\bibfnamefont
  {L.}~\bibnamefont {Martin-Samos}}, \bibinfo {author} {\bibfnamefont
  {N.}~\bibnamefont {Marzari}}, \bibinfo {author} {\bibfnamefont
  {F.}~\bibnamefont {Mauri}}, \bibinfo {author} {\bibfnamefont
  {R.}~\bibnamefont {Mazzarello}}, \bibinfo {author} {\bibfnamefont
  {S.}~\bibnamefont {Paolini}}, \bibinfo {author} {\bibfnamefont
  {A.}~\bibnamefont {Pasquarello}}, \bibinfo {author} {\bibfnamefont
  {L.}~\bibnamefont {Paulatto}}, \bibinfo {author} {\bibfnamefont
  {C.}~\bibnamefont {Sbraccia}}, \bibinfo {author} {\bibfnamefont
  {S.}~\bibnamefont {Scandolo}}, \bibinfo {author} {\bibfnamefont
  {G.}~\bibnamefont {Sclauzero}}, \bibinfo {author} {\bibfnamefont {A.~P.}\
  \bibnamefont {Seitsonen}}, \bibinfo {author} {\bibfnamefont {A.}~\bibnamefont
  {Smogunov}}, \bibinfo {author} {\bibfnamefont {P.}~\bibnamefont {Umari}},\
  and\ \bibinfo {author} {\bibfnamefont {R.~M.}\ \bibnamefont {Wentzcovitch}},\
  }\href {https://doi.org/10.1088/0953-8984/21/39/395502} {\bibfield  {journal}
  {\bibinfo  {journal} {Journal of Physics: Condensed Matter}\ }\textbf
  {\bibinfo {volume} {21}},\ \bibinfo {pages} {395502} (\bibinfo {year}
  {2009})}\BibitemShut {NoStop}%
\bibitem [{\citenamefont {Giannozzi}\ \emph {et~al.}(2017)\citenamefont
  {Giannozzi}, \citenamefont {Andreussi}, \citenamefont {Brumme}, \citenamefont
  {Bunau}, \citenamefont {Nardelli}, \citenamefont {Calandra}, \citenamefont
  {Car}, \citenamefont {Cavazzoni}, \citenamefont {Ceresoli}, \citenamefont
  {Cococcioni}, \citenamefont {Colonna}, \citenamefont {Carnimeo},
  \citenamefont {Corso}, \citenamefont {de~Gironcoli}, \citenamefont {Delugas},
  \citenamefont {DiStasio}, \citenamefont {Ferretti}, \citenamefont {Floris},
  \citenamefont {Fratesi}, \citenamefont {Fugallo}, \citenamefont {Gebauer},
  \citenamefont {Gerstmann}, \citenamefont {Giustino}, \citenamefont {Gorni},
  \citenamefont {Jia}, \citenamefont {Kawamura}, \citenamefont {Ko},
  \citenamefont {Kokalj}, \citenamefont {Küçükbenli}, \citenamefont
  {Lazzeri}, \citenamefont {Marsili}, \citenamefont {Marzari}, \citenamefont
  {Mauri}, \citenamefont {Nguyen}, \citenamefont {Nguyen}, \citenamefont {de-la
  Roza}, \citenamefont {Paulatto}, \citenamefont {Poncé}, \citenamefont
  {Rocca}, \citenamefont {Sabatini}, \citenamefont {Santra}, \citenamefont
  {Schlipf}, \citenamefont {Seitsonen}, \citenamefont {Smogunov}, \citenamefont
  {Timrov}, \citenamefont {Thonhauser}, \citenamefont {Umari}, \citenamefont
  {Vast}, \citenamefont {Wu},\ and\ \citenamefont {Baroni}}]{Giannozzi_2017}%
  \BibitemOpen
  \bibfield  {author} {\bibinfo {author} {\bibfnamefont {P.}~\bibnamefont
  {Giannozzi}}, \bibinfo {author} {\bibfnamefont {O.}~\bibnamefont
  {Andreussi}}, \bibinfo {author} {\bibfnamefont {T.}~\bibnamefont {Brumme}},
  \bibinfo {author} {\bibfnamefont {O.}~\bibnamefont {Bunau}}, \bibinfo
  {author} {\bibfnamefont {M.~B.}\ \bibnamefont {Nardelli}}, \bibinfo {author}
  {\bibfnamefont {M.}~\bibnamefont {Calandra}}, \bibinfo {author}
  {\bibfnamefont {R.}~\bibnamefont {Car}}, \bibinfo {author} {\bibfnamefont
  {C.}~\bibnamefont {Cavazzoni}}, \bibinfo {author} {\bibfnamefont
  {D.}~\bibnamefont {Ceresoli}}, \bibinfo {author} {\bibfnamefont
  {M.}~\bibnamefont {Cococcioni}}, \bibinfo {author} {\bibfnamefont
  {N.}~\bibnamefont {Colonna}}, \bibinfo {author} {\bibfnamefont
  {I.}~\bibnamefont {Carnimeo}}, \bibinfo {author} {\bibfnamefont {A.~D.}\
  \bibnamefont {Corso}}, \bibinfo {author} {\bibfnamefont {S.}~\bibnamefont
  {de~Gironcoli}}, \bibinfo {author} {\bibfnamefont {P.}~\bibnamefont
  {Delugas}}, \bibinfo {author} {\bibfnamefont {R.~A.}\ \bibnamefont
  {DiStasio}}, \bibinfo {author} {\bibfnamefont {A.}~\bibnamefont {Ferretti}},
  \bibinfo {author} {\bibfnamefont {A.}~\bibnamefont {Floris}}, \bibinfo
  {author} {\bibfnamefont {G.}~\bibnamefont {Fratesi}}, \bibinfo {author}
  {\bibfnamefont {G.}~\bibnamefont {Fugallo}}, \bibinfo {author} {\bibfnamefont
  {R.}~\bibnamefont {Gebauer}}, \bibinfo {author} {\bibfnamefont
  {U.}~\bibnamefont {Gerstmann}}, \bibinfo {author} {\bibfnamefont
  {F.}~\bibnamefont {Giustino}}, \bibinfo {author} {\bibfnamefont
  {T.}~\bibnamefont {Gorni}}, \bibinfo {author} {\bibfnamefont
  {J.}~\bibnamefont {Jia}}, \bibinfo {author} {\bibfnamefont {M.}~\bibnamefont
  {Kawamura}}, \bibinfo {author} {\bibfnamefont {H.-Y.}\ \bibnamefont {Ko}},
  \bibinfo {author} {\bibfnamefont {A.}~\bibnamefont {Kokalj}}, \bibinfo
  {author} {\bibfnamefont {E.}~\bibnamefont {Küçükbenli}}, \bibinfo {author}
  {\bibfnamefont {M.}~\bibnamefont {Lazzeri}}, \bibinfo {author} {\bibfnamefont
  {M.}~\bibnamefont {Marsili}}, \bibinfo {author} {\bibfnamefont
  {N.}~\bibnamefont {Marzari}}, \bibinfo {author} {\bibfnamefont
  {F.}~\bibnamefont {Mauri}}, \bibinfo {author} {\bibfnamefont {N.~L.}\
  \bibnamefont {Nguyen}}, \bibinfo {author} {\bibfnamefont {H.-V.}\
  \bibnamefont {Nguyen}}, \bibinfo {author} {\bibfnamefont {A.~O.}\
  \bibnamefont {de-la Roza}}, \bibinfo {author} {\bibfnamefont
  {L.}~\bibnamefont {Paulatto}}, \bibinfo {author} {\bibfnamefont
  {S.}~\bibnamefont {Poncé}}, \bibinfo {author} {\bibfnamefont
  {D.}~\bibnamefont {Rocca}}, \bibinfo {author} {\bibfnamefont
  {R.}~\bibnamefont {Sabatini}}, \bibinfo {author} {\bibfnamefont
  {B.}~\bibnamefont {Santra}}, \bibinfo {author} {\bibfnamefont
  {M.}~\bibnamefont {Schlipf}}, \bibinfo {author} {\bibfnamefont {A.~P.}\
  \bibnamefont {Seitsonen}}, \bibinfo {author} {\bibfnamefont {A.}~\bibnamefont
  {Smogunov}}, \bibinfo {author} {\bibfnamefont {I.}~\bibnamefont {Timrov}},
  \bibinfo {author} {\bibfnamefont {T.}~\bibnamefont {Thonhauser}}, \bibinfo
  {author} {\bibfnamefont {P.}~\bibnamefont {Umari}}, \bibinfo {author}
  {\bibfnamefont {N.}~\bibnamefont {Vast}}, \bibinfo {author} {\bibfnamefont
  {X.}~\bibnamefont {Wu}},\ and\ \bibinfo {author} {\bibfnamefont
  {S.}~\bibnamefont {Baroni}},\ }\href
  {https://doi.org/10.1088/1361-648X/aa8f79} {\bibfield  {journal} {\bibinfo
  {journal} {Journal of Physics: Condensed Matter}\ }\textbf {\bibinfo {volume}
  {29}},\ \bibinfo {pages} {465901} (\bibinfo {year} {2017})}\BibitemShut
  {NoStop}%
\bibitem [{\citenamefont {Kohn}\ and\ \citenamefont
  {Sham}(1965)}]{PhysRev.140.A1133}%
  \BibitemOpen
  \bibfield  {author} {\bibinfo {author} {\bibfnamefont {W.}~\bibnamefont
  {Kohn}}\ and\ \bibinfo {author} {\bibfnamefont {L.~J.}\ \bibnamefont
  {Sham}},\ }\href {https://doi.org/10.1103/PhysRev.140.A1133} {\bibfield
  {journal} {\bibinfo  {journal} {Phys. Rev.}\ }\textbf {\bibinfo {volume}
  {140}},\ \bibinfo {pages} {A1133} (\bibinfo {year} {1965})}\BibitemShut
  {NoStop}%
\bibitem [{\citenamefont {Perdew}\ \emph {et~al.}(1996)\citenamefont {Perdew},
  \citenamefont {Burke},\ and\ \citenamefont
  {Ernzerhof}}]{PhysRevLett.77.3865}%
  \BibitemOpen
  \bibfield  {author} {\bibinfo {author} {\bibfnamefont {J.~P.}\ \bibnamefont
  {Perdew}}, \bibinfo {author} {\bibfnamefont {K.}~\bibnamefont {Burke}},\ and\
  \bibinfo {author} {\bibfnamefont {M.}~\bibnamefont {Ernzerhof}},\ }\href
  {https://doi.org/10.1103/PhysRevLett.77.3865} {\bibfield  {journal} {\bibinfo
   {journal} {Phys. Rev. Lett.}\ }\textbf {\bibinfo {volume} {77}},\ \bibinfo
  {pages} {3865} (\bibinfo {year} {1996})}\BibitemShut {NoStop}%
\bibitem [{\citenamefont {Grimme}\ \emph {et~al.}(2010)\citenamefont {Grimme},
  \citenamefont {Antony}, \citenamefont {Ehrlich},\ and\ \citenamefont
  {Krieg}}]{grimme}%
  \BibitemOpen
  \bibfield  {author} {\bibinfo {author} {\bibfnamefont {S.}~\bibnamefont
  {Grimme}}, \bibinfo {author} {\bibfnamefont {J.}~\bibnamefont {Antony}},
  \bibinfo {author} {\bibfnamefont {S.}~\bibnamefont {Ehrlich}},\ and\ \bibinfo
  {author} {\bibfnamefont {H.}~\bibnamefont {Krieg}},\ }\href
  {https://doi.org/10.1063/1.3382344} {\bibfield  {journal} {\bibinfo
  {journal} {The Journal of Chemical Physics}\ }\textbf {\bibinfo {volume}
  {132}},\ \bibinfo {pages} {154104} (\bibinfo {year} {2010})}\BibitemShut
  {NoStop}%
\bibitem [{\citenamefont {{van Setten}}\ \emph {et~al.}(2018)\citenamefont
  {{van Setten}}, \citenamefont {Giantomassi}, \citenamefont {Bousquet},
  \citenamefont {Verstraete}, \citenamefont {Hamann}, \citenamefont {Gonze},\
  and\ \citenamefont {Rignanese}}]{VANSETTEN201839}%
  \BibitemOpen
  \bibfield  {author} {\bibinfo {author} {\bibfnamefont {M.}~\bibnamefont {{van
  Setten}}}, \bibinfo {author} {\bibfnamefont {M.}~\bibnamefont {Giantomassi}},
  \bibinfo {author} {\bibfnamefont {E.}~\bibnamefont {Bousquet}}, \bibinfo
  {author} {\bibfnamefont {M.}~\bibnamefont {Verstraete}}, \bibinfo {author}
  {\bibfnamefont {D.}~\bibnamefont {Hamann}}, \bibinfo {author} {\bibfnamefont
  {X.}~\bibnamefont {Gonze}},\ and\ \bibinfo {author} {\bibfnamefont {G.-M.}\
  \bibnamefont {Rignanese}},\ }\href
  {https://doi.org/https://doi.org/10.1016/j.cpc.2018.01.012} {\bibfield
  {journal} {\bibinfo  {journal} {Computer Physics Communications}\ }\textbf
  {\bibinfo {volume} {226}},\ \bibinfo {pages} {39} (\bibinfo {year}
  {2018})}\BibitemShut {NoStop}%
\bibitem [{\citenamefont {Fischer}\ and\ \citenamefont
  {Almlof}(1992)}]{fischer1992general}%
  \BibitemOpen
  \bibfield  {author} {\bibinfo {author} {\bibfnamefont {T.~H.}\ \bibnamefont
  {Fischer}}\ and\ \bibinfo {author} {\bibfnamefont {J.}~\bibnamefont
  {Almlof}},\ }\href {https://doi.org/https://doi.org/10.1021/j100203a036}
  {\bibfield  {journal} {\bibinfo  {journal} {The Journal of Physical
  Chemistry}\ }\textbf {\bibinfo {volume} {96}},\ \bibinfo {pages} {9768}
  (\bibinfo {year} {1992})}\BibitemShut {NoStop}%
\bibitem [{\citenamefont {Monkhorst}\ and\ \citenamefont
  {Pack}(1976)}]{PhysRevB.13.5188}%
  \BibitemOpen
  \bibfield  {author} {\bibinfo {author} {\bibfnamefont {H.~J.}\ \bibnamefont
  {Monkhorst}}\ and\ \bibinfo {author} {\bibfnamefont {J.~D.}\ \bibnamefont
  {Pack}},\ }\href {https://doi.org/10.1103/PhysRevB.13.5188} {\bibfield
  {journal} {\bibinfo  {journal} {Phys. Rev. B}\ }\textbf {\bibinfo {volume}
  {13}},\ \bibinfo {pages} {5188} (\bibinfo {year} {1976})}\BibitemShut
  {NoStop}%
\bibitem [{\citenamefont {Onida}\ \emph {et~al.}(2002)\citenamefont {Onida},
  \citenamefont {Reining},\ and\ \citenamefont {Rubio}}]{RevModPhys.74.601}%
  \BibitemOpen
  \bibfield  {author} {\bibinfo {author} {\bibfnamefont {G.}~\bibnamefont
  {Onida}}, \bibinfo {author} {\bibfnamefont {L.}~\bibnamefont {Reining}},\
  and\ \bibinfo {author} {\bibfnamefont {A.}~\bibnamefont {Rubio}},\ }\href
  {https://doi.org/10.1103/RevModPhys.74.601} {\bibfield  {journal} {\bibinfo
  {journal} {Rev. Mod. Phys.}\ }\textbf {\bibinfo {volume} {74}},\ \bibinfo
  {pages} {601} (\bibinfo {year} {2002})}\BibitemShut {NoStop}%
\bibitem [{\citenamefont {Hybertsen}\ and\ \citenamefont
  {Louie}(1986)}]{PhysRevB.34.5390}%
  \BibitemOpen
  \bibfield  {author} {\bibinfo {author} {\bibfnamefont {M.~S.}\ \bibnamefont
  {Hybertsen}}\ and\ \bibinfo {author} {\bibfnamefont {S.~G.}\ \bibnamefont
  {Louie}},\ }\href {https://doi.org/10.1103/PhysRevB.34.5390} {\bibfield
  {journal} {\bibinfo  {journal} {Phys. Rev. B}\ }\textbf {\bibinfo {volume}
  {34}},\ \bibinfo {pages} {5390} (\bibinfo {year} {1986})}\BibitemShut
  {NoStop}%
\bibitem [{\citenamefont {Strinati}(1988)}]{strinati1988application}%
  \BibitemOpen
  \bibfield  {author} {\bibinfo {author} {\bibfnamefont {G.}~\bibnamefont
  {Strinati}},\ }\href {https://doi.org/https://doi.org/10.1007/BF02725962}
  {\bibfield  {journal} {\bibinfo  {journal} {La Rivista del Nuovo Cimento
  (1978-1999)}\ }\textbf {\bibinfo {volume} {11}},\ \bibinfo {pages} {1}
  (\bibinfo {year} {1988})}\BibitemShut {NoStop}%
\bibitem [{\citenamefont {Marini}\ \emph {et~al.}(2009)\citenamefont {Marini},
  \citenamefont {Hogan}, \citenamefont {Grüning},\ and\ \citenamefont
  {Varsano}}]{MARINI20091392}%
  \BibitemOpen
  \bibfield  {author} {\bibinfo {author} {\bibfnamefont {A.}~\bibnamefont
  {Marini}}, \bibinfo {author} {\bibfnamefont {C.}~\bibnamefont {Hogan}},
  \bibinfo {author} {\bibfnamefont {M.}~\bibnamefont {Grüning}},\ and\
  \bibinfo {author} {\bibfnamefont {D.}~\bibnamefont {Varsano}},\ }\href
  {https://doi.org/https://doi.org/10.1016/j.cpc.2009.02.003} {\bibfield
  {journal} {\bibinfo  {journal} {Computer Physics Communications}\ }\textbf
  {\bibinfo {volume} {180}},\ \bibinfo {pages} {1392} (\bibinfo {year}
  {2009})}\BibitemShut {NoStop}%
\bibitem [{\citenamefont {Sangalli}\ \emph {et~al.}(2019)\citenamefont
  {Sangalli}, \citenamefont {Ferretti}, \citenamefont {Miranda}, \citenamefont
  {Attaccalite}, \citenamefont {Marri}, \citenamefont {Cannuccia},
  \citenamefont {Melo}, \citenamefont {Marsili}, \citenamefont {Paleari},
  \citenamefont {Marrazzo}, \citenamefont {Prandini}, \citenamefont {Bonfà},
  \citenamefont {Atambo}, \citenamefont {Affinito}, \citenamefont {Palummo},
  \citenamefont {Molina-Sánchez}, \citenamefont {Hogan}, \citenamefont
  {Grüning}, \citenamefont {Varsano},\ and\ \citenamefont
  {Marini}}]{sangalli2019many}%
  \BibitemOpen
  \bibfield  {author} {\bibinfo {author} {\bibfnamefont {D.}~\bibnamefont
  {Sangalli}}, \bibinfo {author} {\bibfnamefont {A.}~\bibnamefont {Ferretti}},
  \bibinfo {author} {\bibfnamefont {H.}~\bibnamefont {Miranda}}, \bibinfo
  {author} {\bibfnamefont {C.}~\bibnamefont {Attaccalite}}, \bibinfo {author}
  {\bibfnamefont {I.}~\bibnamefont {Marri}}, \bibinfo {author} {\bibfnamefont
  {E.}~\bibnamefont {Cannuccia}}, \bibinfo {author} {\bibfnamefont
  {P.}~\bibnamefont {Melo}}, \bibinfo {author} {\bibfnamefont {M.}~\bibnamefont
  {Marsili}}, \bibinfo {author} {\bibfnamefont {F.}~\bibnamefont {Paleari}},
  \bibinfo {author} {\bibfnamefont {A.}~\bibnamefont {Marrazzo}}, \bibinfo
  {author} {\bibfnamefont {G.}~\bibnamefont {Prandini}}, \bibinfo {author}
  {\bibfnamefont {P.}~\bibnamefont {Bonfà}}, \bibinfo {author} {\bibfnamefont
  {M.~O.}\ \bibnamefont {Atambo}}, \bibinfo {author} {\bibfnamefont
  {F.}~\bibnamefont {Affinito}}, \bibinfo {author} {\bibfnamefont
  {M.}~\bibnamefont {Palummo}}, \bibinfo {author} {\bibfnamefont
  {A.}~\bibnamefont {Molina-Sánchez}}, \bibinfo {author} {\bibfnamefont
  {C.}~\bibnamefont {Hogan}}, \bibinfo {author} {\bibfnamefont
  {M.}~\bibnamefont {Grüning}}, \bibinfo {author} {\bibfnamefont
  {D.}~\bibnamefont {Varsano}},\ and\ \bibinfo {author} {\bibfnamefont
  {A.}~\bibnamefont {Marini}},\ }\href
  {https://doi.org/10.1088/1361-648X/ab15d0} {\bibfield  {journal} {\bibinfo
  {journal} {Journal of Physics: Condensed Matter}\ }\textbf {\bibinfo {volume}
  {31}},\ \bibinfo {pages} {325902} (\bibinfo {year} {2019})}\BibitemShut
  {NoStop}%
\bibitem [{\citenamefont {Larson}\ \emph {et~al.}(2013)\citenamefont {Larson},
  \citenamefont {Dvorak},\ and\ \citenamefont {Wu}}]{PhysRevB.88.125205}%
  \BibitemOpen
  \bibfield  {author} {\bibinfo {author} {\bibfnamefont {P.}~\bibnamefont
  {Larson}}, \bibinfo {author} {\bibfnamefont {M.}~\bibnamefont {Dvorak}},\
  and\ \bibinfo {author} {\bibfnamefont {Z.}~\bibnamefont {Wu}},\ }\href
  {https://doi.org/10.1103/PhysRevB.88.125205} {\bibfield  {journal} {\bibinfo
  {journal} {Phys. Rev. B}\ }\textbf {\bibinfo {volume} {88}},\ \bibinfo
  {pages} {125205} (\bibinfo {year} {2013})}\BibitemShut {NoStop}%
\bibitem [{\citenamefont {Bruneval}\ and\ \citenamefont
  {Gonze}(2008)}]{PhysRevB.78.085125}%
  \BibitemOpen
  \bibfield  {author} {\bibinfo {author} {\bibfnamefont {F.}~\bibnamefont
  {Bruneval}}\ and\ \bibinfo {author} {\bibfnamefont {X.}~\bibnamefont
  {Gonze}},\ }\href {https://doi.org/10.1103/PhysRevB.78.085125} {\bibfield
  {journal} {\bibinfo  {journal} {Phys. Rev. B}\ }\textbf {\bibinfo {volume}
  {78}},\ \bibinfo {pages} {085125} (\bibinfo {year} {2008})}\BibitemShut
  {NoStop}%
\bibitem [{\citenamefont {Dancoff}(1950)}]{PhysRev.78.382}%
  \BibitemOpen
  \bibfield  {author} {\bibinfo {author} {\bibfnamefont {S.~M.}\ \bibnamefont
  {Dancoff}},\ }\href {https://doi.org/10.1103/PhysRev.78.382} {\bibfield
  {journal} {\bibinfo  {journal} {Phys. Rev.}\ }\textbf {\bibinfo {volume}
  {78}},\ \bibinfo {pages} {382} (\bibinfo {year} {1950})}\BibitemShut
  {NoStop}%
\bibitem [{\citenamefont {Arora}\ \emph {et~al.}(2015)\citenamefont {Arora},
  \citenamefont {Koperski}, \citenamefont {Nogajewski}, \citenamefont {Marcus},
  \citenamefont {Faugeras},\ and\ \citenamefont {Potemski}}]{AroraWSe2}%
  \BibitemOpen
  \bibfield  {author} {\bibinfo {author} {\bibfnamefont {A.}~\bibnamefont
  {Arora}}, \bibinfo {author} {\bibfnamefont {M.}~\bibnamefont {Koperski}},
  \bibinfo {author} {\bibfnamefont {K.}~\bibnamefont {Nogajewski}}, \bibinfo
  {author} {\bibfnamefont {J.}~\bibnamefont {Marcus}}, \bibinfo {author}
  {\bibfnamefont {C.}~\bibnamefont {Faugeras}},\ and\ \bibinfo {author}
  {\bibfnamefont {M.}~\bibnamefont {Potemski}},\ }\href
  {https://doi.org/10.1039/C5NR01536G} {\bibfield  {journal} {\bibinfo
  {journal} {Nanoscale}\ }\textbf {\bibinfo {volume} {7}},\ \bibinfo {pages}
  {10421} (\bibinfo {year} {2015})}\BibitemShut {NoStop}%
\bibitem [{\citenamefont {Molas}\ \emph {et~al.}(2017)\citenamefont {Molas},
  \citenamefont {Nogajewski}, \citenamefont {Slobodeniuk}, \citenamefont
  {Binder}, \citenamefont {Bartos},\ and\ \citenamefont {Potemski}}]{MolasWS2}%
  \BibitemOpen
  \bibfield  {author} {\bibinfo {author} {\bibfnamefont {M.~R.}\ \bibnamefont
  {Molas}}, \bibinfo {author} {\bibfnamefont {K.}~\bibnamefont {Nogajewski}},
  \bibinfo {author} {\bibfnamefont {A.~O.}\ \bibnamefont {Slobodeniuk}},
  \bibinfo {author} {\bibfnamefont {J.}~\bibnamefont {Binder}}, \bibinfo
  {author} {\bibfnamefont {M.}~\bibnamefont {Bartos}},\ and\ \bibinfo {author}
  {\bibfnamefont {M.}~\bibnamefont {Potemski}},\ }\href
  {https://doi.org/10.1039/C7NR04672C} {\bibfield  {journal} {\bibinfo
  {journal} {Nanoscale}\ }\textbf {\bibinfo {volume} {9}},\ \bibinfo {pages}
  {13128} (\bibinfo {year} {2017})}\BibitemShut {NoStop}%
\bibitem [{\citenamefont {Kezerashvili}\ and\ \citenamefont
  {Tsiklauri}(2017)}]{kezerashvili2017trion}%
  \BibitemOpen
  \bibfield  {author} {\bibinfo {author} {\bibfnamefont {R.~Y.}\ \bibnamefont
  {Kezerashvili}}\ and\ \bibinfo {author} {\bibfnamefont {S.~M.}\ \bibnamefont
  {Tsiklauri}},\ }\href
  {https://doi.org/https://doi.org/10.1007/s00601-016-1186-x} {\bibfield
  {journal} {\bibinfo  {journal} {Few-Body Systems}\ }\textbf {\bibinfo
  {volume} {58}},\ \bibinfo {pages} {1} (\bibinfo {year} {2017})}\BibitemShut
  {NoStop}%
\bibitem [{\citenamefont {Vaclavkova}\ \emph {et~al.}(2018)\citenamefont
  {Vaclavkova}, \citenamefont {Wyzula}, \citenamefont {Nogajewski},
  \citenamefont {Bartos}, \citenamefont {Slobodeniuk}, \citenamefont
  {Faugeras}, \citenamefont {Potemski},\ and\ \citenamefont
  {Molas}}]{vaclavkova2018singlet}%
  \BibitemOpen
  \bibfield  {author} {\bibinfo {author} {\bibfnamefont {D.}~\bibnamefont
  {Vaclavkova}}, \bibinfo {author} {\bibfnamefont {J.}~\bibnamefont {Wyzula}},
  \bibinfo {author} {\bibfnamefont {K.}~\bibnamefont {Nogajewski}}, \bibinfo
  {author} {\bibfnamefont {M.}~\bibnamefont {Bartos}}, \bibinfo {author}
  {\bibfnamefont {A.}~\bibnamefont {Slobodeniuk}}, \bibinfo {author}
  {\bibfnamefont {C.}~\bibnamefont {Faugeras}}, \bibinfo {author}
  {\bibfnamefont {M.}~\bibnamefont {Potemski}},\ and\ \bibinfo {author}
  {\bibfnamefont {M.}~\bibnamefont {Molas}},\ }\href
  {https://doi.org/10.1088/1361-6528/aac65c} {\bibfield  {journal} {\bibinfo
  {journal} {Nanotechnology}\ }\textbf {\bibinfo {volume} {29}},\ \bibinfo
  {pages} {325705} (\bibinfo {year} {2018})}\BibitemShut {NoStop}%
\bibitem [{\citenamefont {Yang}\ \emph {et~al.}(2015)\citenamefont {Yang},
  \citenamefont {Xu}, \citenamefont {Pei}, \citenamefont {Myint}, \citenamefont
  {Wang}, \citenamefont {Wang}, \citenamefont {Zhang}, \citenamefont {Yu},\
  and\ \citenamefont {Lu}}]{yang2015optical}%
  \BibitemOpen
  \bibfield  {author} {\bibinfo {author} {\bibfnamefont {J.}~\bibnamefont
  {Yang}}, \bibinfo {author} {\bibfnamefont {R.}~\bibnamefont {Xu}}, \bibinfo
  {author} {\bibfnamefont {J.}~\bibnamefont {Pei}}, \bibinfo {author}
  {\bibfnamefont {Y.~W.}\ \bibnamefont {Myint}}, \bibinfo {author}
  {\bibfnamefont {F.}~\bibnamefont {Wang}}, \bibinfo {author} {\bibfnamefont
  {Z.}~\bibnamefont {Wang}}, \bibinfo {author} {\bibfnamefont {S.}~\bibnamefont
  {Zhang}}, \bibinfo {author} {\bibfnamefont {Z.}~\bibnamefont {Yu}},\ and\
  \bibinfo {author} {\bibfnamefont {Y.}~\bibnamefont {Lu}},\ }\href
  {https://doi.org/https://doi.org/10.1038/lsa.2015.85} {\bibfield  {journal}
  {\bibinfo  {journal} {Light: Science \& Applications}\ }\textbf {\bibinfo
  {volume} {4}},\ \bibinfo {pages} {e312} (\bibinfo {year} {2015})}\BibitemShut
  {NoStop}%
\bibitem [{\citenamefont {Xu}\ \emph {et~al.}(2016)\citenamefont {Xu},
  \citenamefont {Zhang}, \citenamefont {Wang}, \citenamefont {Yang},
  \citenamefont {Wang}, \citenamefont {Pei}, \citenamefont {Myint},
  \citenamefont {Xing}, \citenamefont {Yu}, \citenamefont {Fu} \emph
  {et~al.}}]{xu2016extraordinarily}%
  \BibitemOpen
  \bibfield  {author} {\bibinfo {author} {\bibfnamefont {R.}~\bibnamefont
  {Xu}}, \bibinfo {author} {\bibfnamefont {S.}~\bibnamefont {Zhang}}, \bibinfo
  {author} {\bibfnamefont {F.}~\bibnamefont {Wang}}, \bibinfo {author}
  {\bibfnamefont {J.}~\bibnamefont {Yang}}, \bibinfo {author} {\bibfnamefont
  {Z.}~\bibnamefont {Wang}}, \bibinfo {author} {\bibfnamefont {J.}~\bibnamefont
  {Pei}}, \bibinfo {author} {\bibfnamefont {Y.~W.}\ \bibnamefont {Myint}},
  \bibinfo {author} {\bibfnamefont {B.}~\bibnamefont {Xing}}, \bibinfo {author}
  {\bibfnamefont {Z.}~\bibnamefont {Yu}}, \bibinfo {author} {\bibfnamefont
  {L.}~\bibnamefont {Fu}}, \emph {et~al.},\ }\href
  {https://doi.org/https://doi.org/10.1021/acsnano.5b06193} {\bibfield
  {journal} {\bibinfo  {journal} {Acs Nano}\ }\textbf {\bibinfo {volume}
  {10}},\ \bibinfo {pages} {2046} (\bibinfo {year} {2016})}\BibitemShut
  {NoStop}%
\bibitem [{\citenamefont {Ayari}\ \emph {et~al.}(2020)\citenamefont {Ayari},
  \citenamefont {Quick}, \citenamefont {Owschimikow}, \citenamefont
  {Christodoulou}, \citenamefont {Bertrand}, \citenamefont {Artemyev},
  \citenamefont {Moreels}, \citenamefont {Woggon}, \citenamefont {Jaziri},\
  and\ \citenamefont {Achtstein}}]{ayari2020tuning}%
  \BibitemOpen
  \bibfield  {author} {\bibinfo {author} {\bibfnamefont {S.}~\bibnamefont
  {Ayari}}, \bibinfo {author} {\bibfnamefont {M.~T.}\ \bibnamefont {Quick}},
  \bibinfo {author} {\bibfnamefont {N.}~\bibnamefont {Owschimikow}}, \bibinfo
  {author} {\bibfnamefont {S.}~\bibnamefont {Christodoulou}}, \bibinfo {author}
  {\bibfnamefont {G.~H.}\ \bibnamefont {Bertrand}}, \bibinfo {author}
  {\bibfnamefont {M.}~\bibnamefont {Artemyev}}, \bibinfo {author}
  {\bibfnamefont {I.}~\bibnamefont {Moreels}}, \bibinfo {author} {\bibfnamefont
  {U.}~\bibnamefont {Woggon}}, \bibinfo {author} {\bibfnamefont
  {S.}~\bibnamefont {Jaziri}},\ and\ \bibinfo {author} {\bibfnamefont {A.~W.}\
  \bibnamefont {Achtstein}},\ }\href
  {https://doi.org/https://doi.org/10.1039/D0NR03170D} {\bibfield  {journal}
  {\bibinfo  {journal} {Nanoscale}\ }\textbf {\bibinfo {volume} {12}},\
  \bibinfo {pages} {14448} (\bibinfo {year} {2020})}\BibitemShut {NoStop}%
\bibitem [{\citenamefont {Durnev}\ and\ \citenamefont
  {Glazov}(2018)}]{durnev2018excitons}%
  \BibitemOpen
  \bibfield  {author} {\bibinfo {author} {\bibfnamefont {M.~V.}\ \bibnamefont
  {Durnev}}\ and\ \bibinfo {author} {\bibfnamefont {M.~M.}\ \bibnamefont
  {Glazov}},\ }\href {https://doi.org/10.3367/UFNe.2017.07.038172} {\bibfield
  {journal} {\bibinfo  {journal} {Physics-Uspekhi}\ }\textbf {\bibinfo {volume}
  {61}},\ \bibinfo {pages} {825} (\bibinfo {year} {2018})}\BibitemShut
  {NoStop}%
\bibitem [{\citenamefont {Van~der Donck}\ \emph {et~al.}(2018)\citenamefont
  {Van~der Donck}, \citenamefont {Zarenia},\ and\ \citenamefont
  {Peeters}}]{van2018excitons}%
  \BibitemOpen
  \bibfield  {author} {\bibinfo {author} {\bibfnamefont {M.}~\bibnamefont
  {Van~der Donck}}, \bibinfo {author} {\bibfnamefont {M.}~\bibnamefont
  {Zarenia}},\ and\ \bibinfo {author} {\bibfnamefont {F.}~\bibnamefont
  {Peeters}},\ }\href
  {https://doi.org/https://doi.org/10.1103/PhysRevB.97.195408} {\bibfield
  {journal} {\bibinfo  {journal} {Physical Review B}\ }\textbf {\bibinfo
  {volume} {97}},\ \bibinfo {pages} {195408} (\bibinfo {year}
  {2018})}\BibitemShut {NoStop}%
\bibitem [{\citenamefont {Louyer}\ \emph {et~al.}(2011)\citenamefont {Louyer},
  \citenamefont {Biadala}, \citenamefont {Trebbia}, \citenamefont {Fern{\'e}e},
  \citenamefont {Tamarat},\ and\ \citenamefont {Lounis}}]{louyer2011efficient}%
  \BibitemOpen
  \bibfield  {author} {\bibinfo {author} {\bibfnamefont {Y.}~\bibnamefont
  {Louyer}}, \bibinfo {author} {\bibfnamefont {L.}~\bibnamefont {Biadala}},
  \bibinfo {author} {\bibfnamefont {J.-B.}\ \bibnamefont {Trebbia}}, \bibinfo
  {author} {\bibfnamefont {M.~J.}\ \bibnamefont {Fern{\'e}e}}, \bibinfo
  {author} {\bibfnamefont {P.}~\bibnamefont {Tamarat}},\ and\ \bibinfo {author}
  {\bibfnamefont {B.}~\bibnamefont {Lounis}},\ }\href
  {https://doi.org/https://doi.org/10.1021/nl202506c} {\bibfield  {journal}
  {\bibinfo  {journal} {Nano letters}\ }\textbf {\bibinfo {volume} {11}},\
  \bibinfo {pages} {4370} (\bibinfo {year} {2011})}\BibitemShut {NoStop}%
\bibitem [{\citenamefont {Steinhoff}\ \emph {et~al.}(2018)\citenamefont
  {Steinhoff}, \citenamefont {Florian}, \citenamefont {Singh}, \citenamefont
  {Tran}, \citenamefont {Kolarczik}, \citenamefont {Helmrich}, \citenamefont
  {Achtstein}, \citenamefont {Woggon}, \citenamefont {Owschimikow},
  \citenamefont {Jahnke} \emph {et~al.}}]{steinhoff2018biexciton}%
  \BibitemOpen
  \bibfield  {author} {\bibinfo {author} {\bibfnamefont {A.}~\bibnamefont
  {Steinhoff}}, \bibinfo {author} {\bibfnamefont {M.}~\bibnamefont {Florian}},
  \bibinfo {author} {\bibfnamefont {A.}~\bibnamefont {Singh}}, \bibinfo
  {author} {\bibfnamefont {K.}~\bibnamefont {Tran}}, \bibinfo {author}
  {\bibfnamefont {M.}~\bibnamefont {Kolarczik}}, \bibinfo {author}
  {\bibfnamefont {S.}~\bibnamefont {Helmrich}}, \bibinfo {author}
  {\bibfnamefont {A.~W.}\ \bibnamefont {Achtstein}}, \bibinfo {author}
  {\bibfnamefont {U.}~\bibnamefont {Woggon}}, \bibinfo {author} {\bibfnamefont
  {N.}~\bibnamefont {Owschimikow}}, \bibinfo {author} {\bibfnamefont
  {F.}~\bibnamefont {Jahnke}}, \emph {et~al.},\ }\href
  {https://doi.org/https://doi.org/10.1038/s41567-018-0282-x} {\bibfield
  {journal} {\bibinfo  {journal} {Nature Physics}\ }\textbf {\bibinfo {volume}
  {14}},\ \bibinfo {pages} {1199} (\bibinfo {year} {2018})}\BibitemShut
  {NoStop}%
\bibitem [{\citenamefont {Tonndorf}\ \emph {et~al.}(2015)\citenamefont
  {Tonndorf}, \citenamefont {Schmidt}, \citenamefont {Schneider}, \citenamefont
  {Kern}, \citenamefont {Buscema}, \citenamefont {Steele}, \citenamefont
  {Castellanos-Gomez}, \citenamefont {van~der Zant}, \citenamefont
  {de~Vasconcellos},\ and\ \citenamefont {Bratschitsch}}]{tonndorf2015single}%
  \BibitemOpen
  \bibfield  {author} {\bibinfo {author} {\bibfnamefont {P.}~\bibnamefont
  {Tonndorf}}, \bibinfo {author} {\bibfnamefont {R.}~\bibnamefont {Schmidt}},
  \bibinfo {author} {\bibfnamefont {R.}~\bibnamefont {Schneider}}, \bibinfo
  {author} {\bibfnamefont {J.}~\bibnamefont {Kern}}, \bibinfo {author}
  {\bibfnamefont {M.}~\bibnamefont {Buscema}}, \bibinfo {author} {\bibfnamefont
  {G.~A.}\ \bibnamefont {Steele}}, \bibinfo {author} {\bibfnamefont
  {A.}~\bibnamefont {Castellanos-Gomez}}, \bibinfo {author} {\bibfnamefont
  {H.~S.}\ \bibnamefont {van~der Zant}}, \bibinfo {author} {\bibfnamefont
  {S.~M.}\ \bibnamefont {de~Vasconcellos}},\ and\ \bibinfo {author}
  {\bibfnamefont {R.}~\bibnamefont {Bratschitsch}},\ }\href
  {https://doi.org/https://doi.org/10.1364/OPTICA.2.000347} {\bibfield
  {journal} {\bibinfo  {journal} {Optica}\ }\textbf {\bibinfo {volume} {2}},\
  \bibinfo {pages} {347} (\bibinfo {year} {2015})}\BibitemShut {NoStop}%
\bibitem [{\citenamefont {Wang}\ \emph {et~al.}(2014)\citenamefont {Wang},
  \citenamefont {Bouet}, \citenamefont {Lagarde}, \citenamefont {Vidal},
  \citenamefont {Balocchi}, \citenamefont {Amand}, \citenamefont {Marie},\ and\
  \citenamefont {Urbaszek}}]{wang2014valley}%
  \BibitemOpen
  \bibfield  {author} {\bibinfo {author} {\bibfnamefont {G.}~\bibnamefont
  {Wang}}, \bibinfo {author} {\bibfnamefont {L.}~\bibnamefont {Bouet}},
  \bibinfo {author} {\bibfnamefont {D.}~\bibnamefont {Lagarde}}, \bibinfo
  {author} {\bibfnamefont {M.}~\bibnamefont {Vidal}}, \bibinfo {author}
  {\bibfnamefont {A.}~\bibnamefont {Balocchi}}, \bibinfo {author}
  {\bibfnamefont {T.}~\bibnamefont {Amand}}, \bibinfo {author} {\bibfnamefont
  {X.}~\bibnamefont {Marie}},\ and\ \bibinfo {author} {\bibfnamefont
  {B.}~\bibnamefont {Urbaszek}},\ }\href
  {https://doi.org/https://doi.org/10.1103/PhysRevB.90.075413} {\bibfield
  {journal} {\bibinfo  {journal} {Physical Review B}\ }\textbf {\bibinfo
  {volume} {90}},\ \bibinfo {pages} {075413} (\bibinfo {year}
  {2014})}\BibitemShut {NoStop}%
\bibitem [{\citenamefont {Zhang}\ \emph {et~al.}(2017)\citenamefont {Zhang},
  \citenamefont {Wang}, \citenamefont {Li}, \citenamefont {Huang},
  \citenamefont {Li}, \citenamefont {Ji},\ and\ \citenamefont
  {Wu}}]{zhang2017defect}%
  \BibitemOpen
  \bibfield  {author} {\bibinfo {author} {\bibfnamefont {S.}~\bibnamefont
  {Zhang}}, \bibinfo {author} {\bibfnamefont {C.-G.}\ \bibnamefont {Wang}},
  \bibinfo {author} {\bibfnamefont {M.-Y.}\ \bibnamefont {Li}}, \bibinfo
  {author} {\bibfnamefont {D.}~\bibnamefont {Huang}}, \bibinfo {author}
  {\bibfnamefont {L.-J.}\ \bibnamefont {Li}}, \bibinfo {author} {\bibfnamefont
  {W.}~\bibnamefont {Ji}},\ and\ \bibinfo {author} {\bibfnamefont
  {S.}~\bibnamefont {Wu}},\ }\href
  {https://doi.org/https://doi.org/10.1103/PhysRevLett.119.046101} {\bibfield
  {journal} {\bibinfo  {journal} {Physical review letters}\ }\textbf {\bibinfo
  {volume} {119}},\ \bibinfo {pages} {046101} (\bibinfo {year}
  {2017})}\BibitemShut {NoStop}%
\bibitem [{\citenamefont {Pelant}\ and\ \citenamefont
  {Valenta}(2012)}]{pelant2012luminescence}%
  \BibitemOpen
  \bibfield  {author} {\bibinfo {author} {\bibfnamefont {I.}~\bibnamefont
  {Pelant}}\ and\ \bibinfo {author} {\bibfnamefont {J.}~\bibnamefont
  {Valenta}},\ }\href
  {https://doi.org/https://doi.org/10.1093/acprof:oso/9780199588336.001.0001}
  {\emph {\bibinfo {title} {Luminescence spectroscopy of semiconductors}}}\
  (\bibinfo  {publisher} {OUP Oxford},\ \bibinfo {year} {2012})\BibitemShut
  {NoStop}%
\bibitem [{\citenamefont {Lin}\ \emph {et~al.}(2016)\citenamefont {Lin},
  \citenamefont {Carvalho}, \citenamefont {Kahn}, \citenamefont {Lv},
  \citenamefont {Rao}, \citenamefont {Terrones}, \citenamefont {Pimenta},\ and\
  \citenamefont {Terrones}}]{lin2016defect}%
  \BibitemOpen
  \bibfield  {author} {\bibinfo {author} {\bibfnamefont {Z.}~\bibnamefont
  {Lin}}, \bibinfo {author} {\bibfnamefont {B.~R.}\ \bibnamefont {Carvalho}},
  \bibinfo {author} {\bibfnamefont {E.}~\bibnamefont {Kahn}}, \bibinfo {author}
  {\bibfnamefont {R.}~\bibnamefont {Lv}}, \bibinfo {author} {\bibfnamefont
  {R.}~\bibnamefont {Rao}}, \bibinfo {author} {\bibfnamefont {H.}~\bibnamefont
  {Terrones}}, \bibinfo {author} {\bibfnamefont {M.~A.}\ \bibnamefont
  {Pimenta}},\ and\ \bibinfo {author} {\bibfnamefont {M.}~\bibnamefont
  {Terrones}},\ }\href
  {https://doi.org/https://doi.org/10.1021/acsnano.0c09666} {\bibfield
  {journal} {\bibinfo  {journal} {2D Materials}\ }\textbf {\bibinfo {volume}
  {3}},\ \bibinfo {pages} {022002} (\bibinfo {year} {2016})}\BibitemShut
  {NoStop}%
\bibitem [{\citenamefont {Singh}\ \emph {et~al.}(2016)\citenamefont {Singh},
  \citenamefont {Moody}, \citenamefont {Tran}, \citenamefont {Scott},
  \citenamefont {Overbeck}, \citenamefont {Bergh{\"a}user}, \citenamefont
  {Schaibley}, \citenamefont {Seifert}, \citenamefont {Pleskot}, \citenamefont
  {Gabor} \emph {et~al.}}]{singh2016trion}%
  \BibitemOpen
  \bibfield  {author} {\bibinfo {author} {\bibfnamefont {A.}~\bibnamefont
  {Singh}}, \bibinfo {author} {\bibfnamefont {G.}~\bibnamefont {Moody}},
  \bibinfo {author} {\bibfnamefont {K.}~\bibnamefont {Tran}}, \bibinfo {author}
  {\bibfnamefont {M.~E.}\ \bibnamefont {Scott}}, \bibinfo {author}
  {\bibfnamefont {V.}~\bibnamefont {Overbeck}}, \bibinfo {author}
  {\bibfnamefont {G.}~\bibnamefont {Bergh{\"a}user}}, \bibinfo {author}
  {\bibfnamefont {J.}~\bibnamefont {Schaibley}}, \bibinfo {author}
  {\bibfnamefont {E.~J.}\ \bibnamefont {Seifert}}, \bibinfo {author}
  {\bibfnamefont {D.}~\bibnamefont {Pleskot}}, \bibinfo {author} {\bibfnamefont
  {N.~M.}\ \bibnamefont {Gabor}}, \emph {et~al.},\ }\href
  {https://doi.org/https://doi.org/10.1103/PhysRevB.93.041401} {\bibfield
  {journal} {\bibinfo  {journal} {Physical Review B}\ }\textbf {\bibinfo
  {volume} {93}},\ \bibinfo {pages} {041401} (\bibinfo {year}
  {2016})}\BibitemShut {NoStop}%
\bibitem [{\citenamefont {Sangalli}\ \emph {et~al.}(2017)\citenamefont
  {Sangalli}, \citenamefont {Berger}, \citenamefont {Attaccalite},
  \citenamefont {Gr\"uning},\ and\ \citenamefont
  {Romaniello}}]{PhysRevB.95.155203}%
  \BibitemOpen
  \bibfield  {author} {\bibinfo {author} {\bibfnamefont {D.}~\bibnamefont
  {Sangalli}}, \bibinfo {author} {\bibfnamefont {J.~A.}\ \bibnamefont
  {Berger}}, \bibinfo {author} {\bibfnamefont {C.}~\bibnamefont {Attaccalite}},
  \bibinfo {author} {\bibfnamefont {M.}~\bibnamefont {Gr\"uning}},\ and\
  \bibinfo {author} {\bibfnamefont {P.}~\bibnamefont {Romaniello}},\ }\href
  {https://doi.org/10.1103/PhysRevB.95.155203} {\bibfield  {journal} {\bibinfo
  {journal} {Phys. Rev. B}\ }\textbf {\bibinfo {volume} {95}},\ \bibinfo
  {pages} {155203} (\bibinfo {year} {2017})}\BibitemShut {NoStop}%
\bibitem [{\citenamefont {Rohlfing}\ and\ \citenamefont
  {Louie}(1998)}]{PhysRevLett.81.2312}%
  \BibitemOpen
  \bibfield  {author} {\bibinfo {author} {\bibfnamefont {M.}~\bibnamefont
  {Rohlfing}}\ and\ \bibinfo {author} {\bibfnamefont {S.~G.}\ \bibnamefont
  {Louie}},\ }\href {https://doi.org/10.1103/PhysRevLett.81.2312} {\bibfield
  {journal} {\bibinfo  {journal} {Phys. Rev. Lett.}\ }\textbf {\bibinfo
  {volume} {81}},\ \bibinfo {pages} {2312} (\bibinfo {year}
  {1998})}\BibitemShut {NoStop}%
\bibitem [{\citenamefont {Molina-S\'anchez}\ \emph {et~al.}(2013)\citenamefont
  {Molina-S\'anchez}, \citenamefont {Sangalli}, \citenamefont {Hummer},
  \citenamefont {Marini},\ and\ \citenamefont {Wirtz}}]{PhysRevB.88.045412}%
  \BibitemOpen
  \bibfield  {author} {\bibinfo {author} {\bibfnamefont {A.}~\bibnamefont
  {Molina-S\'anchez}}, \bibinfo {author} {\bibfnamefont {D.}~\bibnamefont
  {Sangalli}}, \bibinfo {author} {\bibfnamefont {K.}~\bibnamefont {Hummer}},
  \bibinfo {author} {\bibfnamefont {A.}~\bibnamefont {Marini}},\ and\ \bibinfo
  {author} {\bibfnamefont {L.}~\bibnamefont {Wirtz}},\ }\href
  {https://doi.org/10.1103/PhysRevB.88.045412} {\bibfield  {journal} {\bibinfo
  {journal} {Phys. Rev. B}\ }\textbf {\bibinfo {volume} {88}},\ \bibinfo
  {pages} {045412} (\bibinfo {year} {2013})}\BibitemShut {NoStop}%
\bibitem [{\citenamefont {Makinistian}\ and\ \citenamefont
  {Albanesi}(2006)}]{makinistian2006first}%
  \BibitemOpen
  \bibfield  {author} {\bibinfo {author} {\bibfnamefont {L.}~\bibnamefont
  {Makinistian}}\ and\ \bibinfo {author} {\bibfnamefont {E.}~\bibnamefont
  {Albanesi}},\ }\href
  {https://doi.org/https://doi.org/10.1103/PhysRevB.74.045206} {\bibfield
  {journal} {\bibinfo  {journal} {Physical Review B}\ }\textbf {\bibinfo
  {volume} {74}},\ \bibinfo {pages} {045206} (\bibinfo {year}
  {2006})}\BibitemShut {NoStop}%
\bibitem [{\citenamefont {Oliva}\ \emph
  {et~al.}(2020{\natexlab{b}})\citenamefont {Oliva}, \citenamefont
  {Wo\ifmmode~\acute{z}\else \'{z}\fi{}niak}, \citenamefont {Dybala},
  \citenamefont {To\l{}\l{}oczko}, \citenamefont {Kopaczek}, \citenamefont
  {Scharoch},\ and\ \citenamefont {Kudrawiec}}]{PhysRevB.101.235205}%
  \BibitemOpen
  \bibfield  {author} {\bibinfo {author} {\bibfnamefont {R.}~\bibnamefont
  {Oliva}}, \bibinfo {author} {\bibfnamefont {T.}~\bibnamefont
  {Wo\ifmmode~\acute{z}\else \'{z}\fi{}niak}}, \bibinfo {author} {\bibfnamefont
  {F.}~\bibnamefont {Dybala}}, \bibinfo {author} {\bibfnamefont
  {A.}~\bibnamefont {To\l{}\l{}oczko}}, \bibinfo {author} {\bibfnamefont
  {J.}~\bibnamefont {Kopaczek}}, \bibinfo {author} {\bibfnamefont
  {P.}~\bibnamefont {Scharoch}},\ and\ \bibinfo {author} {\bibfnamefont
  {R.}~\bibnamefont {Kudrawiec}},\ }\href
  {https://doi.org/10.1103/PhysRevB.101.235205} {\bibfield  {journal} {\bibinfo
   {journal} {Phys. Rev. B}\ }\textbf {\bibinfo {volume} {101}},\ \bibinfo
  {pages} {235205} (\bibinfo {year} {2020}{\natexlab{b}})}\BibitemShut
  {NoStop}%
\bibitem [{\citenamefont {Castillo-Mussot}\ \emph {et~al.}(2002)\citenamefont
  {Castillo-Mussot}, \citenamefont {Vazquez},\ and\ \citenamefont
  {Reyes}}]{castillo2002variational}%
  \BibitemOpen
  \bibfield  {author} {\bibinfo {author} {\bibfnamefont {M.~d.}\ \bibnamefont
  {Castillo-Mussot}}, \bibinfo {author} {\bibfnamefont {G.~J.}\ \bibnamefont
  {Vazquez}},\ and\ \bibinfo {author} {\bibfnamefont {J.}~\bibnamefont
  {Reyes}},\ }\href@noop {} {\bibfield  {journal} {\bibinfo  {journal} {Revista
  mexicana de f{\'\i}sica}\ }\textbf {\bibinfo {volume} {48}},\ \bibinfo
  {pages} {504} (\bibinfo {year} {2002})}\BibitemShut {NoStop}%
\bibitem [{\citenamefont {Schindlmayr}(1997)}]{schindlmayr1997excitons}%
  \BibitemOpen
  \bibfield  {author} {\bibinfo {author} {\bibfnamefont {A.}~\bibnamefont
  {Schindlmayr}},\ }\href {https://doi.org/10.1088/0143-0807/18/5/011}
  {\bibfield  {journal} {\bibinfo  {journal} {European Journal of Physics}\
  }\textbf {\bibinfo {volume} {18}},\ \bibinfo {pages} {374} (\bibinfo {year}
  {1997})}\BibitemShut {NoStop}%
\bibitem [{\citenamefont {Carr{\'e}}\ \emph {et~al.}(2021)\citenamefont
  {Carr{\'e}}, \citenamefont {Sponza}, \citenamefont {Lusson}, \citenamefont
  {Stenger}, \citenamefont {Gaufr{\`e}s}, \citenamefont {Loiseau},\ and\
  \citenamefont {Barjon}}]{carre2021excitons}%
  \BibitemOpen
  \bibfield  {author} {\bibinfo {author} {\bibfnamefont {E.}~\bibnamefont
  {Carr{\'e}}}, \bibinfo {author} {\bibfnamefont {L.}~\bibnamefont {Sponza}},
  \bibinfo {author} {\bibfnamefont {A.}~\bibnamefont {Lusson}}, \bibinfo
  {author} {\bibfnamefont {I.}~\bibnamefont {Stenger}}, \bibinfo {author}
  {\bibfnamefont {{\'E}.}~\bibnamefont {Gaufr{\`e}s}}, \bibinfo {author}
  {\bibfnamefont {A.}~\bibnamefont {Loiseau}},\ and\ \bibinfo {author}
  {\bibfnamefont {J.}~\bibnamefont {Barjon}},\ }\href
  {https://doi.org/10.1088/2053-1583/abca81} {\bibfield  {journal} {\bibinfo
  {journal} {2D Materials}\ }\textbf {\bibinfo {volume} {8}},\ \bibinfo {pages}
  {021001} (\bibinfo {year} {2021})}\BibitemShut {NoStop}%
\bibitem [{\citenamefont {Wirtz}\ \emph {et~al.}(2005)\citenamefont {Wirtz},
  \citenamefont {Marini}, \citenamefont {Gruning},\ and\ \citenamefont
  {Rubio}}]{wirtz2005excitonic}%
  \BibitemOpen
  \bibfield  {author} {\bibinfo {author} {\bibfnamefont {L.}~\bibnamefont
  {Wirtz}}, \bibinfo {author} {\bibfnamefont {A.}~\bibnamefont {Marini}},
  \bibinfo {author} {\bibfnamefont {M.}~\bibnamefont {Gruning}},\ and\ \bibinfo
  {author} {\bibfnamefont {A.}~\bibnamefont {Rubio}},\ }\href@noop {}
  {\bibfield  {journal} {\bibinfo  {journal} {arXiv preprint cond-mat/0508421}\
  } (\bibinfo {year} {2005})}\BibitemShut {NoStop}%
\bibitem [{\citenamefont {Molina-S{\'a}nchez}\ \emph
  {et~al.}(2013)\citenamefont {Molina-S{\'a}nchez}, \citenamefont {Sangalli},
  \citenamefont {Hummer}, \citenamefont {Marini},\ and\ \citenamefont
  {Wirtz}}]{molina2013effect}%
  \BibitemOpen
  \bibfield  {author} {\bibinfo {author} {\bibfnamefont {A.}~\bibnamefont
  {Molina-S{\'a}nchez}}, \bibinfo {author} {\bibfnamefont {D.}~\bibnamefont
  {Sangalli}}, \bibinfo {author} {\bibfnamefont {K.}~\bibnamefont {Hummer}},
  \bibinfo {author} {\bibfnamefont {A.}~\bibnamefont {Marini}},\ and\ \bibinfo
  {author} {\bibfnamefont {L.}~\bibnamefont {Wirtz}},\ }\href
  {https://doi.org/https://doi.org/10.1103/PhysRevB.88.045412} {\bibfield
  {journal} {\bibinfo  {journal} {Physical Review B}\ }\textbf {\bibinfo
  {volume} {88}},\ \bibinfo {pages} {045412} (\bibinfo {year}
  {2013})}\BibitemShut {NoStop}%
\bibitem [{\citenamefont {Olsen}\ \emph {et~al.}(2016)\citenamefont {Olsen},
  \citenamefont {Latini}, \citenamefont {Rasmussen},\ and\ \citenamefont
  {Thygesen}}]{olsen2016simple}%
  \BibitemOpen
  \bibfield  {author} {\bibinfo {author} {\bibfnamefont {T.}~\bibnamefont
  {Olsen}}, \bibinfo {author} {\bibfnamefont {S.}~\bibnamefont {Latini}},
  \bibinfo {author} {\bibfnamefont {F.}~\bibnamefont {Rasmussen}},\ and\
  \bibinfo {author} {\bibfnamefont {K.~S.}\ \bibnamefont {Thygesen}},\ }\href
  {https://doi.org/https://doi.org/10.1103/PhysRevLett.116.056401} {\bibfield
  {journal} {\bibinfo  {journal} {Physical review letters}\ }\textbf {\bibinfo
  {volume} {116}},\ \bibinfo {pages} {056401} (\bibinfo {year}
  {2016})}\BibitemShut {NoStop}%
\bibitem [{\citenamefont {Cudazzo}\ \emph {et~al.}(2011)\citenamefont
  {Cudazzo}, \citenamefont {Tokatly},\ and\ \citenamefont
  {Rubio}}]{cudazzo2011dielectric}%
  \BibitemOpen
  \bibfield  {author} {\bibinfo {author} {\bibfnamefont {P.}~\bibnamefont
  {Cudazzo}}, \bibinfo {author} {\bibfnamefont {I.~V.}\ \bibnamefont
  {Tokatly}},\ and\ \bibinfo {author} {\bibfnamefont {A.}~\bibnamefont
  {Rubio}},\ }\href
  {https://doi.org/https://doi.org/10.1103/PhysRevB.84.085406} {\bibfield
  {journal} {\bibinfo  {journal} {Physical Review B}\ }\textbf {\bibinfo
  {volume} {84}},\ \bibinfo {pages} {085406} (\bibinfo {year}
  {2011})}\BibitemShut {NoStop}%
\bibitem [{\citenamefont {Keldysh}(1979)}]{keldysh1979coulomb}%
  \BibitemOpen
  \bibfield  {author} {\bibinfo {author} {\bibfnamefont {L.}~\bibnamefont
  {Keldysh}},\ }\href@noop {} {\bibfield  {journal} {\bibinfo  {journal}
  {Soviet Journal of Experimental and Theoretical Physics Letters}\ }\textbf
  {\bibinfo {volume} {29}},\ \bibinfo {pages} {658} (\bibinfo {year}
  {1979})}\BibitemShut {NoStop}%
\bibitem [{\citenamefont {Chernikov}\ \emph {et~al.}(2014)\citenamefont
  {Chernikov}, \citenamefont {Berkelbach}, \citenamefont {Hill}, \citenamefont
  {Rigosi}, \citenamefont {Li}, \citenamefont {Aslan}, \citenamefont
  {Reichman}, \citenamefont {Hybertsen},\ and\ \citenamefont
  {Heinz}}]{chernikov2014exciton}%
  \BibitemOpen
  \bibfield  {author} {\bibinfo {author} {\bibfnamefont {A.}~\bibnamefont
  {Chernikov}}, \bibinfo {author} {\bibfnamefont {T.~C.}\ \bibnamefont
  {Berkelbach}}, \bibinfo {author} {\bibfnamefont {H.~M.}\ \bibnamefont
  {Hill}}, \bibinfo {author} {\bibfnamefont {A.}~\bibnamefont {Rigosi}},
  \bibinfo {author} {\bibfnamefont {Y.}~\bibnamefont {Li}}, \bibinfo {author}
  {\bibfnamefont {B.}~\bibnamefont {Aslan}}, \bibinfo {author} {\bibfnamefont
  {D.~R.}\ \bibnamefont {Reichman}}, \bibinfo {author} {\bibfnamefont {M.~S.}\
  \bibnamefont {Hybertsen}},\ and\ \bibinfo {author} {\bibfnamefont {T.~F.}\
  \bibnamefont {Heinz}},\ }\href
  {https://doi.org/https://doi.org/10.1103/PhysRevLett.113.076802} {\bibfield
  {journal} {\bibinfo  {journal} {Physical review letters}\ }\textbf {\bibinfo
  {volume} {113}},\ \bibinfo {pages} {076802} (\bibinfo {year}
  {2014})}\BibitemShut {NoStop}%
\bibitem [{\citenamefont {Rytova}(2018)}]{rytova2018screened}%
  \BibitemOpen
  \bibfield  {author} {\bibinfo {author} {\bibfnamefont {N.~S.}\ \bibnamefont
  {Rytova}},\ }\href@noop {} {\bibfield  {journal} {\bibinfo  {journal} {arXiv
  preprint arXiv:1806.00976}\ } (\bibinfo {year} {2018})}\BibitemShut {NoStop}%
\bibitem [{\citenamefont {Taguchi}\ \emph {et~al.}(1988)\citenamefont
  {Taguchi}, \citenamefont {Goto}, \citenamefont {Takeda},\ and\ \citenamefont
  {Kido}}]{taguchi1988magneto}%
  \BibitemOpen
  \bibfield  {author} {\bibinfo {author} {\bibfnamefont {S.}~\bibnamefont
  {Taguchi}}, \bibinfo {author} {\bibfnamefont {T.}~\bibnamefont {Goto}},
  \bibinfo {author} {\bibfnamefont {M.}~\bibnamefont {Takeda}},\ and\ \bibinfo
  {author} {\bibfnamefont {G.}~\bibnamefont {Kido}},\ }\href
  {https://doi.org/https://doi.org/10.1143/JPSJ.57.3256} {\bibfield  {journal}
  {\bibinfo  {journal} {Journal of the Physical Society of Japan}\ }\textbf
  {\bibinfo {volume} {57}},\ \bibinfo {pages} {3256} (\bibinfo {year}
  {1988})}\BibitemShut {NoStop}%
\bibitem [{\citenamefont {Burstein}\ and\ \citenamefont
  {Weisbuch}(2012)}]{burstein2012confined}%
  \BibitemOpen
  \bibfield  {author} {\bibinfo {author} {\bibfnamefont {E.}~\bibnamefont
  {Burstein}}\ and\ \bibinfo {author} {\bibfnamefont {C.}~\bibnamefont
  {Weisbuch}},\ }\href
  {https://doi.org/https://doi.org/10.1007/978-1-4615-1963-8} {\emph {\bibinfo
  {title} {Confined electrons and photons: New physics and applications}}},\
  Vol.\ \bibinfo {volume} {340}\ (\bibinfo  {publisher} {Springer Science \&
  Business Media},\ \bibinfo {year} {2012})\BibitemShut {NoStop}%
\bibitem [{\citenamefont {Stier}\ \emph {et~al.}(1999)\citenamefont {Stier},
  \citenamefont {Grundmann},\ and\ \citenamefont
  {Bimberg}}]{stier1999electronic}%
  \BibitemOpen
  \bibfield  {author} {\bibinfo {author} {\bibfnamefont {O.}~\bibnamefont
  {Stier}}, \bibinfo {author} {\bibfnamefont {M.}~\bibnamefont {Grundmann}},\
  and\ \bibinfo {author} {\bibfnamefont {D.}~\bibnamefont {Bimberg}},\ }\href
  {https://doi.org/https://doi.org/10.1103/PhysRevB.59.5688} {\bibfield
  {journal} {\bibinfo  {journal} {Physical Review B}\ }\textbf {\bibinfo
  {volume} {59}},\ \bibinfo {pages} {5688} (\bibinfo {year}
  {1999})}\BibitemShut {NoStop}%
\bibitem [{\citenamefont {Bir}\ \emph {et~al.}(1974)\citenamefont {Bir},
  \citenamefont {Pikus},\ and\ \citenamefont {Louvish}}]{bir1974symmetry}%
  \BibitemOpen
  \bibfield  {author} {\bibinfo {author} {\bibfnamefont {G.~L.}\ \bibnamefont
  {Bir}}, \bibinfo {author} {\bibfnamefont {G.~E.}\ \bibnamefont {Pikus}},\
  and\ \bibinfo {author} {\bibfnamefont {D.}~\bibnamefont {Louvish}},\
  }\href@noop {} {\emph {\bibinfo {title} {Symmetry and strain-induced effects
  in semiconductors}}},\ Vol.\ \bibinfo {volume} {484}\ (\bibinfo  {publisher}
  {Wiley New York},\ \bibinfo {year} {1974})\BibitemShut {NoStop}%
\end{thebibliography}%

\newpage\hbox{}\thispagestyle{empty}\newpage
%%%%%%%%%% Merge with supplemental materials %%%%%%%%%%
\pagebreak
\widetext
\begin{center}
\textbf{\large Supporting Information:\\ Optical properties of orthorhombic germanium sulfide: Unveiling the Anisotropic Nature of Wannier Exciton}
\end{center}
%%%%%%%%%% Merge with supplemental materials %%%%%%%%%%
%%%%%%%%%% Prefix a "S" to all equations, figures, tables and reset the counter %%%%%%%%%%
\setcounter{equation}{0}
\setcounter{figure}{0}
\setcounter{table}{0}
\setcounter{page}{1}
\makeatletter
\renewcommand{\theequation}{S\arabic{equation}}
\renewcommand{\thefigure}{S\arabic{figure}}
\renewcommand{\bibnumfmt}[1]{[S#1]}
%\renewcommand{\citenumfont}[1]{S#1}
%%%%%%%%%% Prefix a "S" to all equations, figures, tables and reset the counter %%%%%%%%%%

\setcounter{section}{0}
\renewcommand{\thesection}{S.\Roman{section}}

\section{Supporting experimental results}

%%%%%%%%%%%%%%%%%%%%%%%%%%%%%%%%%% Polarization-resolved RC %%%%%%%%%%%%%%%%%%%%%%%%%
\subsection{Polarization-resolved reflectance contrast spectra of GeS}

\begin{figure}[H]
    \centering
    \includegraphics[width=0.4\textwidth]{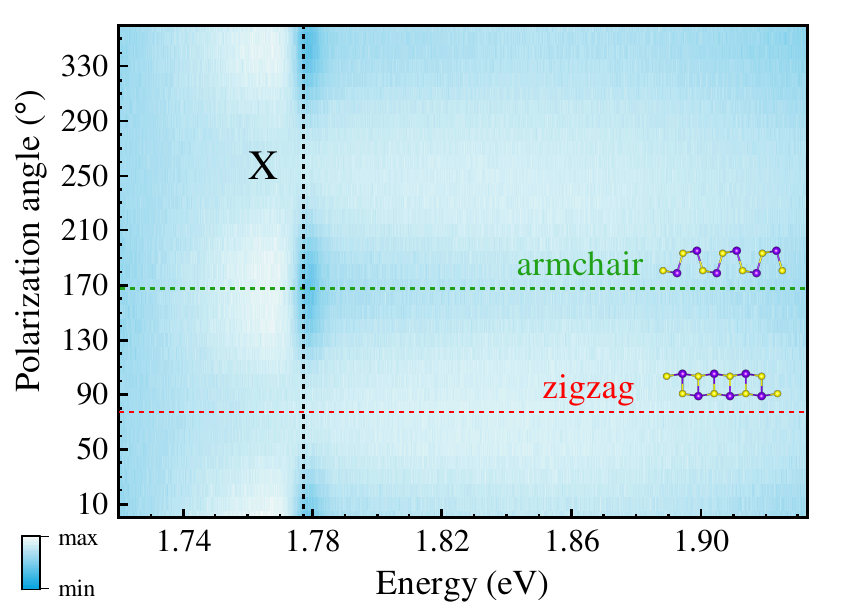}
    \caption{False-colour map of the low-temperature ($T$=5~K) polarization-resolved RC spectra measured on the GeS encapsulated in $h$-BN flakes.}
    \label{fig:rc}
\end{figure}

Figure~\ref{fig:rc} shows the low-temperature ($T$=5~K) polarization-resolved reflectance contrast (RC) spectra measured on the GeS encapsulated in $h$-BN flakes.
In contrast to the several emission lines observed in the corresponding PL spectra shown in Figures~\ref{fig:ple} and \ref{fig:anis_pl}, the RC spectra consist of a single resonance.
Furthermore, its energy of about 1.78~eV coincides with the X emission line.
The polarization dependence of the X resonance indicates that this transition is linearly polarized along the armchair direction, as well as its emission counterpart (see Figure~\ref{fig:anis_pl}).
The obtained RC results confirm that the energetically lowest optical transitions in GeS are dominated by a direct transition polarized along the armchair direction.

%%%%%%%%%%%%%%%%%%%%%%%%%%%%%%%%%% The influence of excitation energy %%%%%%%%%%%%%%
\subsection{The influence of excitation energy on the PL spectra of GeS}

\begin{figure}[H]
    \centering
    \includegraphics[width=0.4\textwidth]{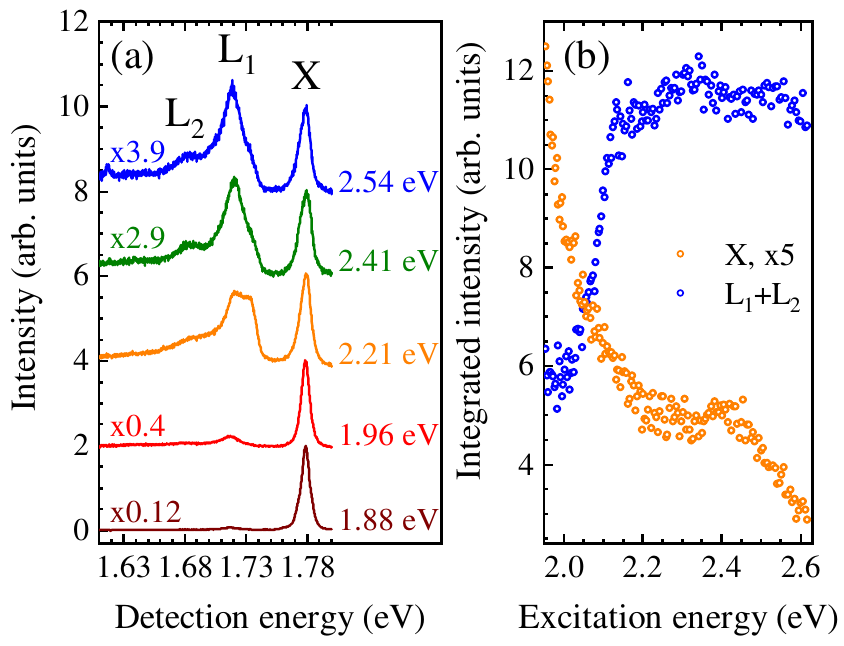}
    \caption{(a) The low-temperature ($T$=5~K) PL spectra of the investigated GeS encapsulated in $h$-BN flakes under five different excitation energies, denoted in the figure.
    (b) The integrated intensities of the free (X) and localized (L$_1$ and L$_2$) excitons as a function of excitation energy.
    Note that the X intensity was multiplied 5 times for clarity.}
    \label{fig:ple}
\end{figure}

As we found that the encapsulation energy can strongly affect the L$_1$ and L$_2$ intensities (compare Figure 1(c) in the main text and the one reported in Ref.~\cite{zawadzka2021anisotropic}), we performed measurements of the low-temperature PL spectra of the GeS encapsulated in the $h$-BN flakes under five different excitation energies, see Figure~\ref{fig:ple}(a).
The modification of the laser energy causes two main effects:
(i) The emission intensity of the X line is significantly reduced by around 30 times with the change of the excitation energy from 1.88~eV to 2.54~eV.
While the 1.88~eV excitation is possible only around the $\Gamma$ point of the Brillouin zone (BZ), the high energy 2.54~eV one may involve transitions originating from different points of the BZ, see Figure~5(a) in the main text.
(ii) The relative intensity of the emission lines due to the localized (L$_1$ and L$_2$) and free (X) excitons is also substantially modified when excitation is changed.
To investigate this effect in detail, we measured PL excitation (PLE) spectra.
Figure~\ref{fig:ple} presents the integrated intensities of the free (X) and localized (L$_1$ + L$_2$) excitons as a function of the excitation energy in the energy range from about 1.97~eV to 2.63~eV.
The measured PLe spectra of the X and L$_1$ + L$_2$ are analogous to their behaviour shown in panel (a) of the Figure.
The emission intensity of the X line experiences a monotonic reduction of more than 4 times.
Simultaneously, the L$_1$ + L$_2$ intensity is enhanced twice when the excitation energy is increased from 1.97~eV to about 2.15~eV, and then stays almost at the same level.
It suggests that at illuminations higher than 2.15~eV, the excited carriers are subjected to more nonradiative processes, which leads to an increase in probability of the L$_1$ and L$_2$ emission, see Figure~5(a) in the main text.
Note that when we combine the results obtained in panels (a) and (b) of Figure~\ref{fig:ple}, the extraordinary reduction of the X intensity in transition from 1.88~eV to 2.63~eV excitation is on the order of 50 times, which is accompanied by an increase in the intensity of L$_1$ + L$_2$.

%%%%%%%%%%%%%%%%%%%%%%%%%%%%%%%%%%%% The polarization temperature PL %%%%%%%%%%%%%
\subsection{The polarization and temperature evolutions of the PL spectra of GeS under the 2.41 eV excitation}

\begin{figure}[H]
    \centering
    \includegraphics[width=0.4\textwidth]{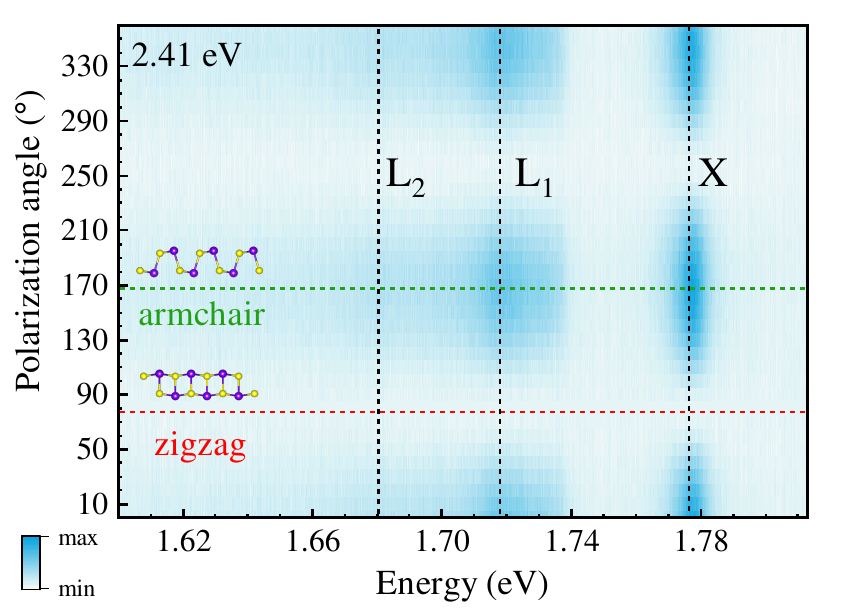}
    \caption{False-colour map of the low-temperature ($T$=5~K) polarization-resolved PL spectra measured on the GeS encapsulated in $h$-BN flakes under excitation of 2.41~eV.
    Note that the intensity scale is linear.}
    \label{fig:anis_pl}
\end{figure}

\begin{figure}[H]
    \centering
    \includegraphics[width=0.9\textwidth]{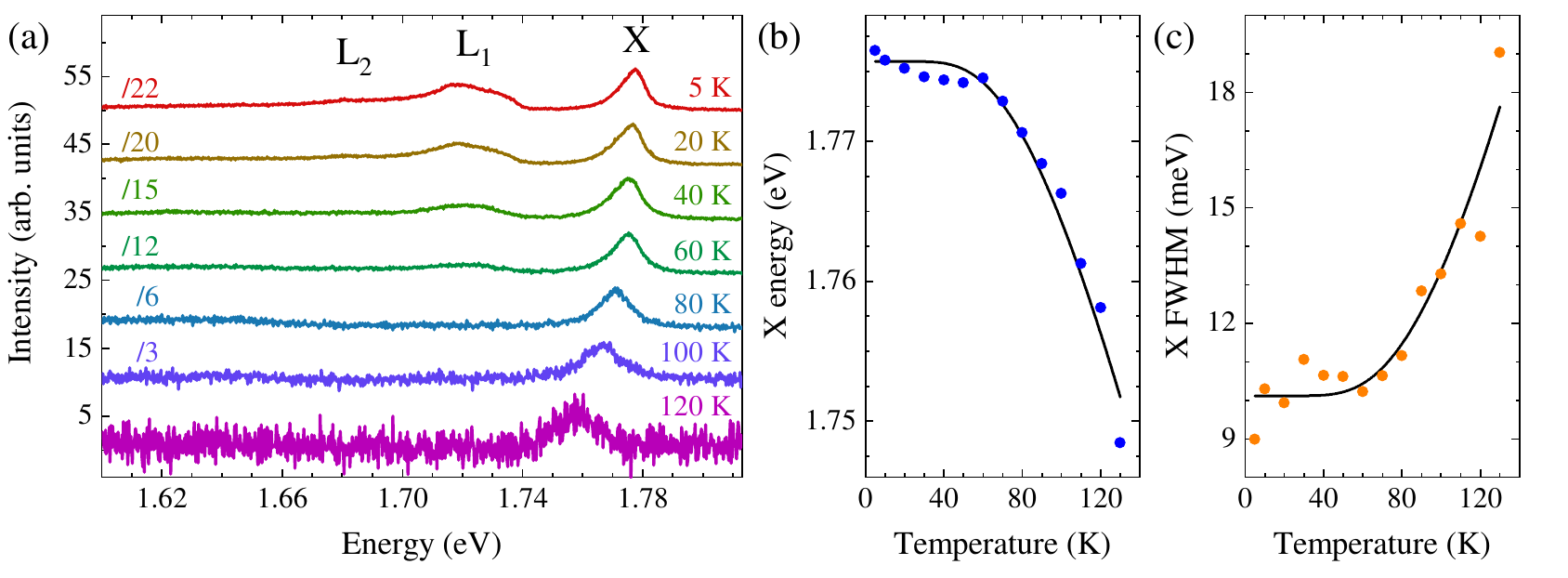}
    \caption{(a) The temperature-dependent PL spectra measured on the GeS flake with 2.41~eV laser light excitation.
    The spectra are vertically shifted and are divided by scaling factors for clarity.
    The determined (b) energy and (c) full width at half maximum (FWHM) of the neutral exciton (X) line.
    The circles represent the experimental results while the curves are fits to the data obtained with the aid of Eq. 1 and 2, described in the main text.}
    \label{fig:temp_pl}
\end{figure}

To investigate the origin of the L$_1$ and L$_2$ emission lines, we measured the polarization and temperature evolutions of the low-temperature ($T$=5~K) PL spectra of the GeS encapsulated in $h$-BN flakes under excitation of 2.41~eV, see Figures~\ref{fig:anis_pl} and \ref{fig:temp_pl}.
The choice of excitation energy is motivated by the relatively high intensity of the L$_1$ and L$_2$ peaks under this laser energy compared to others, see Figure~\ref{fig:ple}.
As can be seen in Figure~\ref{fig:anis_pl}, the L$_1$ and L$_2$ lines are linearly polarized along the same armchair direction as the neutral exciton transition. 
Moreover, with increasing temperature, the low-energy L$_1$ and L$_2$ peaks quickly disappear from the PL spectra, see Figure~\ref{fig:temp_pl}(a). 
At $T$=60~K, only the X emission contributes to the PL spectrum, which is seen up to 130 K.
The possibility of the X observation at smaller temperature range, only to 130~K under the 2.41~eV excitation versus 190~K under the 1.88~eV, can be understood in terms of the significant decrease in the X intensity when the excitation energy increases (see Figure~\ref{fig:ple}). 
The temperature dependence of the X line leads to the typical redshift and the linewidth broadening at higher temperatures, which 
 is presented in Figures.~\ref{fig:temp_pl}(b) and (c) with the corresponding fitted curves.
The temperature evolutions of the energy and the linewidth of the X line are characterized using the Odonnell~\cite{odonnell1991} and Rudin~\cite{rudin1990} relations, introduced in the main text.
We use the same values of phonon related parameters, i.e., $<\hbar \omega>$=26 meV and $\hbar \omega$=30~meV, while the other parameters were free.
The fitted curves reproduce quite well the experimental data.
Note that the observed temperature dependence of the L lines is very similar to the previously reported behavior of the so-called “localized” excitons in monolayers of WS$_2$ and WSe$_2$ exfoliated on Si/SiO$_2$ substrates.~\cite{AroraWSe2, MolasWS2}

%%%%%%%%%%%%%%%%%%%%%%%%%%%%%%%%%%%% Identification of low energy peaks %%%%%%%%%%%%%%%%%%%%%
\section{Identification of low energy peaks in GeS}
In addition to the free neutral exciton peak located around 1.78~eV (see Figure~\ref{fig:anis_pl}), at low temperature ($T$=5~K), PL measurements proved the existence of multiple emission peaks on GeS flake, which are located about $60$-$100$~meV, below the neutral exciton. 
The nature and origin of these low-energy emission bands remain unknown from both theoretical and experimental considerations.  
Different scenarios can be discussed to identify some of those peaks: i) exciton plus negative trion (X$_{-}$) and its fine structure; ii) exciton plus biexciton (XX) and auger recombination process; iii) exciton plus an optical phonon replica; and iv) exciton plus a localized state. 

Starting by testing the assumption that the lower state is an emission of charged excitons (X$_{-}$) and it's a fine structure. 
In the presence of residual charges, the Coulomb interaction could further bind an electron or hole to an exciton to form a charged exciton (negative and/or positive, depending on the doping conditions).   
X$_{-}$ appears frequently in 2D TMDs~\cite{kezerashvili2017trion,vaclavkova2018singlet}, phosphorus~\cite{yang2015optical,xu2016extraordinarily}, and CdSe platelets~\cite{ayari2020tuning}.
In the absorption and emission spectra, X$_{-}$ and its fine structure are observed at energies lower than those of the neutral excitons~\cite{kezerashvili2017trion,vaclavkova2018singlet}. Typically, the binding energy (BE) of a charged exciton, i.e., the energy difference between X and X$_{-}$ is very small compared to the exciton BE~\cite{ayari2020tuning,durnev2018excitons,van2018excitons}, so it is difficult to observe it in semi-3D conductors, which require low temperatures and very high-quality crystals. 
To form an X$_{-}$ complex, the excess carrier added to the generated exciton has to stem from an ionized exciton from the previous laser pulse. Thus, the very low excitation density used rules out an efficient X$_{-}$ generation.
Among the options considered for the lower states, $L_j$, is a Biexciton (XX). 
An XX complex is a bound state of two electrons and two holes.  It is often described approximately as the bound state of two excitons, where the interaction between the excitons is treated as a perturbation~\cite{louyer2011efficient,steinhoff2018biexciton}. 
This approximation is justified if the BE of XX (the energy difference between the two free excitons and the XX) is much smaller than that of the exciton.

An agreement against XXs is the low excitation density $50 \mu W / cm^{-2}$ used in our PL measurements, as Poisson statistics result in a negligible probability for absorption of two photons within one laser pulse. 
In addition to XX and X$_{-}$ fine structure, two other possible mechanisms including exciton–phonon interaction and defect-related scattering can be discussed to give a reasonable explanation for these exceptional multiple emission peaks in the spectra.

Starting with the phonon replicas.
The $D_{2h}^{16}$ space group gives rise to 24 vibrational modes. 
Their representation in the center of the zone is: $\Gamma = 4A_g + 2B_{1g} + 4B_{2g} + 2B_{3g} + 2A_{u} + 4B_{1u} + 2B_{2u} + 4B_{3u}$.  There are seven infrared-active phonons and twelve Raman-active phonons. 
 Raman and infrared modes are split as a result of interlayer interactions \cite{zawadzka2021anisotropic}.  
A replica phonon of the exciton peak at $E_{L_1}$ is one of the options available for the low-peaks. Excitons with a large wavevector $\boldsymbol{K}$ can recombine if a phonon or several phonons are involved that provide the necessary momentum $\boldsymbol{q} = \boldsymbol{K}_1 -\boldsymbol{K}_2$, with $\boldsymbol{K}_1$ ($\boldsymbol{K_2}$) being the wavevector of the initial (intermediate) exciton state. The so-called \textit{zero-phonon} line at energy $E^{X}_{\tilde{1s}}$ is then accompanied by a phonon replica below $E^{X}_{\tilde{1s}}$ at integer multiples of the optical phonon energy $\hbar\omega_{ph}$ with energy $E_n = E^{X}_{\tilde{1s}}- n\hbar\omega_{ph}$. 
This presumption can only be proved if the exciton and the induced replica phonon peak have the same dynamic properties. In Figure~\ref{fig:ple}(a,b), we experimentally investigated the effect of excitation energy on the PL spectra and we found that the exciton peak and the low emission line L$_1$ and L$_2$  have different behavior as a function of the excitation energy, which can be an argument against the replica phonon nature of those low-lying peaks, since it is expected that they follow the same trend as the main exciton peak, as both are related to the excitonic transition. 

Finally, we test the assumption that the lower states L$_1$ and L$_2$ are localized excitons. As a result of thermal equilibrium and the kinetics of processing, all real materials contain structural imperfections that could significantly affect their properties~\cite{tonndorf2015single,wang2014valley,zhang2017defect}. In fact, structural defects such as residual impurities, vacancies, adatoms, interstitials, and anti-sites~\cite{tonndorf2015single, wang2014valley,zhang2017defect, pelant2012luminescence,lin2016defect}, often introduce rich luminescent properties in semiconductor materials. 
When photo-excited electron-hole pairs are trapped in a disorder potential, which may be created by lattice defects, localized states may form within the band gap (BG) with an emission energy below the exciton. 
In temperature-dependent PL spectroscopy measurements (see Figure~\ref{fig:temp_pl}), these low emission peaks were observed at relatively low temperatures ($T<60 K$), in such a way that they vanish faster than the exciton as a function of temperature. 
A possible explanation is that, as compared with exciton, these peaks have smaller BEs, which can be termalized more easily with increasing temperatures. 
In fact, at low temperatures, a certain amount of carriers could be captured by these localized, trapped states. 
As the temperature increased, the trapped carriers could be released again from the localized states and recombine radiatively, leading to the increased PL intensity of the exciton. Furthermore, the dependence of the L$_1$ and L$_2$ peak on the excitation energy (in Figure~\ref{fig:ple}(a, b)) suggests the presence of an effective exciton “mobility edge”, i.e., below (above) a certain energy, the center-of-mass motion of the excitons is localized (delocalized)~\cite{singh2016trion}. 
Depending on this qualitatively study, exciton localized state is the most likely mechanism. 
The latter has also been widely used for explaining the low-energy peak in PL spectroscopy in TMDs and traditional semiconductor materials. 
However, our understanding of the fundamental properties of the multiple emission lines in GeS remains incomplete. 
These suggested possible mechanisms still need to be systematically addressed theoretically and experimentally, which is beyond the scope of this article.

%%%%%%%%%%%%%%%%%%%%%%%%%%%%%%%%%%% Independent Particle Approximation %%%%%%
\section{Independent Particle Approximation for Linear Optical Response }

In the independent particle approximation (IPA), the diagonal elements of the imaginary part of the dielectric tensor in the long-wavelength limit are basically given by~\cite{PhysRevB.95.155203}
\begin{equation}\label{dielectric_im}
\begin{split}
\epsilon_{2_{i,i}}(\omega)&=
\underbrace{
\frac{4 \pi e^2}{\Omega N_{\mathbf{k}} m^2} \sum_{n} \sum_{\mathbf{k}\in BZ} \frac{d f\left(E_{\mathbf{k}, n}\right)}{d E_{\mathbf{k}, n}} \frac{\eta \omega \mathbf{M}_{i, i}^{c, v}(\mathbf{k})}{\omega^4+\eta^2 \omega^2}}_{\text{Intraband}}\\
&+
\underbrace{
\frac{8 \pi e^2}{\Omega N_{\mathbf{k}} m^2} \sum_{n_c \neq n_v} \sum_{\mathbf{k}\in BZ} \frac{\mathbf{M}_{i, i}^{c, v}(\mathbf{k})}{E_{\mathbf{k}, n_c}-E_{\mathbf{k}, n_v}} 
 \frac{\gamma \omega f\left(E_{\mathbf{k}, n_v}\right)}{\left[\left(\omega_{\mathbf{k}, n_c}-\omega_{\mathbf{k}, n_v}\right)^2-\omega^2\right]^2+\gamma^2 \omega^2},
 }_{\text{Interband}}
\end{split}
\end{equation}
while the real part comes from the Kramers-Kronig transformation, which reads
\begin{equation}\label{dielectric_re}
\begin{split}
\epsilon_{1_{i, i}}(\omega)&=1-
\underbrace{
\frac{4 \pi e^2}{\Omega N_{\mathbf{k}} m^2} \sum_{n} \sum_{\mathbf{k}\in BZ} \frac{d f\left(E_{\mathbf{k}, n}\right)}{d E_{\mathbf{k}, n}} \frac{\omega^2 \mathbf{M}_{i, i}^{c, v}(\mathbf{k})}{\omega^4+\eta^2 \omega^2}
}_{\text{Intraband}}\\
&+
\underbrace{\frac{8 \pi e^2}{\Omega N_{\mathrm{k}} m^2} \sum_{n_v \neq n_c} \sum_{\mathrm{k}\in BZ} \frac{\mathbf{M}_{i, i}^{c, v}(\mathbf{k})}{E_{\mathrm{k}, n_c}-E_{\mathbf{k}, n_v}} \frac{\left[\left(\omega_{\mathrm{k}, n_c}-\omega_{\mathrm{k}, n_v}\right)^2-\omega^2\right] f\left(E_{\mathrm{k}, n_v}\right)}{\left[\left(\omega_{\mathrm{k}, n_c}-\omega_{\mathrm{k}, n_v}\right)^2-\omega^2\right]^2+\gamma^2 \omega^2}
}_{\text{Interband}}
\end{split}
\end{equation}
\sloppy 
where $\gamma$ and $\eta$ are the broadening of the interband and intraband transitions, respectively. 
$\mathbf{M}_{i, i}^{c,v}(\mathbf{k})$ is the squared optical transition dipole matrix elements (OME), where $\mathbf{k}$ is the single particle wavevector. 
The OME, $\mathbf{M}_{i, i}^{c,v}(\mathbf{k})$, obtained by the density functional theory (DFT) calculation and determines the optical strength of a transition. 
Furthermore, it contains all symmetry-imposed selection rules. 
$\Omega$ is the volume of the lattice cell, $n_v$ and $n_c$ belong correspondingly to the valence and conduction bands, $E_{\mathbf{k},n}$ are the eigenvalues, and $f(E_{\mathbf{k},n})$ is the Fermi distribution function that accounts for the occupation of the bands with band index $n$. 
The derivative of the Fermi distribution function for interband transitions, described by the first sum in Eq.~\ref{dielectric_im} and Eq.~\ref{dielectric_re}, is substantially zero, except in the region close to the Fermi level.

%%%%%%%%%%%%%%%%%%%%%%%%%%%%%%%%%%% GW method and Bethe-Salpeter Equation %%%%%%
\section{GW method and Bethe-Salpeter Equation\label{bse_appendix}}
 The Bethe-Salpeter equation (BSE) is calculated on top of the GW eigenvalues. 
 Conduction band energies are rigidly shifted by the GW correction, although valence band energies are barely affected. 
 Therefore, the effective masses of the hole and electron states do not differ between DFT and GW. 
 Based on the Kohn–Sham wavefunctions and the quasi-particles correction, the optical spectra are derived at the level of the BSE using the following equation~\cite{PhysRevLett.81.2312}
\begin{equation}
\left(E_{c \mathbf{k}}^{QP}-E_{v \mathbf{k}}^{QP}\right) A_{v c \mathbf{k}}^S+ \sum_{\mathbf{k}^{\prime} v^{\prime} c^{\prime}}\left\langle v c \mathbf{k}\left|K_{e h}\right| v^{\prime} c^{\prime} \mathbf{k}^{\prime}\right\rangle A_{v^{\prime} c^{\prime} \mathbf{k}^{\prime}}^S=\Omega^S A_{v c \mathbf{k}}^S,
\end{equation}
where $E_{c \mathbf{k}}^{QP}$ and $E_{v \mathbf{k}}^{QP}$ are the quasiparticle energies of the valence and conduction band states obtained with the GW method, respectively. 
$A_{v c k}^S$ and $\Omega^S$ correspond to the exciton eigenstates and eigenvalues of the $S^{th}$ exciton. 
The electron-hole interaction kernel, $K_{e h}$, contains the exchange interaction $V$ (repulsive) and the screened Coulomb interaction $W$ (attractive). 
The imaginary part of the dielectric function $\epsilon_2(\hbar \omega)$ is calculated from the excitonic states as follows~\cite{PhysRevB.88.045412}
\begin{equation}
\epsilon_2(\hbar \omega) \propto \sum_S\left|\sum_{c v \mathbf{k}} A_{v c \mathbf{k}}^S \frac{\left\langle c \mathbf{k} \left|\mathbf{p}_i\right| v \mathbf{k}\right\rangle }{\epsilon_{c,\mathbf{k}}-\epsilon_{v,\mathbf{k}}}\right|^2 \delta\left(\Omega^S-\hbar \omega- \gamma \right),
\end{equation}
where $\left\langle c \mathbf{k}\left|\mathbf{p}\right| v \mathbf{k}\right\rangle$ are the dipole matrix elements of the transitions from the valence bands to the conduction bands. $\gamma$ is the broadening energy.
In the main text, the exciton weight represented in k-space is defined as follows $w(\mathbf{k})=\sum_{n_{\mathbf{t}}}\left|\mathbf{A}_{n_{\mathbf{t}} \mathbf{k}}\right|^2$, which sums over transitions at each $\mathbf{k}$ point. 
In addition, The electron-hole (\textit{e-h}) excited state is represented by the expansion~\cite{PhysRevLett.81.2312}
\begin{equation}
    \ket{S}=\sum_{c}^{e}\sum_{v}^{h}\sum_{\mathbf{k}} A_{v c\mathbf{k}} \ket{v c \mathbf{k}}.
\end{equation}
From the solution of the BSE, we obtain the coefficients $A_{v c\mathbf{k}}$. 
Representation of this function is usually done by fixing the hole position and representing the exciton density as a function of the electron position. 

\begin{figure}[H]
    \centering
    \includegraphics[width=0.25\textwidth]{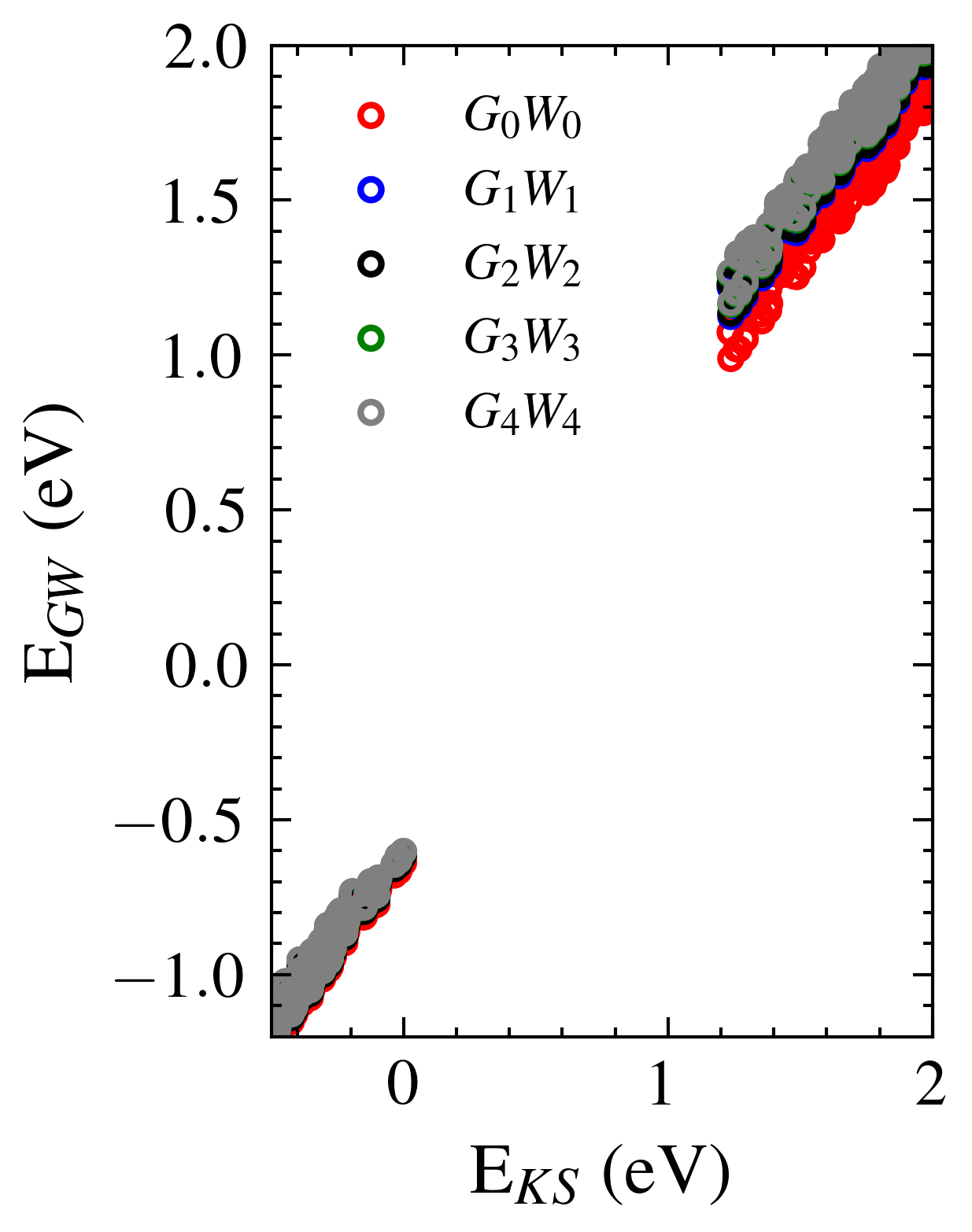}
    \caption{ Quasi-particle bandgap calculation:  Scissor operator correction in QP and renormalization of conduction and valence bands.}
    \label{fig:evgw}
\end{figure}

The single-shot $G_0W_0$ method commonly used in computational physics to calculate the electronic properties of solid systems. 
However, the technique sometimes produces unsatisfactory results because of its reliance on the DFT-starting point, resulting in a BG that is insufficiently small compared to the experimental one.
To overcome this issue, we used the self-consistency of GW only on eigenvalues (evGW), which improves the accuracy of the calculation. It involves an iterative calculation of evGW up to four times, ensuring that the difference between consistent GW values is $0.01$~eV, ultimately leading to convergence of the BG, see Figure~\ref{fig:evgw}. 

%%%%%%%%%%%%%%%%%%%%%%%%%%%%%%%%%%% Projected Density of States %%%%%%
\section{Projected Density of States of GeS\label{appendix:pdos}}
\begin{figure}[H]
    \centering
    \includegraphics[width=\textwidth]{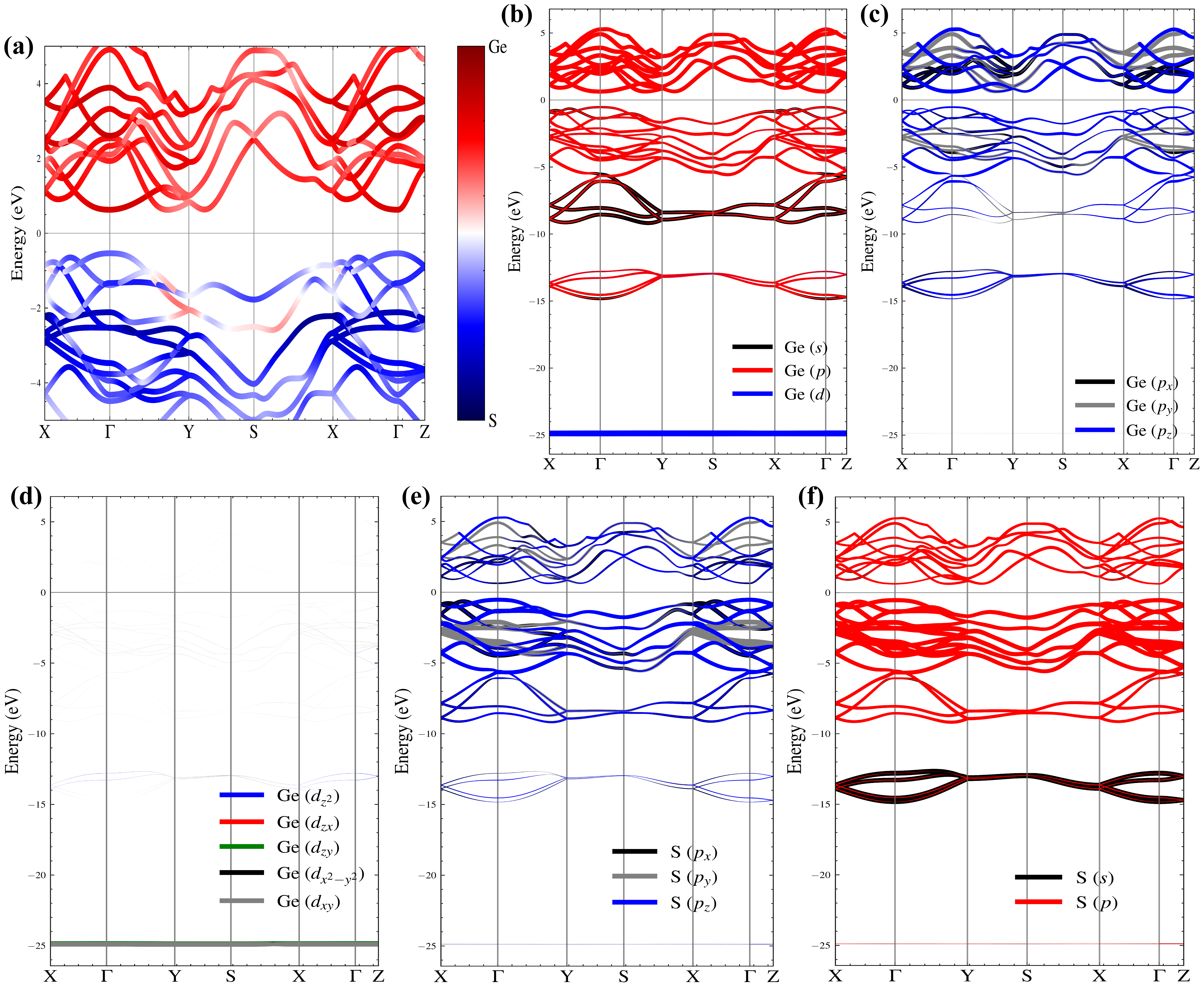}
    \caption{ \textbf{Electronic band structures with orbital projections for GeS}. The orbital projections are depicted with varying radii, with the size of the colored circles representing the relative weight of the orbital contributions to the electronic states.}
    \label{fig:pdos}
\end{figure}

 The projected fatband density of states (PDOS) can be calculated by projecting the wavefunctions onto the atomic orbitals. 
 The resulting fatband PDOS, presented in Figure~\ref{fig:pdos}, provides information on the distribution of the electronic states of  GeS. 
 Each band is assigned a color that represents the $s$-, $p$-, and $d$-orbitals. 
 We report the PDOS on both the Ge and S atoms, as depicted in Figure~\ref{fig:pdos}(a)-(f). 
 The obtained results demonstrate that the valence (conduction) states are contributed predominantly by S (Ge) atoms, with significant contributions of $p$-orbital. 
 These results align well with previous theoretical predictions reported in the Refs.~\cite{PhysRevB.87.245312, makinistian2006first,PhysRevB.101.235205}

%%%%%%%%%%%%%%%%%%%%%%%%%%%%%%%%%%% Effective Mass Approximations %%%%%%
\section{Anisotropic Wannier exciton theory within the effective
mass approximation\label{da_eme}}
Different methods have been used to calculate the optical properties of exciton states in semiconductor nanostructures, such as the variational method,~\cite{castillo2002variational,schindlmayr1997excitons,carre2021excitons} \textit{ab-initio} calculations based on GW+BSE,~\cite{wirtz2005excitonic,molina2013effect} and the effective mass theory (EMA).~\cite{kuzuba1976nearly,baldereschi1970anisotropy,dresselhaus1956effective}
The variational method has limitations and is typically used to calculate low-energy states, making it less powerful than alternative methods for higher-energy states. 
The ab-initio calculations are an accurate way to calculate the exciton states.
However, they are computationally expensive and it is difficult to systematically study some effects by varying a few phenomenological parameters. 
The EMA has the advantage of performing simple and fast computational procedures to accurately calculate the ground and excited states. Moreover, we were able to systematically study the effect of the anisotropy on the optical properties of the exciton in GeS, which helps us to more intuitively understand the role of excitonic effects. 
The anisotropy of excitons in GeS is of great importance for both fundamental research and device applications. 

Here, we employ a variationally optimized diagonalization method based on a hydrogenic basis to calculate the properties of the exciton states in bulk GeS. 
To dig deeper into the excitonic anisotropy in such crystals, we used the EMA. 
The Hamiltonian that describes the exciton anisotropy can be written as follows
\begin{equation}\label{ham_1}
   \mathbf{H}_{X}= \sum_{i=x,y,z} \dfrac{\mathbf{P_{i,e}}^{2}}{2\, m_{eff}^{e,i}}+ \sum_{i=x,y,z} \dfrac{\mathbf {P_{i,h}}^{2}}{2\,m^{h,i}_{eff}}+\mathbf{V}_{Coul}(\vert \mathbf{r_{e}}-\mathbf{r_{h}} \vert),
\end{equation}
where the momentum operator $\mathbf{P_{i,\nu}}=-i\hbar \boldsymbol{\nabla}_{i,\nu}$;  $(i=x,y,z)$ and the subscript $\eta={e,h}$ with $e$ and $h$ refer to the electron and hole, respectively. 
$m_{eff}^{\nu,i}$ is the effective mass of the electron ($\nu$ = e) and the hole ($\nu$ = h) in different directions $i$. 
$\mathbf{r_{\nu}}=(x_{\nu},y_{\nu},z_{\nu})$ are the position vectors of the electron and hole. 
$\mathbf{V}_{Coul}(\vert \mathbf{r_{e}}-\mathbf{r_{h}} \vert)$ represents the Coulomb interaction between the electron and hole screened by the background dielectric constants $\epsilon_i$ of the anisotropic semiconductor. 
Excitonic properties in bulk semiconductors differ from those in monolayers of the same material.
In 2D materials, such as TMDs and black phosphorus, excitons are strongly confined, and dielectric screening is reduced, leading to non-hydrogenic Rytova-Keldysh potentials.~\cite{olsen2016simple,cudazzo2011dielectric, keldysh1979coulomb,chernikov2014exciton,rytova2018screened} 
In contrast, bulk excitons are less confined and less sensitive to the dielectric environment.
Our previous work~\cite{zawadzka2021anisotropic} showed that the low-temperature ($T$=5~K) emission due to the free excitons measured on the GeS exfoliated on the SiO$_2$/Si substrate is similar to those reported here, which is obtained for the GeS encapsulated in the $h$-BN flakes.
This indicates that excitonic optical properties in bulk GeS are independent of the dielectric environment. 
As such, \textit{e-h} interactions in 3D homogeneous dielectric environments can be well described by the Coulomb potential. 
Using the relative and center-of-mass (COM) coordinates,
\begin{align}
   \mathbf{r} = \begin{cases} x = x_{e}-x_{h}, & \\ y = y_{e}-y_{h}, &  \\ z = z_{e}-z_{h}, &  \end{cases},  \mathbf{R}_{CM}= \begin{cases} X_{CM} = \dfrac{ m_{eff}^{e,x} x_{e}+ m_{eff}^{h,x} x_{h}}{m_{eff}^{e,x}+m_{eff}^{h,x}}, & \\ Y_{CM} = \dfrac{ m_{eff}^{e,y} y_{e}+ m_{eff}^{h,y} y_{h}}{m_{eff}^{e,y}+m_{eff}^{h,y}}, &  \\ Z_{CM} = \dfrac{ m_{eff}^{e,z} z_{e}+ m_{eff}^{h,z} z_{h}}{m_{eff}^{e,z}+m_{eff}^{h,z}}, &  \end{cases}.
\end{align}
The Hamiltonian in Eq.~\ref{ham_1} can be separated in COM and relative motion $\mathbf{H}_{X}=\mathbf{H}_{X}^{CM}+ \mathbf{H}_{X}^{rel}$. 
The free COM motion of the exciton is described by 
\begin{equation}
   \mathbf{H}_{X}^{CM}= -\frac{\hbar^2 \boldsymbol{\nabla}_{X_{CM}}^2}{2\,M_{X,x}} -\frac{\hbar^2 \boldsymbol{\nabla}_{Y_{CM}}^2}{2\,M_{X,y}} -\frac{\hbar^2 \boldsymbol{\nabla}_{Z_{CM}}^2}{2\,M_{X,z}},
\end{equation}
where $M_{X,i}=m^{e,i}_{eff}+m^{h,i}_{eff}$ denotes the exciton mass in direction $i$. 

The relative motion of the exciton can be described by the following Schrödinger equation
\begin{equation}
 \Bigg[-\dfrac{\hbar^{2}}{2} \Big(\dfrac{1}{\mu_{x}} \frac{\partial^2}{\partial x^2} +\dfrac{1}{\mu_{y}} \frac{\partial^2}{\partial y^2}+\dfrac{1}{\mu_{z}} \frac{\partial^2}{\partial z^2}\Big)-\dfrac{e^{2}}{\sqrt{\epsilon_{y} \epsilon_{z} x^{2}+\epsilon_{x} \epsilon_{z} y^{2}+\epsilon_{x} \epsilon_{y} z^{2}}}\Bigg] \boldsymbol{\Psi}^{rel}_{\Tilde{n},\Tilde{\ell},\Tilde{m}}(\mathbf{r})=E_{\Tilde{n},\Tilde{\ell},\Tilde{m}}^{\,rel}\boldsymbol{\Psi}^{rel}_{\Tilde{n},\Tilde{\ell},\Tilde{m}}(\mathbf{r}),
\end{equation}
where $E_{\Tilde{n},\Tilde{\ell},\Tilde{m}}^{\,rel}$ represents the relative eigenvalue and $\boldsymbol{\Psi}^{rel}_{\Tilde{n},\Tilde{\ell},\Tilde{m}}(\boldsymbol{r})$ are the corresponding eigenfunctions. 
Here, $\mu_{i}=(m_{eff}^{e,i}m_{eff}^{h,i})/(m_{eff}^{e,i}+m_{eff}^{h,i})$ is the reduced mass of the exciton along the direction $i$ and $e$ is the electron charge. $\epsilon_{i}$ is the dielectric constant in the direction $i$.

If we change the variable $(\,x,\,\,y,\,\,z,\,)$ into ($\xi, \eta, \zeta)=\bigg(\sqrt{\frac{\mu_{x}}{\bar{\mu}}} \,x,\ \sqrt{\frac{\mu_{y}}{\bar{\mu}}} \,y,\, \sqrt{\frac{\mu_{z}}{\bar{\mu}}}\,z\bigg)$, and if we define the anisotropy parameters $A = \dfrac{\bar{\mu}}{\mu_{x}} \dfrac{\epsilon_{y}\epsilon_{z}}{\bar{\epsilon}^{2}}$, $B = \dfrac{\bar{\mu}}{\mu_{y}} \dfrac{\epsilon_{x}\epsilon_{z}}{\bar{\epsilon}^{2}}$ and $C = \dfrac{\bar{\mu}}{\mu_{z}} \dfrac{\epsilon_{x}\epsilon_{y}}{\bar{\epsilon}^{2}}$,~\cite{taguchi1988magneto} we obtain the following equation:
\begin{equation}
 \mathbf{H}_{X}^{rel}=-\dfrac{\hbar^{2}}{2\bar{\mu}} \Big(\frac{\partial^2}{\partial \xi^2} + \frac{\partial^2}{\partial \eta^2}+ \frac{\partial^2}{\partial \zeta^2}\Big)- \dfrac{e^{2}}{ \bar{\epsilon}\sqrt{A \xi^{2}+B \eta^{2}+C \zeta^{2}}},
\end{equation}
where the anisotropy is now concentrated in the potential instead of the kinetic term. 
The anisotropy parameters A, B, and C are supposed to be real and positive. 
\begin{equation}
\frac{1}{\bar{\mu}}=\frac{\bar{\epsilon}}{3} \bigg( \frac{1}{\epsilon_{x}\mu_{x}}+\frac{1}{\epsilon_{y}\mu_{y}}+\frac{1}{\epsilon_{z}\mu_{z}}\bigg),  \bar{\epsilon}=\sqrt[3]{\epsilon_x \epsilon_y \epsilon_z}    
\end{equation}
are the average-reduced mass and the average-dielectric constant, respectively. 
We treat the problem in spherical coordinates, where $\rho =\sqrt{\xi^{2}+\eta^{2}+\zeta^{2}}$. 

When an isotropic Coulomb potential is integrated into the Hamiltonian, the resultant relative Hamiltonian, $\mathbf{H}_{X}^{rel}$, can be decomposed into two distinct components $\mathbf{H}_{X}^{rel}=\mathbf{H}_{hyd}+ \mathbf{H}_{per}$. 
The first term is the unperturbed Hamiltonian $\mathbf{H}_{hyd}$, which has a spherical symmetry and defines the effective 3D hydrogenic Hamiltonian. 
The eigenstates of $\mathbf{H}_{hyd}$ are exact and correspond to the 3D hydrogenic eigenenergies $E_{n}=-\bar{R_{y}}/n^{2}$ and eigenfunction~\cite{baldereschi1970anisotropy}
\begin{equation}
\boldsymbol{\Phi}_{n,\ell,m}(\boldsymbol{\rho},\theta,\phi)=\sqrt{\left(\dfrac{2}{na_{b}}\right)^{3} \dfrac{(n-\ell-1)!}{2n(n+\ell)!}} \, e^{-\dfrac{\boldsymbol{\rho}}{na_{b}}} \bigg(\dfrac{2 \boldsymbol{\rho}}{n a_{b}}\bigg)^{\ell} \mathbf{L}_{n-\ell-1}^{2l+1} \bigg(\dfrac{2\boldsymbol{\rho}}{n a_b}\bigg) \mathbf{Y}_{\ell}^{m}(\theta,\phi),
\end{equation}
where $\mathbf{Y}_{\ell}^{m}(\theta,\phi) $ is the spherical harmonic function. $\mathbf{L}_{\beta}^{\alpha}$ is the generalized Laguerre polynomials. $a_{b}=\frac{\bar{\epsilon} \hbar^{2}}{\bar{\mu} e^{2}}$ is the $3D$-exciton effective Bohr radius and $\bar{R_{y}}= \frac{e^{4} \bar{\mu}}{2\, \bar{\epsilon}^{2}\hbar^{2}}$ is the $3D$-effective Rydberg energy. 
In this notation, $n$, $\ell$ and $m$ are the principal, the azimuthal, and the magnetic quantum number, respectively. 
The degeneracy of the states is quantified as $n^{2}$-fold. 
The labeling convention for these states, according to their $\ell$, is as follows: $s$ for $\ell=0$, $p$ for $\ell=1$, and $d$ for $\ell=2$. 
The spherical coordinate representation of the perturbed Hamiltonian, $\mathbf{H}_{per}$, is expressed as
\iffalse
\begin{equation}
 H_{per}= \dfrac{e^{2}}{ \bar{\epsilon}\sqrt{\xi^{2}+\eta^{2}+\zeta^{2}}}-  \dfrac{e^{2}}{ \bar{\epsilon}\sqrt{A \xi^{2}+B \eta^{2}+C \zeta^{2}}}   
\end{equation}
\fi
\begin{equation}
     \mathbf{H}_{per}= \frac{e^2}{\bar{\epsilon}\boldsymbol{\rho}}\Bigg(1-\frac{1}{\sqrt{\Big(A \cos(\phi)^2+B \sin (\phi)^2\Big)\sin(\theta)^2+ C \cos(\theta)^2 }}\Bigg).
\end{equation}
\sloppy The unperturbed Hamiltonian belongs to the infinite group of isotropic space, which is invariant under all symmetry operations. 
The perturbation reduces the symmetry of the system, leads to increase the degeneracy. 
The original problem of solving $\mathbf{H}_{X}^{rel}$ is now transformed into a matrix diagonalization problem, where the corresponding components are $\bra{\mathbf{\Phi}_{n,\ell,m}}  \mathbf{H}_{X}^{rel} \ket{\mathbf{\Phi}_{n,\ell,m}}$ and the basis 
$
\mathbf{B}_{n,\ell,m}=\left\{ \Phi_{n,\ell,m}(\boldsymbol{\rho}, \theta, \phi), n \in \mathbf{N}^{*}, 0 \leq \ell \leq n-1, -\ell \leq m \leq \ell \right\}.
$
To evaluate the magnitude order of the perturbation, in the following, we compute the perturbed matrix elements $E_{\substack{n,\ell,m\\n^{\prime}, \ell^{\prime},m^{\prime}}}^{\,per}$ for different states
\begin{align}\label{eq:indila}
E_{\substack{n,\ell,m\\ n^{'},\ell^{'},m^{'}}}^{\,per}(A,B,C)= \bra{\Phi_{n,\ell,m}}  \mathbf{H}_{per} \ket{\Phi_{n^{'},\ell^{'},m^{'}}} =- \bar{R}_{y} \,G_{\substack{n,\ell\\n^{'},\ell^{'}}} \bigg(\mathbf{F}_{\substack{\ell,m\\\ell^{'},m^{'}}}(A,B,C)+\mathbf{H}_{\substack{\ell,m\\\ell^{'},m^{'}}}\bigg)
\end{align}
where the integral of the radial part is given by 
\begin{equation}
G_{\substack{n,\ell\\ n^{\prime}, \ell^{\prime}}}=\int_0^{\infty}   R_{n \ell}(\boldsymbol{\rho}) \cdot R_{n^{\prime} \ell^{\prime}}(\boldsymbol{\rho}) \cdot \rho \mathrm{~d} \rho.
\end{equation}
In particular, if $n=n^{\prime}$
\begin{align}
G_{\substack{n,\ell\\ n^{\prime}, \ell^{\prime}}}= \frac{1}{2 n^2}\left[\delta_{\ell,\ell^{\prime}} +\left(\frac{(n-\ell)(n-\ell+1)}{(n+\ell)(n+\ell-1)}\right)^{1 / 2} \delta_{\ell, \ell^{\prime}+2}\right].
\end{align}
The angular part is given by the following two matrix elements
\begin{equation}
\mathbf{H}_{\substack{\ell,m\\\ell^{'},m^{'}}}=\int_{0}^{2 \pi} \int_{0}^{\pi} \mathbf{Y}_{\ell{'}}^{m^{'}}(\theta,\phi) \mathbf{Y}_{\ell}^{m}(\theta,\phi)^{*} sin(\theta) d\theta d\phi=\delta_{\ell,\ell^{\prime}} \delta_{m,m^{\prime}},
\end{equation}

\begin{align}
\mathbf{F}_{\substack{\ell,m\\\ell^{'},m^{'}}}(A,B,C)=\int_{0}^{2 \pi} \int_{0}^{\pi}  \dfrac{\mathbf{Y}_{\ell{'}}^{m^{'}}(\theta,\phi) \mathbf{Y}_{\ell}^{m}(\theta,\phi)^{*} sin(\theta) d\theta d\phi}{\sqrt{\big(A cos^{2}(\phi)+B sin^{2}(\phi\big) sin^{2}(\Theta)+C cos^{2}(\theta)}},
\end{align}
where $\mathbf{F}_{\ell,m,\ell^{'},m^{'}}(A,B,C)$ depend on the anisotropic parameters and on the angular quantum numbers $\ell$, $m$, $\ell^{'}$ and $m^{'}$. 
The diagonalization of the Hamiltonian matrix constructed from the anisotropic Coulomb potential leads to the following eigenstates
$
\mathbf{\Psi}^{rel}_{\Tilde{n},\Tilde{\ell},\Tilde{m}}(\boldsymbol{\rho}, \theta, \phi)=\sum_{n,\ell,m}C_{n,\ell,m} \mathbf{\Phi}_{n,\ell,m}(\boldsymbol{\rho}, \theta, \phi)
$
where the coefficients $C_{n,\ell,m}$ are obtained by solving the matrix problem. 
The subscripts on the coefficient $C_{n,\ell,m}$ noted $\Tilde{n}$,$\Tilde{\ell}$, and $\Tilde{m}$ refers now to the dominant contribution of the coefficients to the excitonic function, corresponding to the coefficient of the highest weight. 
In the actual numerical calculations, the number of basis functions $\mathbf{B}_{n,\ell,m}$ should be finite. 
The basis functions are usually chosen, provided that they are the low-lying energy states of the Hamiltonian. 
We have computed eigenvalues and eigenfunctions that extend the basic set of wavefunctions until convergence is reached. 
To achieve the necessary precision for calculating the ground state energy and to ensure system convergence in our numerical computations, we considered hydrogen states with principal quantum numbers, $n$, up to $15$ along with their corresponding orbital quantum numbers. The numerical procedure becomes stable for $n = 5$. 
The solutions of the resulting Schrödinger equation of the system satisfy the eigenequation  
\begin{equation}
\Big(\mathbf{H}_X^{CM}+ \mathbf{H}_X^{\mathrm{rel}}\Big) \mathbf{\Upsilon}_{\tilde{n},\tilde{\ell},\tilde{m}}(\mathbf{R}_{CM}, \boldsymbol{\rho})=E_{\tilde{n},\tilde{\ell},\tilde{m}}^{X} 
\mathbf{\Upsilon}_{\tilde{n},\tilde{\ell},\tilde{m}}(\mathbf{R}_{CM}, \boldsymbol{\rho})
\end{equation}
are given by the eigenfunction 
\begin{equation}
\mathbf{\Upsilon}_{\tilde{n},\tilde{\ell},\tilde{m}}(\mathbf{R}_{CM}, \boldsymbol{\rho})=\mathbf{\Phi}^{CM}(\mathbf{R}_{CM}) \mathbf{\Psi}^{rel}_{\Tilde{n},\Tilde{\ell},\Tilde{m}}(\boldsymbol{\rho})
\end{equation}
and the eigenvalue  
\begin{equation}
    E^X_{\tilde{n},\tilde{\ell},\tilde{m}} (\mathbf{K})= E_g^{GW} - E_{\tilde{n},\tilde{\ell},\tilde{m}}^{B}+\sum_{i={x,y,z}}\frac{\hbar^2 \mathbf{K}_i^2}{{2\,M_{X,i}}},
\end{equation}
where $E_{\tilde{n},\tilde{\ell},\tilde{m}}^{B} =- E_{\tilde{n},\tilde{\ell},\tilde{m}}^{rel} $ is the exciton BE. 
$\mathbf{\Phi}^{CM}(\mathbf{R}_{CM})=\frac{1}{\sqrt{V}}\exp^{i\mathbf{R}_{CM} \mathbf{K}}$ is COM wavefunction. 
$E_g^{GW} = E_c^{QP}-E_v^{QP}$ and $\mathbf{K}=(K_x, K_y, K_z)$ denote the QP BG and COM wavevector, respectively. 
$V = N\Omega$ is the volume of bulk semiconductors. 
$N$ and $\Omega$ represent the total number of primitive cells in the crystal and the volume of the unit cell, respectively. 
The exciton has an energy dispersion as a function of the COM wavevector $\mathbf{K}$, which describes the translational motion of an exciton with quasi-momentum $\hbar \mathbf{K}$. 
Thus, only the exciton with $\mathbf{K}=0$ can recombine by emitting a photon, which is termed a coherent exciton. 
Consequently, excitons with $\mathbf{K}\neq 0$ cannot recombine directly to emit a photon and therefore are dark excitons. 
However, recombination emission for $\mathbf{K}\neq 0$ (in particular incoherent excitons) is possible and can be generated by a further phonon-assisted process, which we call an indirect transition. 
In the following section, since we are considering coherent optical excitation of excitons by photons, the excited excitons are predominantly those with zero COM momenta.

Once the exciton energy and wave functions have been obtained by diagonalizing the Hamiltonian, using the dipole matrix elements relevant to interband optical transitions, in this section we calculate the oscillator strength (OS) of the free neutral exciton. 
Because we work at low temperatures and under low-density excitation, we do not consider thermalization processes. 
The light-matter interaction in the Coulomb gauge under low-density excitation can be described by the Hamiltonian as follows~\cite{burstein2012confined}
\begin{equation}
H_{\mathrm{opt}}=\frac{e \mathbf{A}(\mathbf{r})\cdot \mathbf{p}}{m_0 c}
\end{equation}
where $\mathbf{p}$ is the electron momentum operator, $e$ is the elementary charge. 
In the Fock representation, the vector potential operator in the second quantization can be written as
\begin{equation}
\boldsymbol{A}_{\mathbf{q}}(\mathbf{r}, t)=\mathbf{A}_{\mathbf{q}}(\mathbf{r}) e^{-i \omega_\mathbf{q} t}+\mathbf{A}_\mathbf{q}^{+}(\mathbf{r}) e^{i \omega_\mathbf{q} t}
\end{equation}
with 
\begin{equation}
\mathbf{A}_{\mathbf{q}} (\mathbf{r})= \sqrt{\frac{2 \pi \hbar c}{n_0 \mathbf{q} V}} \times \boldsymbol{\alpha_\mathbf{q}} e^{i\mathbf{q}\mathbf{r}} a_{\mathbf{q}}
\end{equation}
 The operator $a_{\mathbf{q}}^{+}\left(a_{\mathbf{q}}\right)$ creates (annihilates) a photon with the wavevector $\mathbf{q}$. 
$\boldsymbol{\alpha_\mathbf{q}} $ denotes the photon polarization unit vector. 
The optical angular frequency is defined by $\omega_\mathbf{q}=\frac{c}{n_0}|\mathbf{q}|$, where $n_0$ is the effective optical refraction index of the crystal environment.
$c$ is the light velocity and $V$ is the normalized volume. 
The light-matter coupling can be evaluated on the basis $\left\{\left|\ldots, n_{\mathbf{q}}, \ldots\right\rangle \otimes|\zeta_X^{\tilde{n},\tilde{\ell},\tilde{m}}\rangle\right\}$. Here, $\left.\left\{\mid \ldots, n_{\mathbf{q}}, \ldots.\right\rangle\right\}$ are the electromagnetic field states in the Fock representation. 
The solution of the Schrödinger equation for an anisotropic exciton in a solid is as follows
\begin{equation}
\zeta_X^{\tilde{n},\tilde{\ell},\tilde{m}}\left(\mathbf{r_{e}}, \mathbf{r_{h}}\right)=\Upsilon_{\tilde{n},\tilde{\ell},\tilde{m}}(\mathbf{R_{CM}}, \boldsymbol{\rho})  u_{c, \mathbf{k}}\left(\mathbf{r}_{\mathbf{e}}\right) u_{v, \mathbf{k}}^*\left(\mathbf{r}_{\mathbf{h}}\right)
\end{equation}
where $u_{c, \mathbf{k}}\left(\mathbf{r}_e\right)$ and $u_{v, \mathbf{k}} \left(\mathbf{r}_{\mathbf{h}}\right)$ are the Bloch functions of the valence (v) and conduction (c) bands, respectively. 
$\Upsilon_{\tilde{n},\tilde{\ell},\tilde{m}}(\mathbf{R_{CM}}, \boldsymbol{\rho})$ is the envelope wavefunction.  
In our work, the initial state consists of an excitonic state without a photon $|i\rangle=\left|\zeta_X^{\tilde{n},\tilde{\ell},\tilde{m}}\right\rangle \otimes\left|0_{\mathbf{q}}\right\rangle$, while the final state consists of the crystal ground state $|\emptyset\rangle$ with one photon $|f\rangle=|\emptyset\rangle \otimes$ $\left|1_{\mathbf{q}}\right\rangle$. 
The OME can be written as
\begin{equation}\label{eq:ome}
\Big(\langle 1_{\mathbf{q}} |\otimes\langle  \emptyset|\Big) 
\mathbf{A}^{+}_\mathbf{q}(\mathbf{r}) \cdot 
\mathbf{p}
\Big(|\zeta_X^{\tilde{n},\tilde{\ell},\tilde{m}} \rangle \otimes\left|0_{\mathbf{q}}\right\rangle \Big)= \sqrt{\left(\frac{2 \pi \hbar c}{n_0 \mathbf{q} V}\right)} \bar{F}_{\mathbf{q},c,v}^{\tilde{n},\tilde{\ell},\tilde{m}}
\end{equation}
where
$
\bar{F}_{\mathbf{q},c,v}^{\tilde{n},\tilde{\ell},\tilde{m}}= \boldsymbol{\alpha}_{\mathbf{q}}\cdot \left\langle \emptyset \left|e^{i \mathbf{q} \cdot \mathbf{r}} \mathbf{p}\right| \zeta_X^{\tilde{n},\tilde{\ell},\tilde{m}}\right\rangle
$
is the OME between the crystal ground state $\ket{\emptyset}$ and the excited states $\ket{\zeta_X^{\tilde{n},\tilde{\ell},\tilde{m}}}$ corresponding to the direct exciton state in bulk GeS, which is derived from the electron-photon coupling. 
The linear optical properties of three-dimensional excitons have been investigated theoretically. 
In fact, the OS of the optical interband transition for exciton states is defined by~\cite{stier1999electronic}
\begin{equation}
    f^{\alpha_\mathbf{q}}_{\tilde{n},\tilde{\ell},\tilde{m}}=
    \frac{2}{m_0 \hbar \omega_0}\left|\bar{F}_{\mathbf{q},c,v}^{\tilde{n},\tilde{\ell},\tilde{m}}\right|^{2}
\end{equation}
here, the optical transition frequency is $\omega_0=\left|\omega_i-\omega_f\right|=\omega^X$, with $E^X_{\tilde{n},\tilde{\ell},\tilde{m}}=\hbar \omega^X$. 
To calculate the OME obtained considering the interaction with the first-order electromagnetic field, it is convenient to work with the Fourier transforms of the envelope wavefunction 
\begin{align}
\zeta_X^{\tilde{n},\tilde{\ell},\tilde{m}} (\mathbf{R_{CM}},\boldsymbol{\rho}) =\dfrac{1}{(2\pi)^{6}}\int \int d\mathbf{K} d\mathbf{k} \,
e^{i \big(\mathbf{K}.\mathbf{R_{CM}}+\mathbf{k}.\boldsymbol{\rho}\big)}  \Bar{\Phi}^{CM}(\mathbf{K}) \Bar{\Psi}_{\tilde{n},\tilde{\ell},\tilde{m}}^{rel}(\mathbf{k})
u_{c,\mathbf{k}} (\mathbf{r_{e}})u_{v,\mathbf{k}}^{*}(\mathbf{r_{h}})
\end{align}
where the COM and relative wavefunction, in momentum-space, reads
\begin{subequations}
\begin{eqnarray}
\Bar{\Phi}^{CM}(\mathbf{K})=\int e^{i\mathbf{K}.\mathbf{R_{CM}}} \Phi^{CM}(\textbf{R}_{CM}) d\mathbf{R_{CM}}\\
\Bar{\Psi}_{\tilde{n},\tilde{\ell},\tilde{m}}^{rel}(\mathbf{k})=\int e^{i\mathbf{k}.\boldsymbol{\rho}} \Psi_{\tilde{n},\tilde{\ell},\tilde{m}}^{rel}(\boldsymbol{\boldsymbol{\rho}}) d\boldsymbol{\rho}
\end{eqnarray}
\end{subequations}
here, $\mathbf{K}=\mathbf{k}_e+\mathbf{k}_{\mathbf{h}}$ and $\mathbf{k}=\frac{m_{i}^{h} \mathbf{k}_e-m_{i}^{e} \mathbf{k}_h}{M_{X,i}}$ are the COM and the relative wavevectors, respectively. 
By considering the following change of variables, $\left(\mathbf{K}, \mathbf{k}\right) \Rightarrow  \left(\mathbf{k}_e, \mathbf{k}_h\right)$, with  $d \mathbf{K} d \mathbf{k}=$ $d \mathbf{k}_e d \mathbf{k}_h$,  the envelope wavefunction, $\zeta_X^{\tilde{n},\tilde{\ell},\tilde{m}}$, can be written as follows
\begin{align}\label{eq:wf}
\zeta_X^{\tilde{n},\tilde{\ell},\tilde{m}}\left(\mathbf{r_{e}}, \mathbf{r_{h}}\right)= \frac{1}{(2 \pi)^6} \int \int d \mathbf{k}_e d \mathbf{k}_h e^{i\left( \mathbf{k_e} \cdot \mathbf{r_e}+ \mathbf{k_h} \cdot \mathbf{r_h}\right)}
\Bar{\Phi}\left(\mathbf{K}\right) \Bar{\Psi}_{\tilde{n},\tilde{\ell},\tilde{m}}^{r e l}(\mathbf{k}) u_{c,\mathbf{k}} \left(\mathbf{r}_e\right) u_{v,\mathbf{k}}^{*}\left(\mathbf{r}_h\right)
\end{align}
By converting the integral into a sum ($\frac{1}{V} \sum_{\mathbf{k} \in 1 B Z} \approx \int_{\mathbf{k}  \in 1 B Z} \frac{d \mathbf{k} }{(2 \pi)^3}$), we can rewrite 
$\zeta_X^{\tilde{n},\tilde{\ell},\tilde{m}}\left(\mathbf{r_{e}}, \mathbf{r_{h}}\right)$ as follows
\begin{align}
\zeta_X^{\tilde{n},\tilde{\ell},\tilde{m}}\left(\mathbf{r_{e}}, \mathbf{r_{h}}\right)=& \frac{1}{V^{2}} \sum_{\mathbf{k}_e\in BZ} \sum_{\mathbf{k}_h\in BZ}  e^{i (\mathbf{k}_e \cdot  \mathbf{r}_e+ \mathbf{k}_h \cdot  \mathbf{r}_h)}  \bar{\Phi}^{CM}(\mathbf{K}) \bar{\Psi}_{\tilde{n},\tilde{\ell},\tilde{m}}^{rel}(\mathbf{k}) u_{c, \mathbf{k}}\left(\mathbf{r}_{\mathbf{e}}\right)  \times u_{v, \mathbf{k}}^*\left(\mathbf{r}_h\right)
\end{align}
The Bloch functions $u_{c, \mathbf{k}}\left(\mathbf{r}_e\right)$, $ u_{v, \mathbf{k}}^*\left(\mathbf{r}_h\right)= u_{h, -\mathbf{k}} \left(\mathbf{r}_h\right)$ vary slowly when $\mathbf{k}_e$  ($\mathbf{k}_h$) vary around the $\mathbf{k}$-points. We can therefore write
\begin{align}
\zeta_X^{\tilde{n},\tilde{\ell},\tilde{m}}\left(\mathbf{r_{e}}, \mathbf{r_{h}}\right)=& \frac{1}{V} \sum_{\mathbf{k}_e, \mathbf{k}_h \in BZ} \bar{\Phi}^{CM}(\mathbf{K}) \bar{\Psi}_{\tilde{n},\tilde{\ell},\tilde{m}}^{rel}(\mathbf{k}) \Upsilon_{c, \mathbf{k}_e}\left(\mathbf{r}_{\mathbf{e}}\right)  \times \Upsilon_{v, \mathbf{k}_h}^*\left(\mathbf{r}_h\right)
\end{align}
where the electron and hole wavefunctions reads
\begin{subequations}
\begin{eqnarray}
\Upsilon_{c, \mathbf{k}_e}\left(\mathbf{r}_e\right)=\frac{1}{\sqrt{V}} e^{i  \mathbf{k}_e \cdot \mathbf{r}_e} u_{c,  \mathbf{k}_e}\left(\mathbf{r}_e\right)\\
\Upsilon_{v,  \mathbf{k}_v}\left(\mathbf{r}_h\right)=\frac{1}{\sqrt{V}} e^{i  \mathbf{k}_v \cdot  \mathbf{r}_h} u_{v,  \mathbf{k}_v}\left(\mathbf{r}_h\right).
\end{eqnarray}
\end{subequations}
Here $\mathbf{k}_h=-\mathbf{k}_v$. 
In fact, the hole state is related to the valence electron state by $\Upsilon_{\mathbf{k}_h}(\mathbf{r})=\mathcal{K} \Upsilon_{\mathbf{k}_v}(\mathbf{r})$ with $\mathcal{K}$ being the time-reversal operator.~\cite{bir1974symmetry} 
By substituting Eq.~\ref{eq:wf} in Eq.~\ref{eq:ome}, and since $\mathbf{k}_e, \mathbf{k}_h \in \mathrm{BZ}$, the OME reads
\begin{align}\label{fnlm_}
\bar{F}_{\mathbf{q},c,v}^{\tilde{n},\tilde{\ell},\tilde{m}}=\frac{1}{V} \sum_{\mathbf{k}_e, \mathbf{k}_h} \bar{\Phi}^{CM}(\mathbf{K}) \bar{\Psi}_{\tilde{n},\tilde{\ell},\tilde{m}}^{rel}(\mathbf{k}) \left\langle u_{c, \mathbf{k}_e}\left|\alpha_\mathbf{q} \cdot \mathbf{p}\right| u_{v, \mathbf{k}_v}\right\rangle  \delta_{\mathbf{k}_e, \mathbf{k}_v+\mathbf{q}}.
\end{align}
The COM and relative wavevectors can be rewritten as
\begin{equation}\label{kekh}
 \left\{
\begin{array}{rcr}
\mathbf{K} =\mathbf{k}_e+\mathbf{k}_h=\mathbf{k}_e-\mathbf{k}_v =\mathbf{q}\\ 
\mathbf{k}=\mathbf{k}_e-\frac{m_{eff}^{e, i}}{M_X^{i}} \mathbf{q}
\end{array}
\right.
\end{equation}

Using Eq.~\ref{kekh} and Eq.~\ref{fnlm_}, we can rewrite $\bar{F}_{\mathbf{q},c,v}^{\tilde{n},\tilde{\ell},\tilde{m}}$ as follows

\begin{equation}
\bar{F}_{\mathbf{q},c,v}^{\tilde{n},\tilde{\ell},\tilde{m}}=\frac{1}{V} \sum_{\mathbf{k}_e} 
 \bar{\Phi}^{C M}(\mathbf{q}) \bar{\Psi}_{\tilde{n},\tilde{\ell},\tilde{m}}^{rel}\left(\mathbf{k}_e\right) \alpha_\mathbf{q} \cdot\left\langle u_{c, \mathbf{k}}|\mathbf{p}| u_{v, \mathbf{k}}\right\rangle
\end{equation}
At the scale of $\mathbf{q}$,  $\bar{\Phi}^{C M}(\mathbf{q})$ is slowly varying, so  $\bar{\Phi}^{C M}(\mathbf{q})=\bar{\Phi}^{C M}(0)$, hence,

\begin{equation}
\bar{F}_{\mathbf{q},c,v}^{\tilde{n},\tilde{\ell},\tilde{m}}=\frac{1}{V} \bar{\Phi}^{C M}(\mathbf{K}=0) \sum_{\mathbf{k}_e} \bar{\Psi}_{\tilde{n},\tilde{\ell},\tilde{m}}^{rel}\left(\mathbf{k}_e\right) \alpha_\mathbf{q} \cdot \left\langle u_{c, \mathbf{k}}|\mathbf{p}| u_{v, \mathbf{k}}\right\rangle
\end{equation}
with
\begin{subequations}
\begin{eqnarray}
\frac{1}{V}\sum_{\mathbf{k}_e} \bar{\Psi}_{\tilde{n},\tilde{\ell},\tilde{m}}^{rel}\left(\mathbf{k}_e\right)=\frac{1}{(2 \pi)^3} \int d \mathbf{k}_e \bar{\Psi}_{\tilde{n},\tilde{\ell},\tilde{m}}\left(\mathbf{k}_e\right)=V \Psi_{\tilde{n},\tilde{\ell},\tilde{m}}^{rel}(\rho=0)\\
\Bar{\Phi}^{CM}(\mathbf{K} =0)=\int   d\mathbf{R_{CM}} \,\Phi^{CM}(\textbf{R}_{CM})
\end{eqnarray}
\end{subequations}

In the case of bulk GeS in which we assume the exciton COM motion is free and can be described by a plane wavefunction, the OS is rewritten as follows
\begin{equation}
f^{\alpha_\mathbf{q}}_{\tilde{n},\tilde{\ell},\tilde{m}}= \frac{2V}{m_0 E^X_{\tilde{n},\tilde{\ell},\tilde{m}}} \left|\Psi_{\tilde{n},\tilde{\ell},\tilde{m}}^{rel}(\rho=0)\right|^2 
\times\left|\left\langle u_{c, \mathbf{k}}\left|\boldsymbol{\alpha}_{\mathbf{q}} \cdot \mathbf{p}\right| u_{v, \mathbf{k}}\right\rangle\right|^2.
\end{equation}

\end{document}